\documentclass[apj]{emulateapj}
\long\def\***#1{{\scshape ***#1***}}

\singlespace



\shortauthors{SONGAILA \& COWIE}
\shorttitle{THE EVOLUTION OF LYMAN LIMIT ABSORPTION SYSTEMS TO REDSHIFT SIX}

\begin{document}
\singlespace
\title{The Evolution of Lyman Limit Absorption Systems to Redshift Six\altaffilmark{*}}
\author{
Antoinette Songaila,\altaffilmark{1}\email{acowie@ifa.hawaii.edu}
Lennox L. Cowie\altaffilmark{1}\email{cowie@ifa.hawaii.edu}
}

\affil{Institute for Astronomy, University of Hawaii,
  2680 Woodlawn Drive, Honolulu, HI 96822}

\altaffiltext{*}{Based in part on data obtained from the Multimission Archive at the Space Telescope Science Institute (MAST).  STScI is operated by the Association of Universities for Research in Astronomy, Inc., under NASA contract NAS5--26555.  Support for MAST for non-{\it HST\/} data is provided by the NASA Office of Space Science via grant NAG5--7584 and by other grants and contracts.}
  
\altaffiltext{1}{Visiting Astronomer, W. M. Keck Observatory, which is jointly
  operated by the California Institute of Technology, the University of
  California, and the National Aeronautics and Space Administration}

\slugcomment{To be published in {\it Astrophysical Journal}}

\begin{abstract}
We have measured the redshift evolution of the density of Lyman limit systems (LLS) in the intergalactic medium over the redshift range $0<z<6$.  We have used two new quasar samples to (1) improve coverage at $z \sim 1$, with GALEX grism spectrograph observations of 50 quasars with $0.8 < z_{em} < 1.3$, and (2) extend coverage to $z \sim  6$, with Keck ESI spectra of 25 quasars with $4.17 < z_{em} < 5.99$.   Using these samples together with published data, we find that the number density of LLS per unit redshift, $n(z)$, can be well fit by a simple evolution of the form $n(z) = n_{3.5}[(1+z)/4.5]^{\gamma}$\ 
with $n_{3.5} = 2.80\pm 0.33$\ and $\gamma = 1.94^{+0.36}_{-0.32}$\ for the entire range $0<z<6$.  We have also reanalyzed the evolution of damped Lyman alpha systems (DLAs) in the redshift range $4<z<5$\ using our high-redshift quasar sample.  We find a total of 17 DLAs and sub-DLAs, which we have analyzed in combination with published data. The DLAs with $\log {\rm HI\ column\ density} > 20.3$ show the same redshift evolution as the LLS. 
When combined with previous results, our DLA sample is also consistent with a constant $\Omega_{\rm DLA} = 9 \times 10^{-4}$\ from $z = 2$\ to $z = 5$.  We have used the LLS number density evolution to compute the evolution in the mean free path of ionizing photons.  We find a smooth evolution to $z \sim 6$, very similar in shape to that of Madau, Haardt \& Rees (1999) but about a factor of two higher.  Recent theoretical models roughly match to the $z<6$ data but diverge from the measured power law at $z>6$ in different ways, cautioning against extrapolating the fit to the mean free path outside the measured redshift range.
\end{abstract}


\newpage


\section{Introduction}
\label{secintro}

Intergalactic absorption lines that are optically thick
to radiation at the Lyman continuum edge are generally referred to as
Lyman limit systems (LLS). The LLS are perhaps the single most
valuable tracer of the evolution of the intergalactic medium since
they provide a nearly direct measure of the mean free path of ionizing
radiation at high redshifts. They are therefore a key ingredient in
estimating the required sources of ionizing radiation needed to
maintain the ionization in the intergalactic gas (Madau, Haardt \& Rees, 1999 [MHR]).  The expected rapid
positive evolution in the number of such systems as the intergalactic
gas becomes more opaque means that they could, in principle, be an
indicator of the approach of the epoch of reionization.

  LLS are also the easiest systems to identify in quasar spectra since
the abrupt break in the spectrum produced by the absorption systems
can easily be seen in relatively noisy and low resolution
observations.  Since Tytler's pioneering work on the cosmological
evolution of the LLS (Tytler 1982), numerous groups have measured LLS
evolution over a range of redshifts extending from low redshift
systems, measured with HST, to $z \sim 5$ (Bechtold et al. 2002;
Sargent, Steidel \& Boksenberg 1989 [SSB]; Lanzetta 1991; Storrie-Lombardi
et al. 1994; Stengler-Larrea et al. 1995 [SL95]; Lanzetta et al. 1995; 
P\'eroux 2001; P\'eroux et al.\ 2003 [P03]). The simplicity of identifying LLS means that we can
extend this type of measurement yet further, out to $z \sim 6$, and
this is the primary goal of the present paper. Indeed LLS are the {\it
only\/} neutral hydrogen line systems that we can readily trace beyond
$z\sim 5$.  At these high redshifts, the lower column density
Ly$\alpha$ forest is too blended and the quasar continuum too
ill-defined for these systems to be studied, and we can no longer determine the radiation damping wings that allow us to identify the stronger systems such as the damped Ly$\alpha$ aborbers (DLAs).
The LLS are therefore the one way we have to normalize the
neutral hydrogen column density function at these redshifts, a crucial
input to any modelling of the intergalactic medium.


\begin{deluxetable*}{lcccccccc}
\tabletypesize{\small}
\tablewidth{500pt}
\tablenum{1}
\tablecaption{High redshift quasars\label{tbl:ztable}}
\tablehead{
\colhead{QSO} & \colhead{$z_{\rm em}$}  & 
\colhead{Exp (hr)} & \colhead{$z_{\rm LLS}$}  & 
\colhead{$\tau_{\rm LLS}$} &
\colhead{$z_{\rm metal}$} &
\colhead{$z_{\rm DLA}$} &
\colhead{$\log N_{\rm DLA}$}  &
\colhead{Ref.$^a$} 
}
\startdata

FIRS0747+2739   & 4.17000 & 4.42 & \nodata & \nodata & \nodata & 3.4228 & 20.80 & 20.85 (PW)\\
                &&&&&                                 & 3.9012 & 20.50 & 20.50 (PW)\\
PSS0209+0517    & 4.17000 & 3.00 & 3.99060 & 4.6 & 3.98828 & 3.6661 & 20.48 & 20.45 (PW)\\
                &&&&&                                & 3.8635 & 20.43 & 20.55 (PW)\\
PSS0926+3055    & 4.19000 & 3.00 & \nodata & \nodata & \nodata & \nodata & \nodata & \nodata\\
BRJ0426$-$2202    & 4.32000 & 3.00 & \nodata & \nodata & \nodata & \nodata & \nodata & \nodata\\
BR0334$-$1612     & 4.36300 & 3.00 & 4.23037 & 2.8 & 4.23454 & 3.556 & 21.15 & 21.0 (P03)\\
PSS0747+4434    & 4.39000 & 3.00 & 4.29520 & 1.1 & 4.27546 & 3.762 & 20.15 & 20.3 (P03)\\
                &&&&&                                 & 4.020 & 21.0 & 20.6 (P03)\\
BRI0952$-$0115    & 4.42000 & 4.25 & 4.37745 & 1.1 & 4.37813 & 4.0239 & 20.60 & 20.55 (PW)\\
PC0953+4749     & 4.46000 & 2.00 & 4.32691 & $> 4.2$ & 4.34002 & 3.405 & 21.05 & 21.15 (PW)\\
                &&&&&                                & 3.889 & 21.03 & 21.20 (PW)\\
                &&&&&                                 & 4.2444 & 20.95 & 20.90 (PW)\\
BR1033$-$0327     & 4.51000 & 2.00 & 4.18764  & $>5.2$ & 4.17502 & 4.174 & 20.08 & \nodata\\
BR2237$-$0607     & 4.55000 & 11.56 & 4.25832 & $> 5.7$ & 4.24740 & 4.080 & 20.53 & 20.52 (PW)\\
PSS1347+4956    & 4.56500 & 1.20 & 4.28753 & 1.9 & 4.27248 & \nodata & \nodata & \nodata\\
BR0353$-$3820     & 4.58000 & 2.50 & 4.36619 & 1.5 & 4.35953 & \nodata & \nodata & \nodata\\
BR1202$-$0725     & 4.61000 & 3.00 & 4.48567 & $> 6.2$ & 4.48012 & 4.383 & 20.62 & 20.60 (PW)\\
SDSS2200+0017   & 4.78000 & 6.83 & 4.62166 & $> 4.4$ & 4.65974 & \nodata & \nodata & \nodata\\
SDSS0206+1216   & 4.81000 & 3.00 & 4.34606 & $> 4.3$ & 4.32516 & 4.3255 & 20.78 & \nodata\\
SDSS1737+5828   & 4.85000 & 9.00 & 4.74902 & $> 5.0$ & 4.74314 & 4.7415 & 20.70 & 20.65 (PW)\\
SDSS0211$-$0009   & 4.90000 & 6.00 & 4.67670 & $> 4.1$ & 4.64357 & 4.644 & 20.18 & \nodata\\
SDSS0338+0021   & 5.01000 & 7.25 & 4.97626 & 1.7 & 4.97778 & \nodata & \nodata & \nodata\\
SDSS1204$-$0021   & 5.07000 & 3.00 & \nodata & \nodata & \nodata & \nodata & \nodata & \nodata\\
SDSS0231$-$0728   & 5.42000 & 4.33 & 5.40306 & $> 4.6$ & 5.38874 & \nodata & \nodata & \nodata\\
SDSS1044$-$0125   & 5.74500 & 5.75 & 5.64344 & $> 2.6$ & \nodata & \nodata & \nodata & \nodata\\
SDSS0002+2550   & 5.80000 & 5.75 & 5.72026 & $> 3.6$ & \nodata & \nodata & \nodata & \nodata\\
SDSS0836+0054   & 5.82000 & 11.83 & 5.47923 & $> 2.5$ & \nodata & \nodata & \nodata & \nodata\\
SDSS1411+1217   & 5.93000 & 6.50 & 5.89278 & 2.3 & 5.89602 & \nodata & \nodata & \nodata\\
SDSS1306+0356   & 5.99000 & 9.75 & 5.97264 & 2.6 & \nodata & \nodata & \nodata & \nodata\\

\enddata

\tablenotetext{a\ }{Previously published values of $\log N_{\rm DLA}$.  P03:  P\'eroux et al.\ (2003); PW:  Prochaska et al.\ (2007)}

\end{deluxetable*}

In the present paper we combine data on the LLS from a high redshift 
($z=4-6$) sample with a new low redshift sample ($z<1$) and with pre-existing
observations at intermediate redshifts to compute the evolution of the
LLS over the redshift interval $z=0-6$. The high redshift sample is based
on high signal to noise obervations of a sample of 25 quasars (section 2.1).
The new $z<1$ sample is based on 
a large sample of $z\sim1$ quasars selected from
grism spectra taken with the GALEX satellite.  We use this to reinvestigate the
low redshift density of LLS since the GALEX sample should be more
homogeneous and unbiased than the previous HST and IUE samples
used for this purpose. We find a lower value
for the LLS density at this redshift with the new sample.  
Remarkably we find that the evolution over the full reshift range can
be fit by a single power law of the form  $2.80[(1+z)/4.5]^{1.94}$\  and that this extends smoothly
to near $z=6$, beyond which the Lyman forest portions of the quasar spectra become too opaque to continue the measurement.

We have also measured the DLAs ($\log N > 20.3$)
and sub-DLAs (here defined as $20.0 < \log N < 20.3$) at $z = 4 - 5$ in our high-redshift quasar sample, and used these in combination with previous work to test if the form
of the neutral hydrogen column density distribution is changing as a function
of redshift. The combined data
are consistent with an invariant form of the function. 

We also compute the neutral hydrogen density in the range 
$0<z<5$\ and find that $\Omega_{\rm DLA}$\ is invariant over the redshift range $2<z<5$.  The LLS density evolution implies that this result may extend to $z=6$\ if the neutral hydrogen column density distribution function remains invariant.

We use the form of the LLS evolution to compute the mean free path for ionizing photons, showing
that the present observations decrease this compared with recent analyses (e.g. Faucher-Gigu\`ere et al.\ 2008). Finally we compare the evolution
of the LLS and the mean free path to recent theoretical models. We assume $H_0 = 70$, $\Omega_m = 0.3$\ and $\Omega_{\Lambda} = 0.7$\ throughout.


\tablenum{2}
\begin{deluxetable*}{lclcc}
\tabletypesize{\scriptsize}
\tablecaption{GALEX quasars\label{tbl:gtable}}
\tablehead{
\colhead{QSO} & \colhead{R.A. (J2000.0)}  & 
\colhead{Decl. (J2000.0)} & \colhead{$z_{\rm em}$}  & 
\colhead{$z_{\rm LLS}$}
}
\startdata

GALEX1713+6001  &  258.29000  &  + 60.024200  &  0.8179  &  \nodata  \\
GALEX1238+6143  &  189.70300  &  + 61.73180  &  0.8323  &  \nodata  \\
GALEX1438+3446  &  219.54100  &  + 34.76800  &  0.8331  &  \nodata  \\
GALEX1050+5819  &  162.50100  &  + 58.31790  &  0.8350  &  \nodata  \\
GALEX1421+5241  &  215.30099  &  + 52.69649  &  0.8438  &  \nodata  \\
GALEX1436+3539  &  219.17599  &  + 35.65800  &  0.8450  &  \nodata  \\
GALEX0040-4403  &   10.17800  &  $- 44.061401$  &  0.8524  &  0.7784  \\
GALEX0040-4340  &   10.11710  &  $- 43.67160$  &  0.8726  &  \nodata  \\
GALEX1434+3529  &  218.68699  &  + 35.48809  &  0.8729  &  \nodata  \\
GALEX1437+3508  &  219.25599  &  + 35.13760  &  0.8800  &  \nodata  \\
GALEX1237+6211  &  189.49800  &  + 62.18399  &  0.9129  &  \nodata  \\
GALEX1051+5838  &  162.77200  &  + 58.64239  &  0.9150  &  \nodata  \\
GALEX1434+3456  &  218.69799  &  + 34.93420  &  0.9269  &  0.8864  \\
GALEX0038-4352  &    9.53954  &  $- 43.87810$  &  0.9359  &  \nodata  \\
GALEX1052+5851  &  163.072998  &  + 58.85630  &  0.9700  &  \nodata  \\
GALEX1001+0200  &  150.46299  &  +  2.0090301  &  0.9705  &  \nodata  \\
GALEX1417+5230  &  214.39999  &  + 52.50840  &  0.9877  &  \nodata  \\
GALEX0333-2817  &   53.41400  &  $- 28.29000$  &  0.9906  &  \nodata  \\
GALEX1723+5951  &  260.79299  &  + 59.85160  &  0.9964  &  0.7935  \\
GALEX1238+6202  &  189.56700  &  + 62.0356  &  0.9976  &  \nodata  \\
GALEX1423+5246  &  215.76699  &  + 52.77500  &  0.9993  &  \nodata  \\
GALEX1712+6005  &  258.14498  &  + 60.091800  &  0.9993  &  \nodata  \\
GALEX1237+6217  &  189.27900  &  + 62.28400  &  1.0190  &  \nodata  \\
GALEX0334-2743  &   53.53450  &  $- 27.72710$  &  1.0281  &  \nodata  \\
GALEX0332-2740  &   53.11040  &  $- 27.67630$  &  1.0396  &  \nodata  \\
GALEX1717+5947  &  259.38699  &  + 59.79650  &  1.0482  &  \nodata  \\
GALEX1420+5216  &  215.035003  &  + 52.27980  &  1.0600  &  \nodata  \\
GALEX1719+6013  &  259.96398  &  + 60.22259  &  1.0600  &  \nodata  \\
GALEX1436+3525  &  219.10299  &  + 35.42699  &  1.0648  &  \nodata  \\
GALEX1436+3453  &  219.13200  &  + 34.89020  &  1.0676  &  \nodata  \\
GALEX1724+5921  &  261.21398  &  + 59.35920  &  1.0770  &  0.7770  \\
GALEX1436+3537  &  219.074005  &  + 35.62390  &  1.0776  &  0.8428  \\
GALEX1240+6223  &  190.13900  &  + 62.39090  &  1.1115  &  \nodata  \\
GALEX1437+3442  &  219.43200  &  + 34.71490  &  1.1146  &  \nodata  \\
GALEX1435+3457  &  218.78199  &  + 34.95679  &  1.1174  &  \nodata  \\
GALEX1418+5223  &  214.66000  &  + 52.39989  &  1.1202  &  1.1100   \\
GALEX1234+6233  &  188.50500  &  + 62.55410  &  1.1231  &  \nodata  \\
GALEX0333-2725  &   53.35220  &  $- 27.43070$  &  1.1400  &  \nodata  \\
GALEX1435+3504  &  218.86799  &  + 35.075801  &  1.1472  &  \nodata  \\
GALEX1418+5252  &  214.50700  &  + 52.86690  &  1.1600  &  \nodata  \\
GALEX0334-2745  &   53.57030  &  $- 27.75120$  &  1.1634  &  0.8325  \\
GALEX1720+5932  &  260.13501  &  + 59.53530  &  1.1691  &  \nodata  \\
GALEX1002+0230  &  150.54499  &  +  2.50739  &  1.1691  &  0.7651  \\
GALEX1053+5855  &  163.29200  &  + 58.92580  &  1.1800  &  \nodata  \\
GALEX1054+5852  &  163.58700  &  + 58.87179  &  1.1835  &  \nodata  \\
GALEX1239+6206  &  189.88099  &  + 62.10520  &  1.2008  &  \nodata  \\
GALEX1712+6007  &  258.13198  &  + 60.12210  &  1.2037  &  0.8201   \\
GALEX1416+5259  &  214.20100  &  + 52.98410  &  1.2325  &  \nodata  \\
GALEX0040-4409  &   10.24390  &  $- 44.16410$  &  1.2325  &  \nodata  \\
GALEX1419+5232  &  214.94299  &  + 52.54639  &  1.2469  &  \nodata  
\enddata

\end{deluxetable*}


\section{Data}
\label{data}

\subsection{High redshift quasars}
\label{highz}

We used a data set (Table~\ref{tbl:ztable}) consisting of spectra of 25 quasars
selected to span the redshift range
from  $4.17 < z_{\rm em} < 5.99$. The sample was chosen from among the
brightest quasars at these redshifts based only on accessibility at
the time of the observations.  One additional BALQSO was observed in the program but has been eliminated from the present sample.  
While color-based quasar selection can introduce a bias toward quasars with strong LLS at lower redshifts, the L$\alpha$\ forest is sufficiently strong at these redshifts to ensure that the quasars would be picked out irrespective of the presence of an LLS (see P\'eroux 2001).
The spectra were taken on the Keck II telescope with the ESI
instrument in echellette mode with exposure times ranging from just
under 2 hours to just under 12 hours.  The resolution in this configuration
is comparatively low, $\sim 5300$\ for the $0.75^{\prime\prime}$\ slit
width used, but the red sensitivity is high and the wavelength
coverage is complete from $4000~{\rm \AA}$\ to $10,000~{\rm \AA}$.  At
the red wavelengths, precision sky subtraction is required since the
sky lines can be more than two orders of magnitude brighter than the
quasars. In order to deal with this issue, individual half-hour
exposures were made, stepped along the slit, and the median was used
to provide the primary sky subtraction. The frames were then
registered, filtered to remove cosmic rays and artifacts, and then
added. At this point a second linear fit to the slit profile in the
vicinity of the quasar was used to remove any small residual sky. This procedure gives an extremely accurate sky subtraction and a precise zero level in the spectra.  The
observations were made at the parallactic angle and flux calibrated
using observations of white dwarf standards scattered through the
night. These standards were also used to remove telluric absorption
features in the spectra, including the various atmospheric bands. The
final extractions of the spectra were made using a weighting provided
by the profile of the white dwarf standards. Wavelength calibrations
were obtained using third-order fits to CuAr and Xe lamp observations
at the begining and end of each night, and the absolute wavelength
scale was then obtained by registering these wavelength solutions to
the night sky lines. The wavelengths and redshifts are given in the
vacuum heliocentric frame.  A sample of the final spectra of the 25 quasars is shown in Figure~\ref{fig:spectra1} and all of the spectra are given in the Appendix.

\begin{figure*}[h]
\includegraphics[width=6.5in,angle=90,scale=0.8]{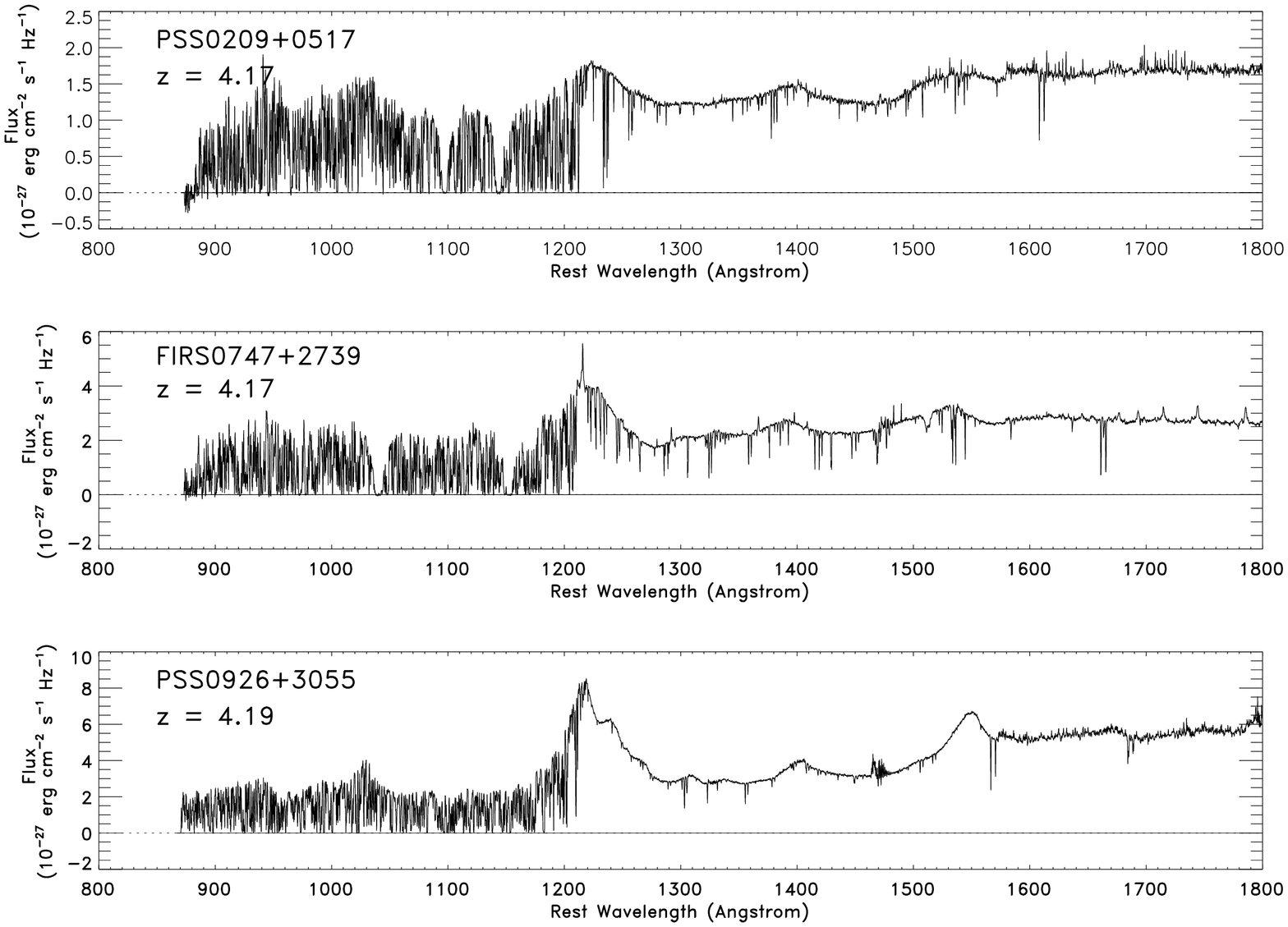}
  \caption{ESI spectra of the QSOs used in this investigation.
\label{fig:spectra1}
}
\figurenum{1}
\end{figure*}

\begin{figure}[h]
\figurenum{2}
     \includegraphics[width=3in,angle=90,scale=0.9]{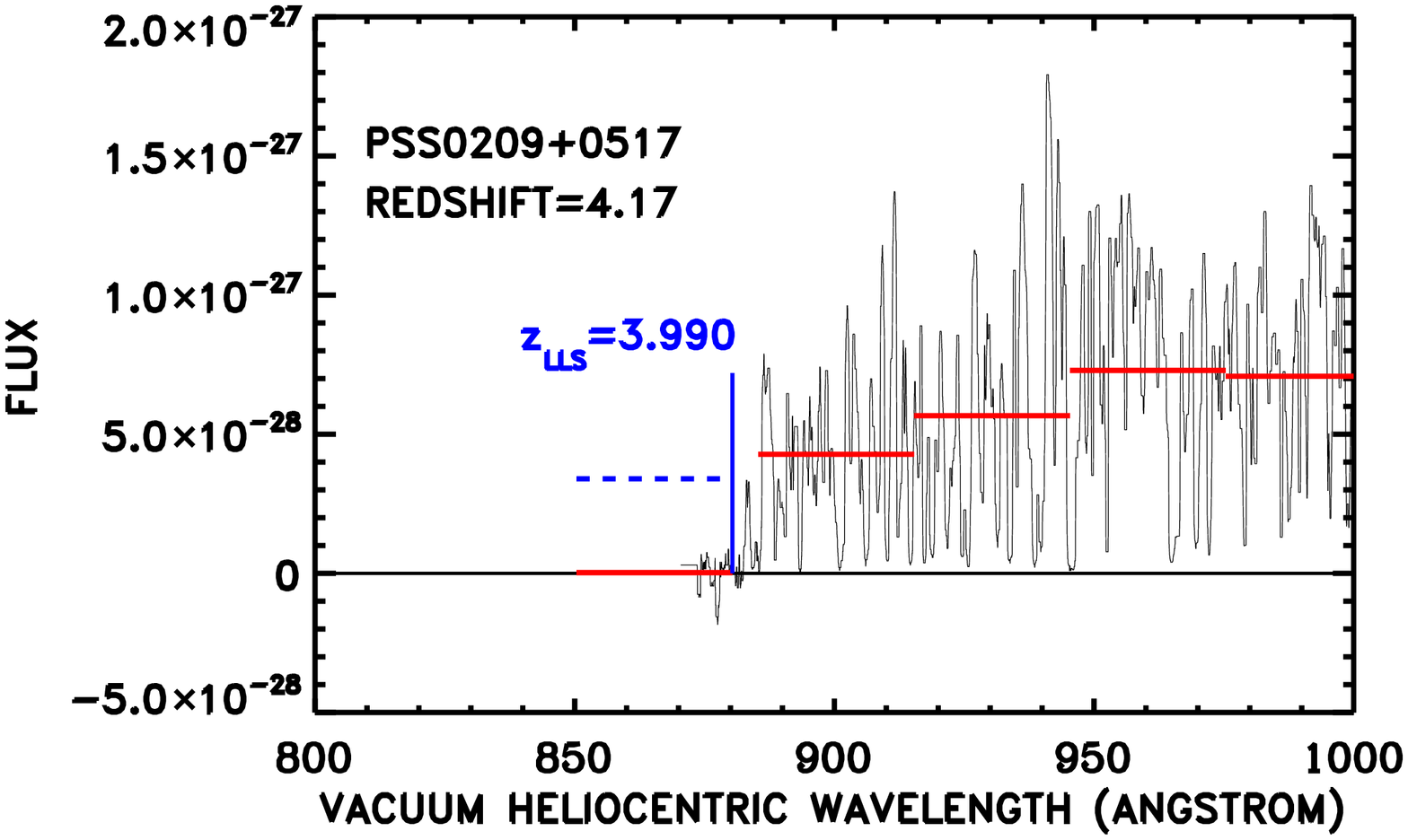}
  \includegraphics[width=3in,angle=90,scale=0.9]{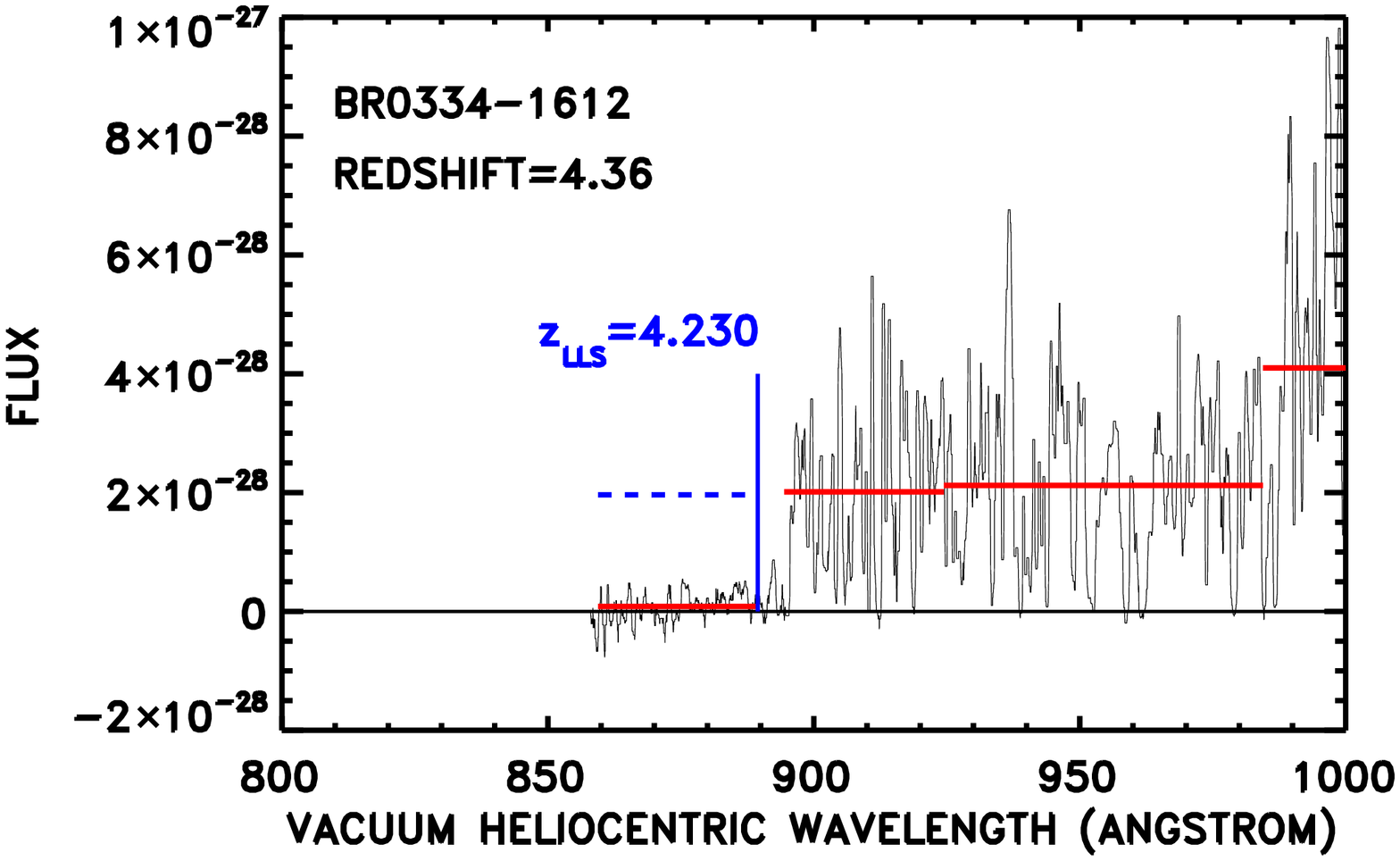}
\caption{Lyman limit systems with $\tau > 1$\ in quasars at intermediate redshift.  The spectra are plotted in the rest frame of the quasar.  The blue vertical line marks the wavelength $\lambda_{\rm LLS}$\ of the LLS and the redshift $z_{\rm LLS}$\ is marked alongside.  The red horizontal bars show the average flux in $30~\rm\AA$\ bins below $\lambda_{\rm LLS}$\ and to redward of $\lambda_{\rm LLS} + 10~{\rm\AA}$.  The blue dashed line shows the predicted continuum level below $\lambda_{\rm LLS}$\ based on a power-law fit to the redward continuum.   
\label{fig:tau_displaya}
}
\end{figure}

Each spectrum was visually inspected to determine the wavelength
($\lambda_{\rm LLS}$) below which the spectrum dropped by more than a factor of 2.7 ($\tau = 1$). We adopt the
redshift $z_{\rm LLS}=\lambda_{\rm LLS}/912-1$ as the Lyman limit redshift.
The procedure used is shown in Figure~\ref{fig:tau_displaya} (see also  Appendix) where we plot the quasar spectra in the vicinity of the LLS and show the average fluxes in wavelength intervals above and below the position of the LLS.
The results are summarized in Table~\ref{tbl:ztable}, where we give the quasar name, its emission redshift $z_{em}$, the ESI exposure time and the LLS redshift (columns 1 -- 4). Where no LLS is detected
to our adopted wavelength limit of $4500\rm\AA$ ($z_{lim}=3.923$) we show a blank
entry in Table~\ref{tbl:ztable}. For each Lyman limit
system we searched for corresponding metal-line absorption and where
this was detected we also give the metal-line redshift (column 6).   The optical depth was now measured by computing the actual flux for a rest frame interval of $30~{\rm\AA}$\ immediately below the LLS wavelength and taking the log of the ratio of this quantity to the flux obtained by extrapolating a power-law fit to the quasar spectrum at 10 -- $160~{\rm\AA}$\ longward of the LLS, excluding the region around the \ion{O}{6} + Ly$\beta$\ quasar emission.  The procedure is illustrated in Figure~\ref{fig:tau_displaya} and the optical depths are tabulated in column 5 of Table~\ref{tbl:ztable}.

We tested the accuracy of the LLS redshift determinations and optical depth measurements by simulating LLS in the spectra and then measuring the recovered redshift and optical depth.  For each quasar for which there was sufficient wavelength space between the first real LLS and the quasar L$\alpha$\ emission line, we inserted a false LLS at a randomly generated wavelength and with a specified optical depth.  Below the wavelength of the false LLS we added in artificial noise matched to the quasar exposure time.  We then remeasured the LLS redshift and optical depth using our standard procedure.  We repeated this twenty times for each quasar for three optical depths: $\tau = 1.0$, 1.5 and 2.0.

\begin{figure}[h]
\figurenum{3}
     \includegraphics[width=3in,angle=0,scale=1]{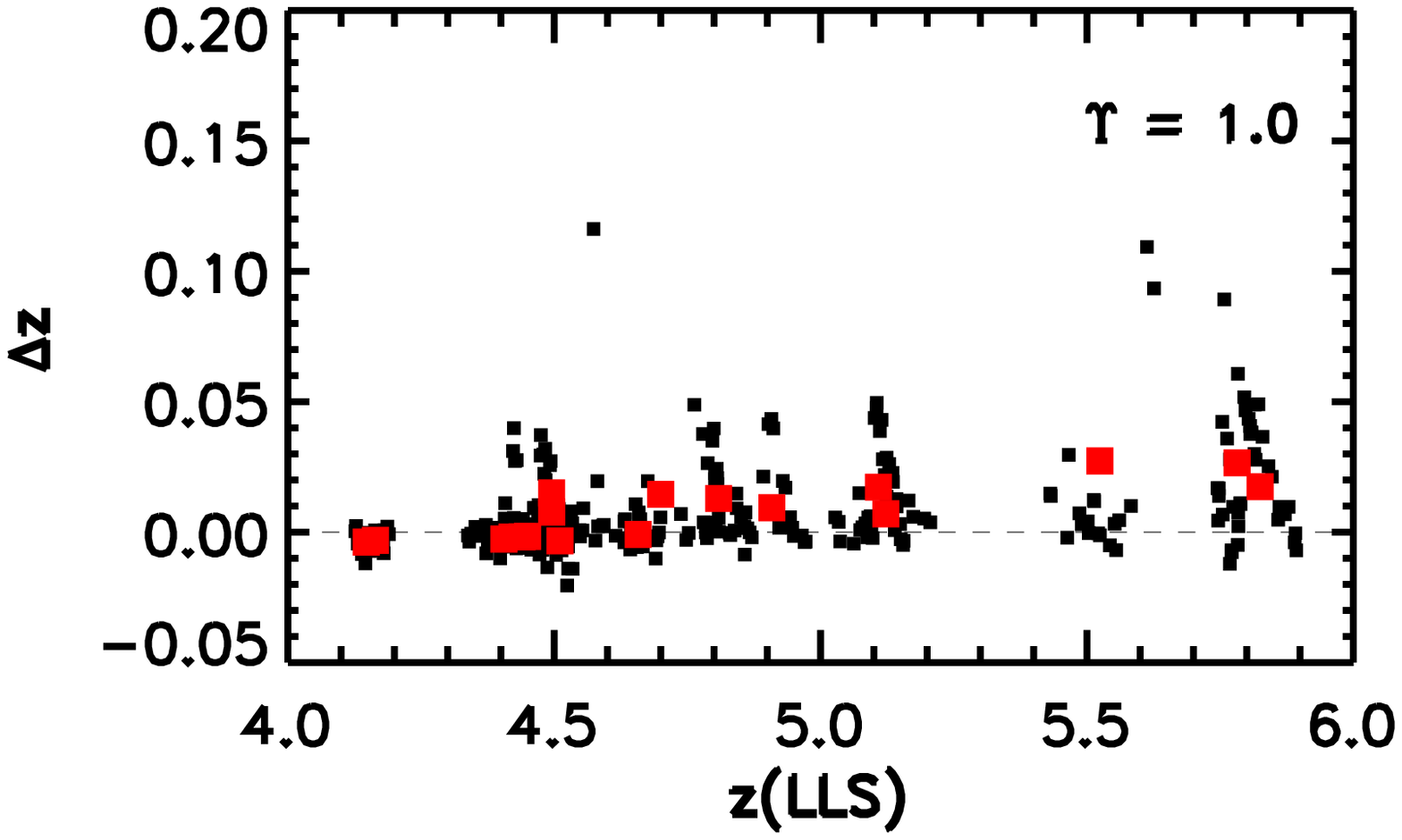}
   \includegraphics[width=3in,angle=0,scale=1]{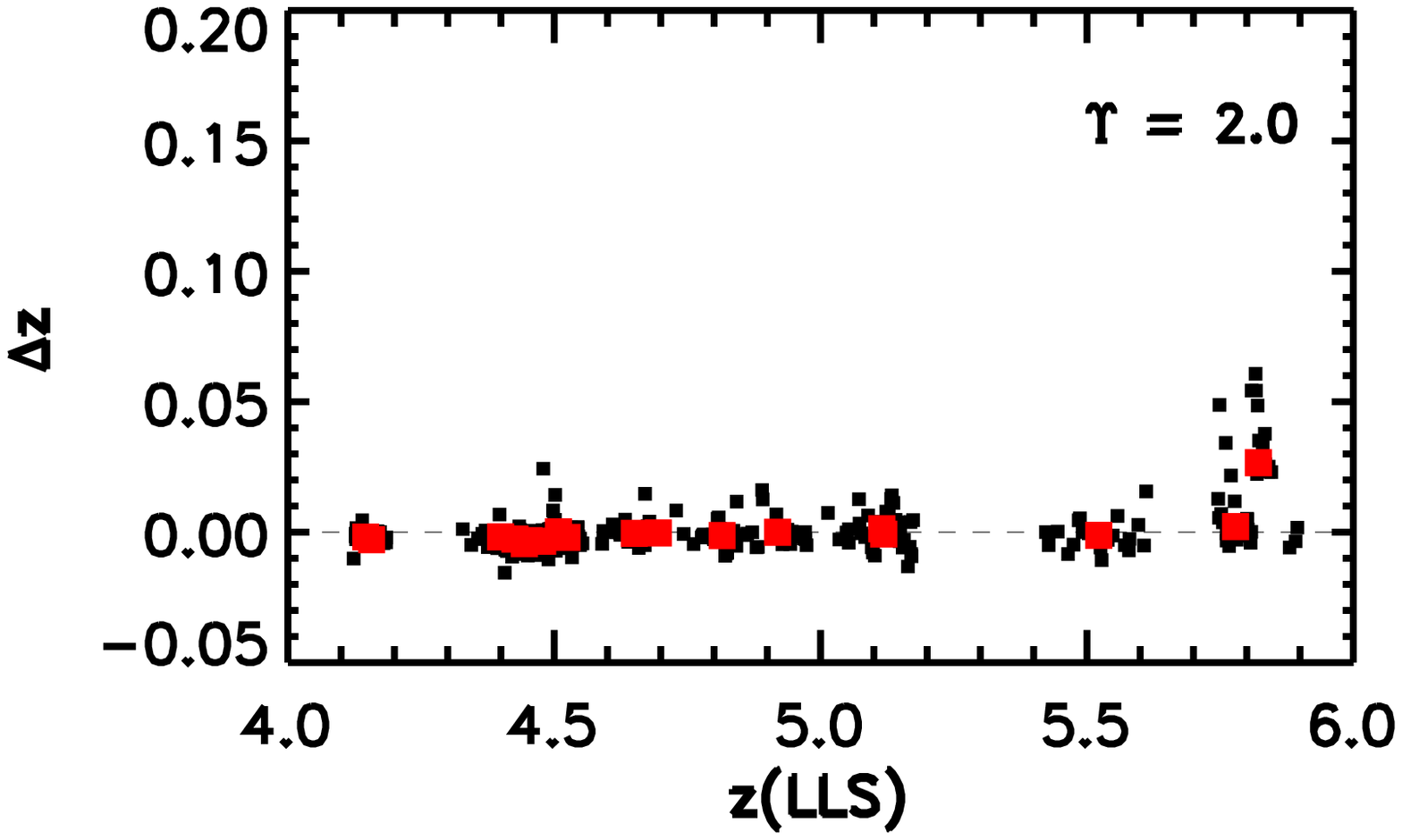}
\caption{Redshift offset of  simulated Lyman limit systems as a function of redshift for simulated LLS optical depth, $\tau = 1.0$\ ({\it top}) and 2.0 ({\it bottom}).  Black squares show individual test points for the quasars in our high redshift sample.  Large red squares are the mean for each quasar.
\label{fig:deltaz}
}
\end{figure}

The measured redshift offset as a function of LLS redshift is shown in Figure~\ref{fig:deltaz} for $\tau = 1.0$\ and 2.0.  In both cases we have shown the individual offsets of each simulation with smaller symbols, and the average in a particular quasar with the larger symbol.  As would be expected, offsets tend to be positive, i.e. the measured redshift is very slightly higher than the real one.  This is a consequence of blending with the forest.  The effect is larger for lower optical depth and at higher redshift, where the forest becomes richer.  Typical offsets in the $\tau=1.0$\ case where the effect is largest are $\sim 0.01$\ below $z = 5$, rising to $\sim 0.02$ -- 0.03 in the highest redshift quasars.  The errors are generally smaller for the $\tau > 1$\ systems.  We shall consider the systematic errors which can be introduced by these offsets in the next section.  However, the systematic errors introduced by the redshift uncertainty are smaller than the statistical uncertainty.  We also used the simulation to calibrate the measurements of the optical depth and examine the systematic errors in the determination of this quantity.  We find that the errors are dominated by the continuum fitting, with a dispersion of 0.3 in $\tau$, and that there is a systematic offset of $+0.3$\ in $\tau$.  This is introduced by the use of a single power law fit which systematically overestimates the continuum.  We have corrected the values of $\tau$\ given in Table~\ref{tbl:ztable} to allow for this offset.  The statistical errors and other systematic errors are completely dominated by the continuum fitting error.  We consider the effects of the continuum fitting error on the determination of the LLS number density in the next section.

Finally
we searched for all damped and sub-damped Lyman alpha systems in each spectrum and
give the Lyman-alpha redshift and column density for each
detected system with $N_{\rm H} > 10^{20}~{\rm cm}^{-2}$\ (columns 7 and 8). Where more than one DLA was detected we give the
properties of all the lines.  Where the DLA has a previously measured value in the literature we give this in column (9) together with the reference.

\subsection{$z\sim1$ GALEX quasars}
\label{lowz}

To improve the determination of the LLS density at low
redshift we selected a sample of $z=0.8-1.3$ quasars from
the table of Cowie et al. (2010) who tabulated all of
the Lyman alpha emission-line objects in the 9 deepest
GALEX blank-field grism spectrograph observations covering
8.2 square degrees of sky.   The GALEX spectra were obtained from the Multimission Archive at STScI (MAST).  Morrissey et al.\ (2007) detail the spectral extraction techniques used by the GALEX team in extracting the spectra from the grism data.  We
restricted ourselves to quasars lying at redshifts $z=0.8-1.3$
and with fluxes above $8\times10^{-17}$ erg cm$^{-2}$
s$^{-1}$ A$^{-1}$ where the Lyman breaks can easily be picked out.
The sample is NUV (1900--2800~${\rm\AA}$) magnitude selected (see Cowie et al.\ 2010 for a detailed description of the selection function) and we do not expect any significant biasing from the presence of a LLS at $z < 1.3$ ($2090~{\rm\AA}$) which would only minimally reduce the NUV magnitude.
The GALEX sample used is given in Table~\ref{tbl:gtable}, where we give the name of the object, the quasar redshift $z_{em}$, its RA(2000) and Dec(2000) and the redshift of the LLS if one is detected.  The observed wavelength range means that only systems with $z_{\rm LLS} > 0.54$\ are included.

\begin{figure}[h]
\figurenum{4}
 \includegraphics[width=3in,angle=0,scale=1]{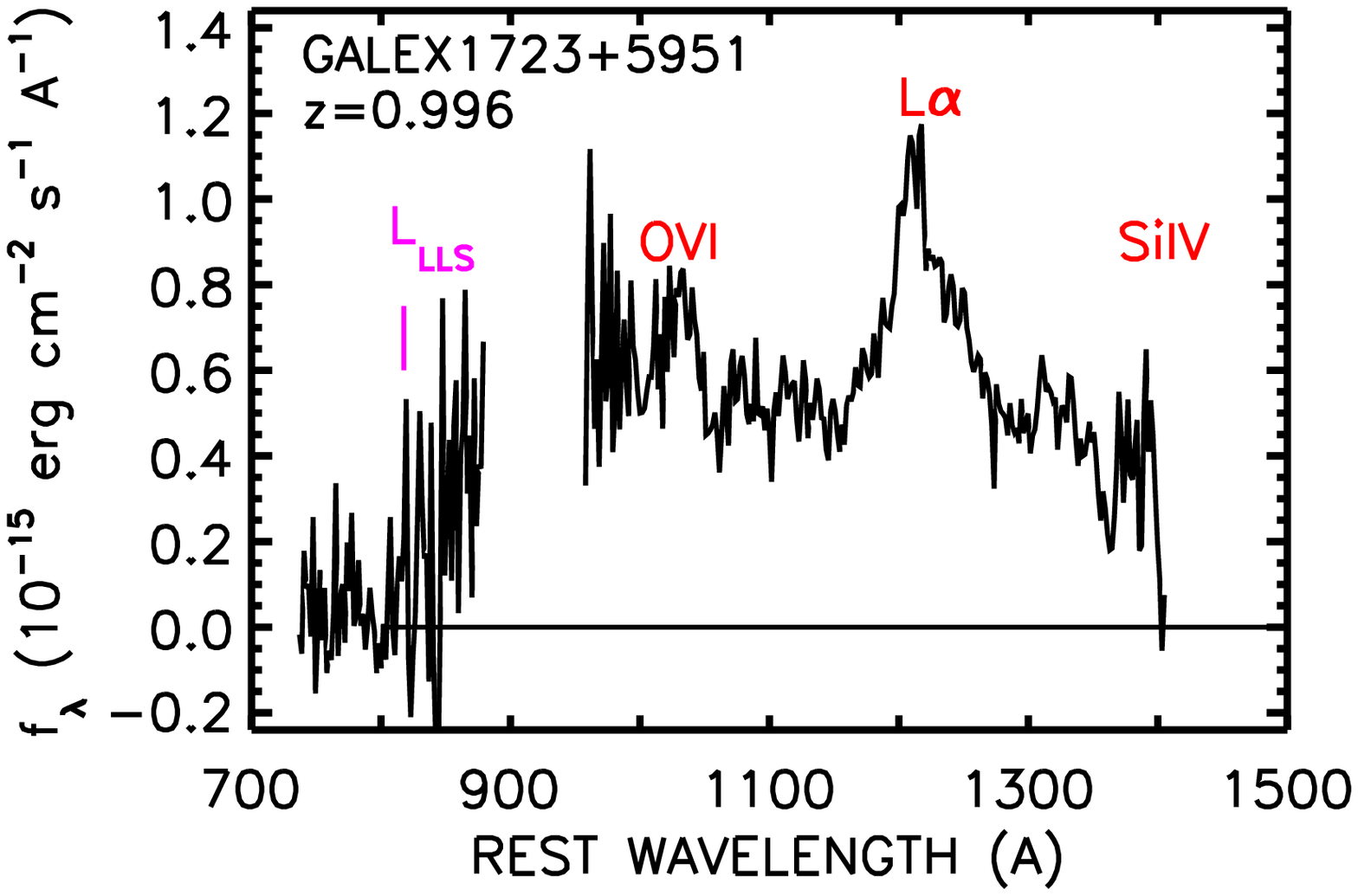}
 \includegraphics[width=3in,angle=0,scale=1]{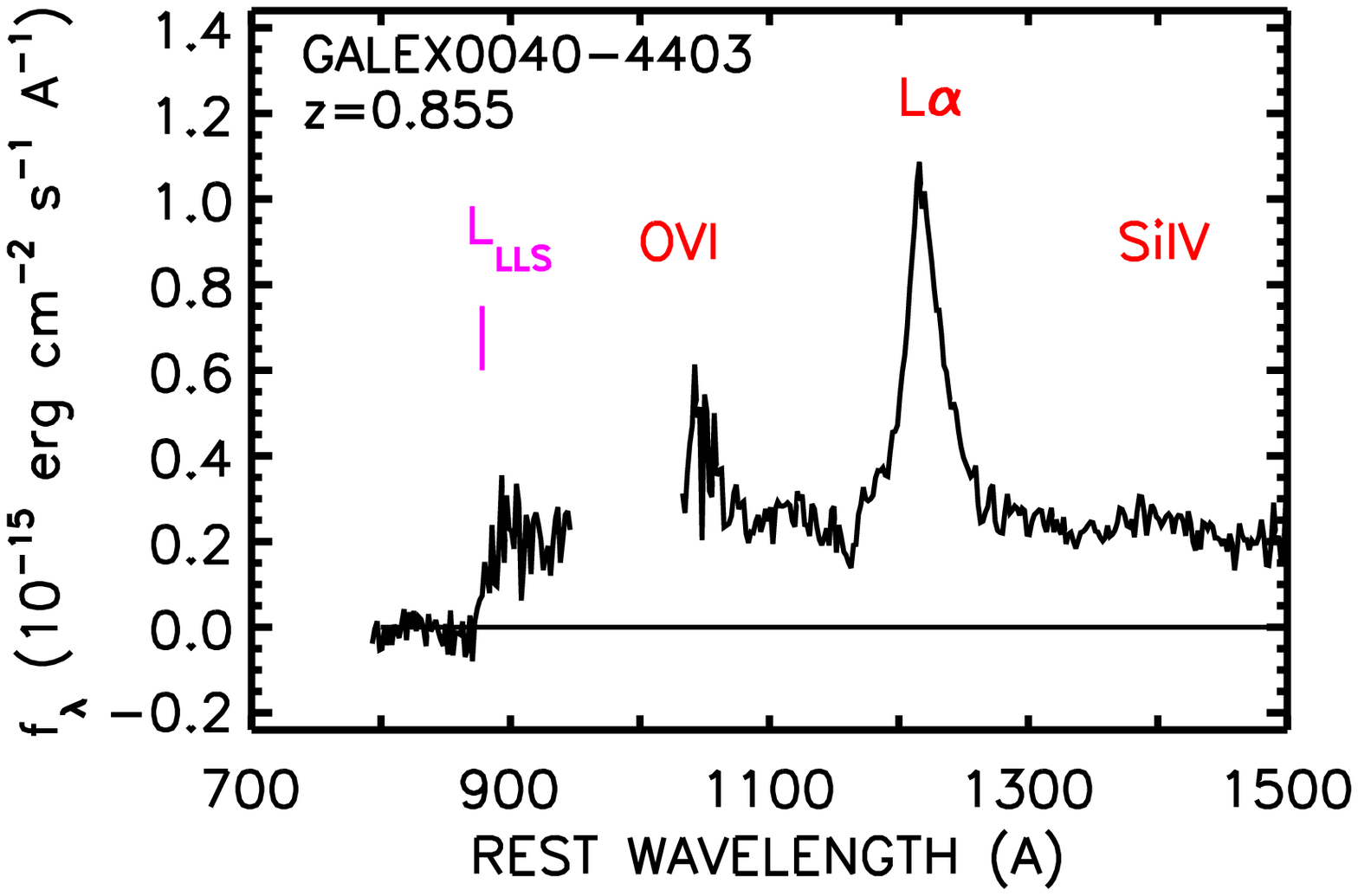}
  \caption{GALEX quasars with Lyman limit systems. The spectrum of the quasar 
   is shown in the rest frame
   with the position of the quasar emission lines marked in red.
   The short wavelength portion of the spectrum is from the GALEX
   FUV grism and the long wavelength portion from the GALEX
   NUV grism. The position of the LLS is marked in purple.  The figures are ordered by the red continuum flux above the Ly$\alpha$\  line.
\label{fig:galex_lls_sample}
}
\end{figure}

 We inspected all of the GALEX spectra for these
objects and picked out those with strong LLS ($\tau > 1$). 
To do this we fitted a power law continuum to the spectrum above $2100~{\rm\AA}$\ and then found the first location at which the spectrum dropped by more than a factor of 2.7 and did not subsequently recover at shorter wavelengths.  The GALEX spectra in which we detected such a strong LLS  are shown in
Figure~\ref{fig:galex_lls_sample} (see also Appendix), where we order the sample by the red continuum flux. 
An important point to note is that
the GALEX spectra consist of two wavelength segments: one
corresponding to the FUV grism ($\sim~1400-1800$~\AA) and one to the
NUV grism ($\sim~1900-2800$~\AA). 
We can still select LLS
in the $z=0.97-1.08$ range since the flux break across the gap is
easily seen;  consequently, the search pathlength is the redshift range including the gap.  In the case where a LLS did fall in the gap, 
we would not be able to make a precise measurement of the
break redshift and we would have to address the redshift uncertainty  in the analysis.  
In fact we do not find any LLS in our sample whose redshift lies in the gap.
LLS systems were
found in 9 of the 50 quasars at redshifts above the $z=0.54$ limit
where we could have detected such a system, though one system (in GALEX 1418+5223) lies close to this region and we consider the changes in the results that would be produced if it lies at a lower redshift.  The LLS in GALEX 1724+5921 is somewhat uncertain because of the noisy spectrum.  We examine the reduction in the number density which would be produced by removing it in the analysis.


\tablenum{3}
\begin{deluxetable}{cccccc}
\tablecaption{Parameters of Figure~\ref{fig:llsbinned}\label{tbl:figparams}}
\tablehead{
\colhead{Redshift bin} & \colhead{\# QSO} & \colhead{\# LLS}  & 
\colhead{$n(z)$} & \colhead{Error}  & \colhead{$\langle z\rangle$}
}
\startdata
(0.36,1.1)$^a$ &  51 & 11 & 0.63 & 0.19 & 0.64 \\
(0.58,1.25)$^b$ & 49 & 7 & 0.30 & 0.12 & 0.82 \\
(0.36,1.25)$^c$ & 100 & 18 & 0.44 & 0.10 & 0.72 \\
(2.4,2.9)  &  77 & 29 & 1.84 & 0.33 & 2.74 \\
(2.9,3.8)  & 106 & 52 & 2.43 & 0.33 & 3.20 \\
(3.8,4.2)  &  52 & 31 & 3.77 & 0.64 & 4.04 \\
(4.2,5.0)  &  61 & 33 & 4.04 & 0.70 & 4.39 \\
(5.0,6.0)  &   8 &  6 & 8.91 & 3.49 & 5.69 \\

\enddata
\tablenotetext{a\ }{Stengler-Larrea et al.\ 1995 sample}
\tablenotetext{b\ }{GALEX  sample}
\tablenotetext{c\ }{Combined Stengler-Larrea and GALEX samples}

\end{deluxetable}

\section{Lyman Limit Systems}
\label{llsys}

\begin{figure}[h]
\figurenum{5}
   \includegraphics[width=3.5in,angle=0,scale=1]{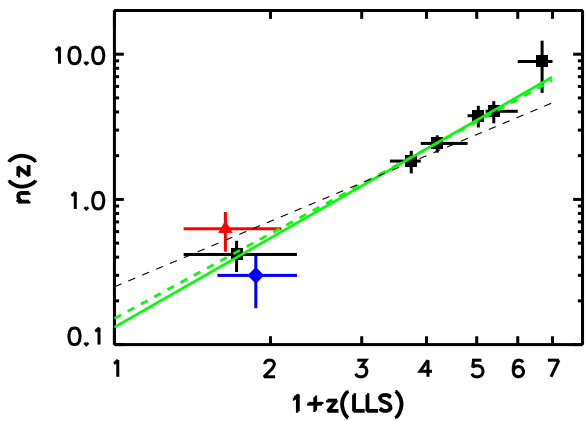}
 \caption{{\it Filled squares\/}: Density of Lyman limit systems as a
  function of mean Lyman limit system redshift for the complete sample
  of quasars.  Error bars are $1~\sigma$.  For each
  bin, the number of quasars, number of LLS, the density of LLS and
  the mean LLS redshift are given in Table 3.  Maximum likelihood
  fits are shown for $1.5 < z < 6$ ({\it green solid line}) and $0 < z < 6$
  ({\it green dashed line}). At low redshifts we show the values inferred
  from the GALEX sample alone
  ({\it blue diamond}) and the SL95 sample alone ({\it red triangle}) 
  as well as the value from the 
  combined data sets, which is shown by the black square.  The black dashed line is the fit of SL95.
\label{fig:llsbinned}
}
\end{figure}

We first computed the number density of LLS per unit 
redshift path d$z$,  which we write as $n(z)$.
As is conventional, we excluded LLS lying within $3000~{\rm km~s}^{-1}$\ of the quasar and computed the redshift path accordingly.
The data can be analysed assuming
a constant density of sources in a redshift interval
or as a fit to an assumed redshift evolution over
the entire redshift interval. To determine $n(z)$ and its
$1~\sigma$\ error
in each redshift interval we used the maximum likelihood
fit (MLF) in the form given by SSB. The
corresponding results are shown in Figure~\ref{fig:llsbinned}. For the higher
redshift interval we have shown only the full samples
(the present data and all previous data taken from the compilation of P03). However,
for $z<1$  we have shown the results
obtained separately from the present GALEX sample (blue diamond)
and
the older SL95 sample (red triangle) as well as that
for the combined sample. We have chosen to show the independent
values since
it is possible that the difference could reflect systematic
selection effects in the archival UV observed quasar samples
analysed by SL95 which might result in an upward
bias in the LLS density. However, the SL95 and GALEX
samples are consistent within the statistical uncertainties
and we shall use the combined value  in the following text. 
The present data roughly doubles the  low redshift sample and provides a corresponding reduction in the error over the previous SL95 result (Table~\ref{tbl:figparams}).  
The
values of $n(z)$ in each redshift interval are summarized in 
Table~\ref{tbl:figparams}, where we give the redshift bin, the number of QSOs used, the number of LLS detected, the value of $n(z)$\ and its $1~\sigma$\ error, and the average redshift of the LLS.  For the combined sample, we find a value of $n(z) = 0.46 \pm 0.11$.   If the LLS in GALEX 1724+5921 is eliminated the values in the $z = 0.36 - 1.25$\ interval would reduce to $0.44 \pm 0.10$, while if the redshift of the LLS in GALEX 1418+5223 is lowered to 0.97, the value is unchanged at the quoted level of significance.  We have eliminated the LLS in GALEX 1724+5921 in the subsequent analysis and the numbers in Table~\ref{tbl:figparams} correspond to the sample excluding this quasar.

We also examined the effects of the estimated redshift and optical depth errors derived from the simulation and discussed in the previous section and shown in Figure~\ref{fig:deltaz}.  For $z = 4.2$ -- 5, reducing the LLS redshift by the offset $-0.01$\ for the $\tau = 1$\ case would reduce the measured $n(z)$\ from $4.04 \pm 0.70$\ to 3.77 and at $z = 5$ -- 6, reducing the LLS redshift by $-0.03$\ would change $n(z)$\ from $8.91 \pm 3.49$\ to 7.03.  The effect is smaller for higher optical depth systems and the systematic changes are smaller than the statistical uncertainty.  The 0.3 scatter in optical depth introduces an Eddington bias since there are more low-$\tau$\ systems to scatter up than scatter down.  However, this effect is also very small, a few percent in $n(z)$, compared with the statistical error.

In analysing the redshift evolution of the LLS it is usual
to parameterize  the evolution in $n(z)$ as

\begin{equation}
n(z)=n_{0} (1+z)^{\gamma}
\label{eq_n0}
\end{equation}

\noindent 
where $n_{0}$ is the value of $n(z)$ at $z = 0$, and to estimate the parameters using a maximum-likelihood fit to the full data set in a given redshift interval.  The motivation
for this form lies in the historical description, based on assuming
that LLS lie in galactic haloes, but it is also a simple parameterization which
provides a reasonable choice of form fitting for the sample
and is useful for comparing with previously derived results. However, the
use of $n_{0}$ results in a strong coupling between this parameter
and $\gamma$, reflecting the lever arm in the power law. This
may be avoided, and the display of the parameter fit simplified,
by normalizing nearer to the center of the redshift range and
we use the form

\begin{equation}
n(z)= n_{3.5} \left ({{1+z} \over {4.5}} \right )^{\gamma}\ ,
\label{eq_n3}
\end{equation}

\noindent
where $n_{3.5}$ is the value of $n(z)$ at $z = 3.5$, to show the error contours on the parameters.

We calculated the log-likelihood function 

\begin{eqnarray}
\ln L = x\ln n_0 + \gamma \sum_{i=1}^x\ln (1+z_{\rm LLS}^i) - \left ({{n_0} \over {1+\gamma}}\right )  \nonumber \\
\times \sum_{j=1}^y\left [ (1+z_{\rm em}^j)^{\gamma +1} - (1+z_{\rm min}^j)^{\gamma +1}\right ]
\label{eq_like}
\end{eqnarray}

\noindent
where the notation follows that of P\'eroux (2001) and $x$\ is the number of LLS and $y$\ is the number of quasars, and used this to estimate $\gamma$\ and $n_{3.5}$\ and their errors in the redshift intervals $1.5 < z < 6$\ and $0 < z < 6$.  The results are summarized in Table~\ref{tbl:likeparams}, where we also list the parameters and fitting ranges of previous fits from the literature.


\tablenum{4}
\begin{deluxetable*}{lcccccc}
\tablecaption{Parameters of Maximum Likelihood Fit \label{tbl:likeparams}}
\tablehead{\colhead{\rm Sample$^a$} &
\colhead{Redshift range} & \colhead{QSO} & \colhead{LLS$^b$}  & 
\colhead{$\gamma$} & \colhead{$n_{3.5}$} & \colhead{$n_0$}
}
\startdata
This paper & 1.5 -- 6 &  216 & 157 & $2.04^{+0.62}_{-0.68}$ & $2.84^{+0.29}_{-0.37}$ & 0.13\\
& 0.0 -- 6  &  315 & 175 & $1.94^{+0.36}_{-0.32}$  & $2.80^{+0.33}_{-0.33}$ & 0.15\\
Ref. (1) & 0.40 -- 4.93 & 148 & 84 & $2.55^{+0.95}_{-0.85}$  & \nodata & $0.06^{+0.15}_{-0.05}$ \\
Ref. (2) & 0.32 -- 4.11 & 169 & 80 & $1.50 \pm 0.39$ &    \nodata & $0.25^{+0.17}_{-0.10}$\\
Ref. (3) & 0.40 -- 4.69 &90 & 48 & $1.55\pm 0.45$  &    \nodata & $0.27^{+0.20}_{-0.13}$ \\
Ref. (4) & 0.67 -- 3.58 & 90 & 54 & $0.68 \pm 0.54$  &    \nodata & 0.76 \\
Ref. (5) & 3.3 -- 4.4 & 469 & 192 & $5.2 \pm 1.5$ &   $1.5 \pm 0.2$ & \nodata \\

\enddata
\noindent
\tablenotetext{a }{{\bf References}: 
(1) P\'eroux (2001);
(2) Stengler-Larrea et al.\ (1995);
(3) Storrie-Lombardi et al.\ (1994);
(4) SSB (1989);
(5) Prochaska et al.\ (2010).
}
\tablenotetext{b }{Refers to number of LLS with $\tau > 1$, except for Ref.~4 ($\tau > 1.5$) and Ref. 5 ($\tau > 2$).
}

\end{deluxetable*}

\begin{figure}[h]
\figurenum{6}
  \includegraphics[width=3in,angle=90,scale=1]{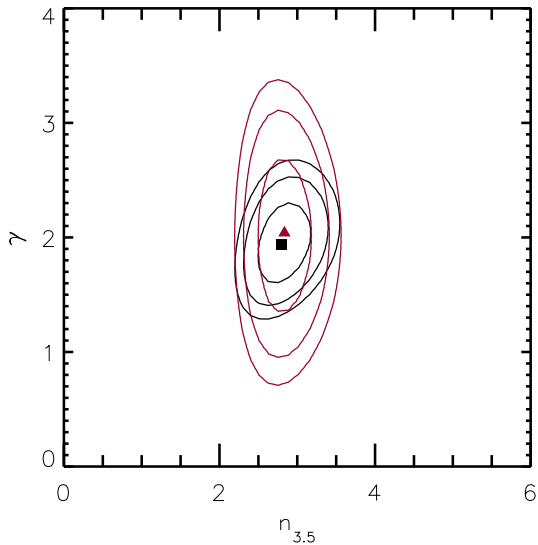}
  \caption{$1~\sigma$, $2~\sigma$, and  $3~\sigma$\ maximum likelihood contours of $n_{3.5}$, $\gamma$\ for LLS
  in the redshift intervals $0 < z_{LLS} < 6$ ({\it black filled square,
  contours}) and $1.5 < z_{LLS} < 6$ ({\it red triangle, contours}).
\label{fig:likeplot}
}
\end{figure}

The fits are shown in Figure~\ref{fig:llsbinned} as the solid  ($1.5 < z < 6$)\ and dashed ($0 < z < 6$) green lines, compared with the redshift evolution of the density of LLS in the complete quasar sample (filled black squares) in suitable redshift bins.  In contrast with previous results, and primarily as a consequence of the new low-redshift $n(z)$\ value, 
the fits over the two redshift ranges are practically indistinguishable and give consistent values for $\gamma$\ and $n_{3.5}$, as is seen in Figure~\ref{fig:likeplot}, which  plots the 1, 2, and $3~\sigma$\ error contours.  A single parameterizaton ($n_{3.5} = 2.80 \pm 0.33$, $\gamma = 1.94^{+0.36}_{-0.32}$) fits well over the entire redshift range, $0 < z < 6$.  The evolution inferred with this data set is generally steeper than previously measured (see Table~\ref{tbl:likeparams}), except for the result of P\'eroux (2001).  In particular, the evolution is steeper than that measured by SL95 for $0.32 < z < 4.11$, (the black dashed line in Figure~\ref{fig:llsbinned}), as a result of the improved low-redshift determination from the large combined sample.

\begin{figure}[h]
\figurenum{7}
 \includegraphics[width=3in,angle=90,scale=0.85]{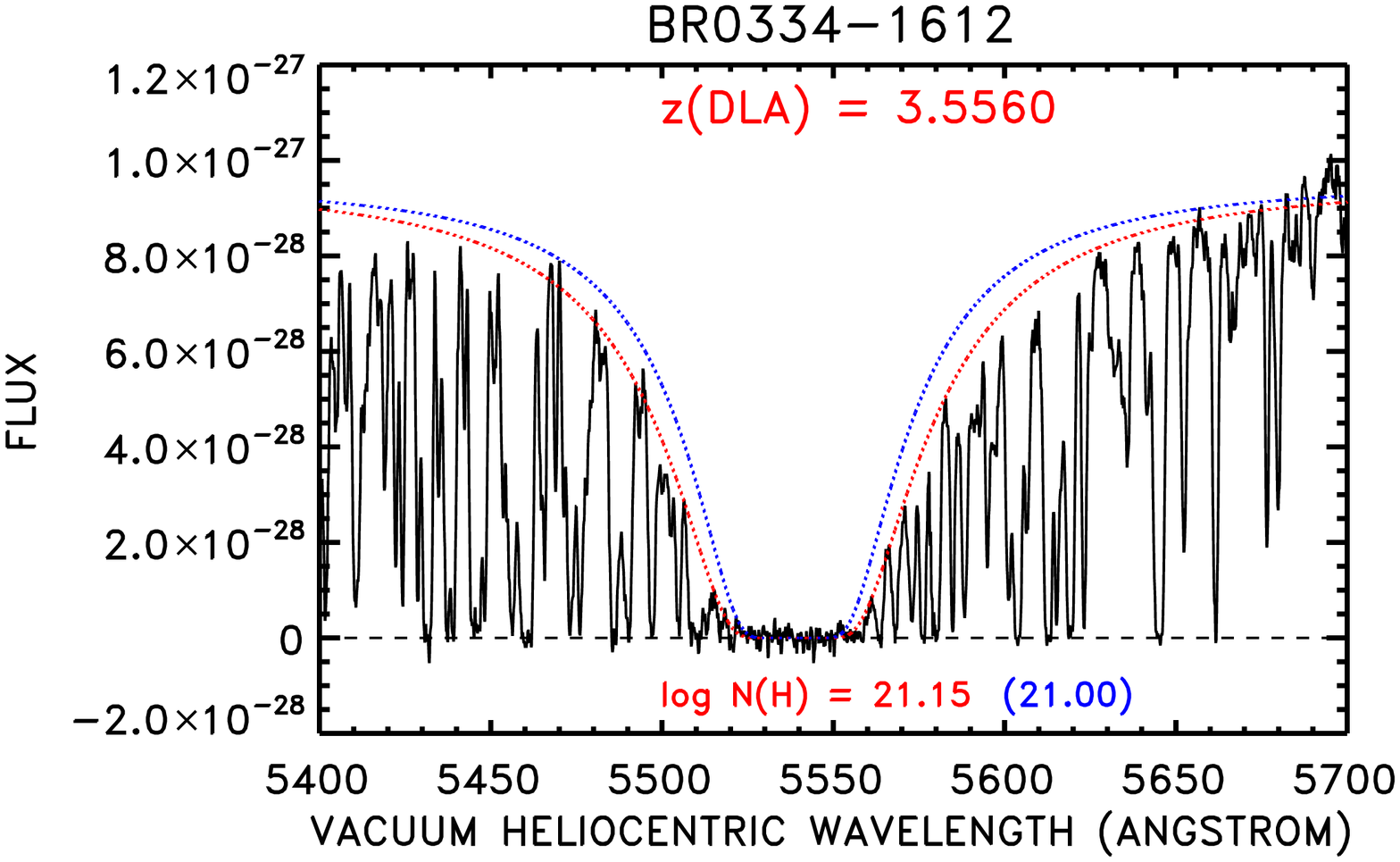}
 \includegraphics[width=3in,angle=90,scale=0.85]{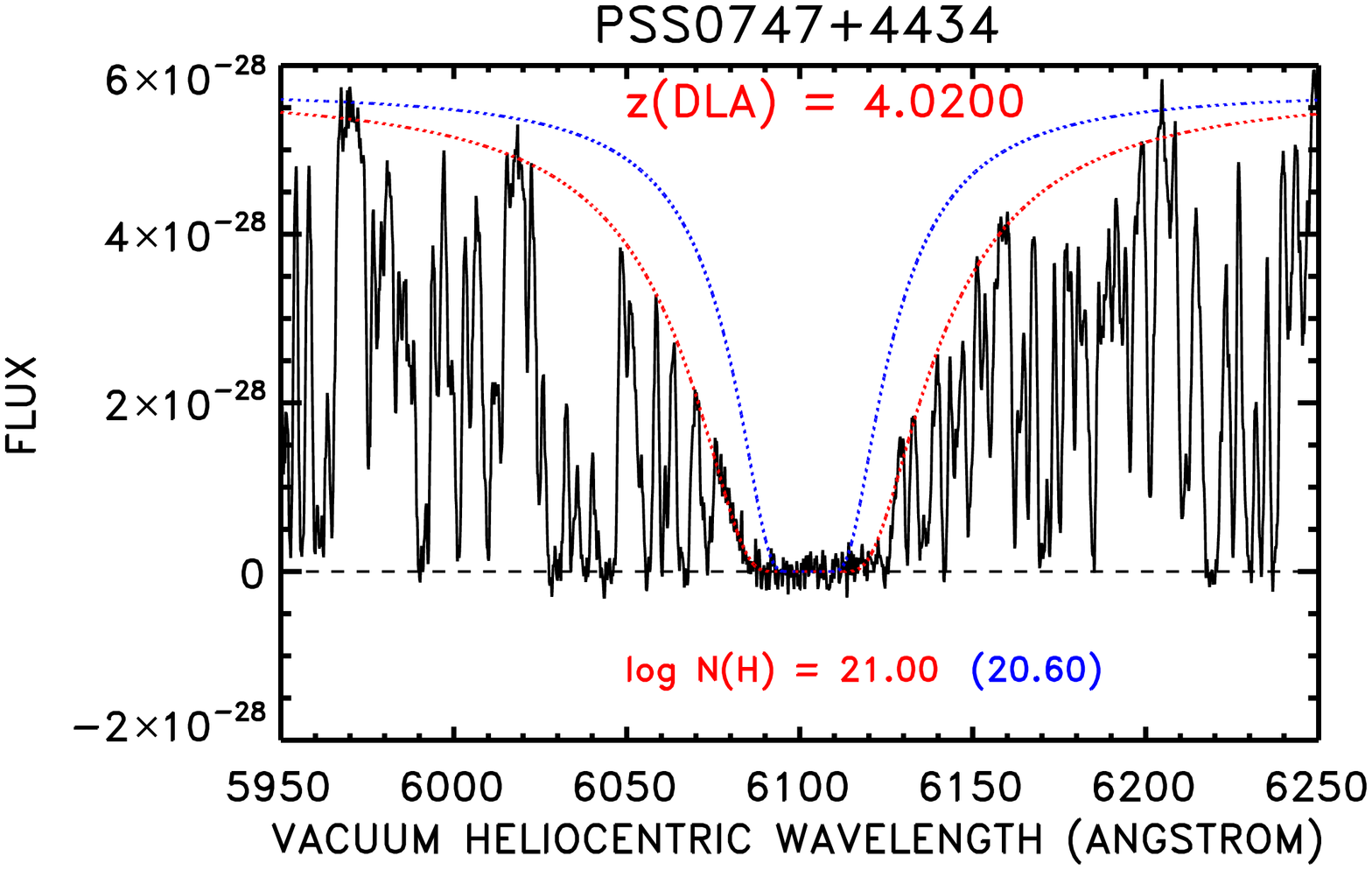}
  \caption{Damped Lyman alpha systems in the high redshift sample.  The red dotted line shows the fit corresponding to $\log N_{\rm H}$\ from Table 1.  Where present, the blue dotted line shows the fit from P03 (see Table 5).
\label{fig:dlas_a}
}
\end{figure}

\section{Damped Lyman Alpha Systems}
\label{dlasys}

We searched the spectra for absorption lines with damping wings and 
found a total of 17 DLAs ($\log N({\rm HI}) > 20.3$) and sub-DLAs ($20.0 < \log N({\rm HI}) < 20.3$) within our $z<5.1$\ quasar sample, 14 DLAs and 3 subDLAs.  We believe the sample is complete to the level $\log N({\rm HI}) = 20$.  However, in the highest redshift quasars the lower column density systems are hard to measure and we have not attempted to extend the measurements below $\log N({\rm HI}) = 20$. DLAs in general become too hard to measure beyond $z = 5.1$\ because of the thickness of the forest.  
We measured $N({\rm HI})$\ for these systems by fitting to the damping wings (Figure~\ref{fig:dlas_a} and Appendix), and column~8 of Table~\ref{tbl:ztable} compares our measured values of $\log N({\rm HI})$\ with previous measurements from Prochaska et al.\ (2007) and P03.  Fourteen of the systems have existing measurements.  All of the DLAs in the present sample are shown in Figure~\ref{fig:dlas_a} together with the damped ${\rm L}_{\alpha}$\ fits (red solid lines).   Nine of the DLAs overlap with DLAs in the P03 sample and the P03 fits are shown with blue dotted lines in Figure~\ref{fig:dlas_a}. One DLA in the P03 sample is classified as a sub-DLA here (the $z=3.762$\ system in PSS0741+4434), and one DLA in the P03 sample (the $z=4.06$\ system in SDSS0338+002) is not confirmed.  However, in general the measured column densities agree well with previous measurements.  The present sample substantially increases the path length above $z = 4.5$.  We summarize the present results combined with $z>4.5$\ quasars from P03 and Guimar\~aes et al.\ (2009) in Table~\ref{tbl:deldla}.  We have used these combined samples to compute the redshift evolution of DLAs out to $z = 5.1$\ to compare with that of the LLS.


\begin{figure}[h]
\figurenum{8}
    \includegraphics[width=3in,angle=90,scale=0.9]{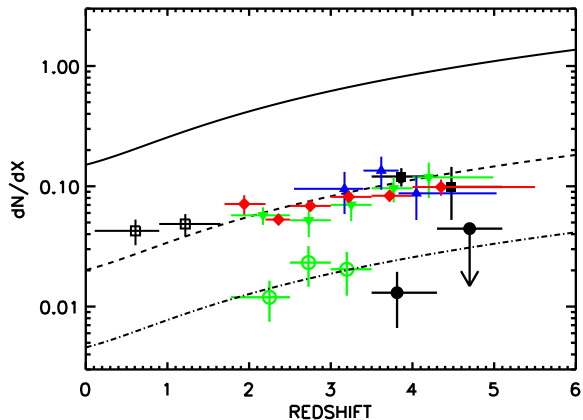}
  \caption{Number density of DLAs as a function of redshift for various samples and computed for two values of the limiting column density, $N_{\rm H}$.  For $N_{\rm H} > 20.3$, colored symbols denote: Rao et al.\ (2006; black open squares); Prochaska et al.\ (2005; red filled diamonds);  P\'eroux et al.\ (2005;  green filled triangles);  Guimara\~es et al.\ (2009; blue filled triangles).  
For $N_{\rm H} > 21.0$, the green open circles show values computed from P\'eroux et al.\ 2005.  The filled black symbols show the high redshift sample of this paper (Table 1) combined with all previous high redshift samples for $N_{\rm H} > 20.3$\ (squares) and $N_{\rm H} > 21.0$\ (circles).  The solid black line is the maximum likelihood fit to the LLS with $0 < z < 6$ from the combined sample (see Table 4).  The dashed and dot-dashed black lines are this relationship scaled by a factor of 7.5 and 33, respectively.
\label{fig:dlanum}
}
\end{figure}

In Figure~\ref{fig:dlanum} we compare the redshift evolution of the DLAs
with those of the LLS. The black line shows the maximum likelihood
fit to the evolution of ${\rm d}n/{\rm d}X$\ in the LLS over the 
range $0 < z < 6$, where d$X$ is defined as

\begin{equation}
{{{\rm d}X} \over {{\rm d}z}} = 
(1+z)^2 \left [ (1+z)^2 (1+z\Omega_{\rm M}) - 
z(2+z)\Omega_{\Lambda} \right ]^{-1/2} \ \ 
\label{eq_dx}
\end{equation}

\noindent
and d$n$\ is the number of $\tau > 1$\ LLS in the interval d$X$.
The black squares show the observed number
of DLAs with $\log N({\rm HI}) > 20.3$\ in the redshift intervals [3.5,4.3], and [4.3,5.1], computed from the present data set combined with preceding measurements.  The error bars are $\pm1~\sigma$
based on the Poisson errors corresponding to the number
of systems in the bin (Gehrels 1986). We have also computed
the errors using a jackknife estimation. These agree well for
bins with higher numbers of systems but underestimate the
error in bins with only a small number of DLAs.
We also show lower redshift DLA number density measurements from the literature (Prochaska et al.\ 2005, P\'eroux et al.\ 2005, Rao et al.\ 2006 and Guimar\~aes et al.\ 2009.) The black circles 
show the distribution for the $\log N({\rm HI})>21$ systems with the P03
results shown with green open circles.

If the neutral hydrogen column density distribution function
is invariant with redshift the redshift evolution of the
LLS and the DLAs would have the same form in Figure~\ref{fig:dlanum}. This appears
to be the case for the $\log N({\rm HI})>20.3$ systems where a simple
scaling down of the LLS numbers by a factor of 7.5 replicates
the evolution of the DLAs, at least for $z>1$. The situation is less clear for
the higher column density ($\log N({\rm HI})>21$) DLAs. At low
redshifts ($z < 3.5$) the LLS can be scaled down by a factor Êof
33 to match the high column density DLAs but the errors
at high redshift could permit a substantial deviation from
this form.  We adopt a single power-law fit, $d^2N({\rm HI})/dN\,dX \propto N({\rm HI})^{-\beta}$, to the column density distribution as a function of redshift between the LLS  ($\log N({\rm HI}) = 17.2$) and DLA ($\log N({\rm HI}) = 20.3$\ and 21.0) column densities.  For $\log N({\rm HI}) = 20.3$\ the individual points are consistent with a single index of $\beta = 1.28$\ for $1<z<5$, whereas an index of $\beta = 1.40$\ is required for $\log N({\rm HI}) = 21.0$.  The steeper index over the wider column density range is a consequence of the well known turnover in the column density distribution function at high column density (e.g. Petitjean et al.\ 1993; Storrie-Lombardi \& Wolfe 2000; Prochaska et al.\ 2005; O'Meara et al.\ 2007; Guimar\~aes et al.\ 2009).


\begin{figure}[h]
\figurenum{9}
  \includegraphics[width=3in,angle=90,scale=0.9]{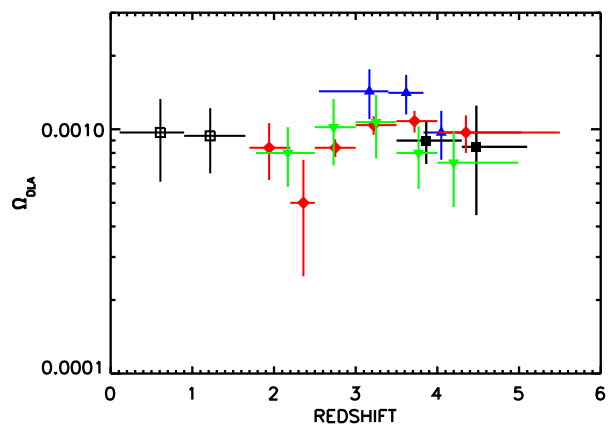}
  \caption{$\Omega_{\rm DLA}$\ as a function of redshift.  Symbols have the same meaning as in Figure~\ref{fig:dlanum}.
\label{fig:omega_plot}
}
\end{figure}

Finally, we have computed $\Omega_{\rm DLA}$\ for the combined sample for $3.7<z<4.4$\ and $4.4<z<5.1$\ and compare these values with results from the literature in Figure~\ref{fig:omega_plot} ($\Omega$\ here includes both hydrogen and  helium).  The present data (black squares and $1~\sigma$\ error bars) show no evolution in $\Omega_{\rm DLA}$\ to $z = 5$ (see also Prochaska et al.\ 2005; P\'eroux et al.\ 2005; Noterdaeme et al.\ 2009; Guimar\~aes et al.\ 2009).  This remains true if we expand the sample to include the subDLAs, which raises  $\Omega_{\rm DLA}$\ by about 30\% in the highest redshift bin.  The evolution is consistent with a constant $\Omega$\ over the observed redshift range.


\tablenum{5}
\begin{deluxetable*}{lcccccl}
\small
\addtolength{\tabcolsep}{-6pt}
\tablecaption{DLAs in $z > 4.5$\ quasars\label{tbl:deldla}}
\tablehead{
\colhead{Quasar} & \colhead{$z_{em}$} & \colhead{$z_{min}$}  & 
\colhead{$z_{\rm DLA}$} & 
\colhead{$\log N({\rm H})$} & \colhead{Source$^a$}  & \colhead{Other$^b$}
}
\startdata

PSS1140+6205 & 4.500 & 3.113 & 0.000 & 0.00 & 1 & \nodata \\
PSS0808+5215 & 4.510 & 3.181 & 3.114 & 20.60 & 1 & \nodata \\
BR1033$-$0327 & 4.510 & 2.888 & 4.174 & 20.08 & 0 & \nodata \\
PSS1723+2243 & 4.515 & 3.062 & 3.697 & 20.35 & 1 & \nodata \\
PSS1715+3809 & 4.520 & 3.004 & 3.341 & 21.05 & 1 & \nodata \\
BR0019$-$1522 & 4.528 & 2.970 & 3.437 & 20.92 & 2 & \nodata \\
PSS0134+3307 & 4.536 & 2.976 & 3.760 & 20.60 & 1 & P 20.6 \\
BR2237$-$0607 & 4.550 & 2.941 & 4.080 & 20.53 & 0 & P 20.50 \\
PSS1347+4956 & 4.565 & 2.963 & 0.000 & 0.00 & 0 &  \nodata \\
BR0353$-$3820 & 4.580 & 3.022 & 0.000 & 0.00 & 0 &  \nodata \\
BR1202$-$0725 & 4.610 & 3.112 & 4.383 & 20.62 & 0 & P 20.49 \\
SDSS0210$-$001 & 4.700 & 3.177 & 0.000 & 0.00 & 1 & P 20.5 (3.420)$^*$ \\
BRJ0307$-$4945 & 4.728 & 3.138 & 4.460 & 20.80 & 2 & \nodata \\
SDSS2200+001  & 4.780 & 3.214 & 0.000 & 0.00 & 0 & \nodata \\
SDSS 0206+121 & 4.810 & 3.007 & 4.326 & 20.78 & 0 & \nodata \\
SDSS0941+594 & 4.820 & 3.322 & 0.000 & 0.00 & 1 & \nodata \\
PSS1247+3406 & 4.897 & 3.314 & 0.000 & 0.00 & 1 & \nodata \\
SDSS0211$-$000 & 4.900 & 3.255 & 4.644 & 20.18 & 0 & \nodata \\
SDSS1737+582 & 4.850 & 3.309 & 4.741 & 20.70 & 0 & PW 20.65 \\
SDSS0338+002 & 5.010 & 3.480 & 0.000 & 0.00 & 0 & P 20.4 (4.060)$^*$ \\
SDSS1204$-$002 & 5.070 & 3.437 & 0.000 & 0.00 & 0 & \nodata \\
SDSS0756+410 & 5.090 & 3.551 & 4.360 & 20.15 & 1& \nodata 

\enddata

\tablenotetext{a}{Source for cols. 1--5: 
0:  Present work;
1:  Guimar\~aes et al.\ (2009);
2:  P\'eroux et al.\ (2003)}

\tablenotetext{b}{Other published values of $\log N({\rm H})$\ (redshift):
P:  P\'eroux et al.\ (2003);
PW:  Prochaska et al.\ (2007)}

\tablenotetext{*}{Not confirmed.}

\end{deluxetable*}

\section{Discussion}
\label{discuss}

The density evolution of the LLS is most closely related
to the attenuation length of ionizing photons in the intergalactic
medium since it is systems with column densities near the value
of the LLS that dominate the neutral hydrogen opacity. Thus the conversion
of $\epsilon$, the source function of ionizing photons, to $\Gamma$,
the ionization rate in the intergalactic medium, can be
determined if we know the evolution of the LLS and the
shape of the column density distribiution in the neighborhood of
the LLS column density.

Following the fundamental work of Madau, Haardt and Rees (1999; MHR),
there have been a number of revisits to the problem, updating 
the calculation using recent values of the evolution
parameters and the column density shape. The most recent
version is given by Faucher-Gigu\`ere et al. (2008). At $z>2$ these
calculations are enormously simplified by the short mean free
path of the photons which means the calculation is essentially local (MHR). In this limit,  if we can approximate the column density distribution function in the neighborhood of the LLS column density as $N^{- \beta}$, 
the mean free path is simply related to the LLS system density $n(z)$ as

\begin{equation}
\Delta l(\nu_0,z) n(z) \propto 
{ {\int_a^{\infty} \tau^{-\beta} {\rm d}\tau} \over
{{\int_0^{\infty} \tau^{-\beta} (1-e^{-\tau})} }}
\label{mfp1}
\end{equation}

\noindent
or

\begin{equation}
\Delta l(\nu_0,z) = 
{{a^{1 - \beta}} \over {\Gamma (2 - \beta )}}\ \ 
{{c} \over {H(z) (1+z) n(z)}}
\label{mfp2}
\end{equation}

\noindent
where $\tau$\ is the optical depth at the Lyman edge, $a$\ is the optical depth above which $n(z)$\ is measured (here $a \approx 1$) and $\nu_0$\ is the frequency of the Lyman continuum edge 
(see for example the equations in Miralda-Escud\'e (2003) or Choudhury (2009),
and note that $\Gamma$ in this equation is the gamma
function and not the ionization rate). MHR adopted redshift-invariant values of $\beta=1.5$ and $\gamma=2$ to calculate the values, basing these
parameters on the work of Petitjean et al. (1993), Hu et al. (1995), and Kim et al.\ (1997) for the $\beta$\ index. Faucher-Gigu\`ere et al. (2008) updated the parameter
$\beta$ to $1.39$ using the results of Misawa et al. (2007), used Stengler-Larrea's fit to the LLS population, and again
assumed a redshift-invariant set of parameters.

The present work supports the assumption of redshift invariance,
and our power law index fit of $0.28$ to the cumulative
distribution between the LLS and the $\log N_{\rm H} > 20.3$\ DLA column
densities  is not substantially different  (in the differential form of $1.28$)
from the Misawa et al. (2007) value adopted by Faucher-Gigu\`ere.
However, the present data implies a higher value of the
LLS density at $z>2$ than was adopted by Faucher-Gigu\`ere,
with a corresponding decrease in the mean free path since, 
as was already known from the work of P03, the SL95 fit to the high redshift LLS is low because their sample did
not extend to higher redshifts. Combining the evolutionary form of the LLS with equation~(\ref{mfp2}), and using the high redshift approximation, $H(z) = H_0 \Omega_m^{1/2} (1+z)^{3/2}$, gives

\begin{equation}
\Delta l(\nu_0,z) = 
{ {a^{1 - \beta} (4.5)^{-2.5}} \over {\Gamma (2 - \beta )} }
{ {c} \over {H_0 \Omega_m^{1/2}} } \ 
{ n_{3.5}^{-1} } \ 
{ \left [ {{1+z} \over {4.5}} \right ] ^{-(2.5 + \gamma )} } 
\label{mfp3}
\end{equation} 

\noindent
where we have used the normalization at $z = 3.5$ which places the errors primarily in the $\gamma$\ exponent (see Figure~\ref{fig:likeplot}).  For $\beta = 1.3$, the maximum likelihood fit for $0<z<6$\ gives

\begin{equation}
\Delta l(\nu_0,z) = 
50 { \left [ {{1+z} \over {4.5}} \right ] ^{-4.44^{+0.36}_{-0.32}} } \ {\rm Mpc.}
\label{mfp4}
\end{equation}

\noindent
For $\beta = 1.5$, the normalization becomes 37~Mpc and for $\beta = 1.1$, 61~Mpc.

\begin{figure}[h]
\figurenum{10}
    \includegraphics[width=3in,angle=90,scale=0.85]{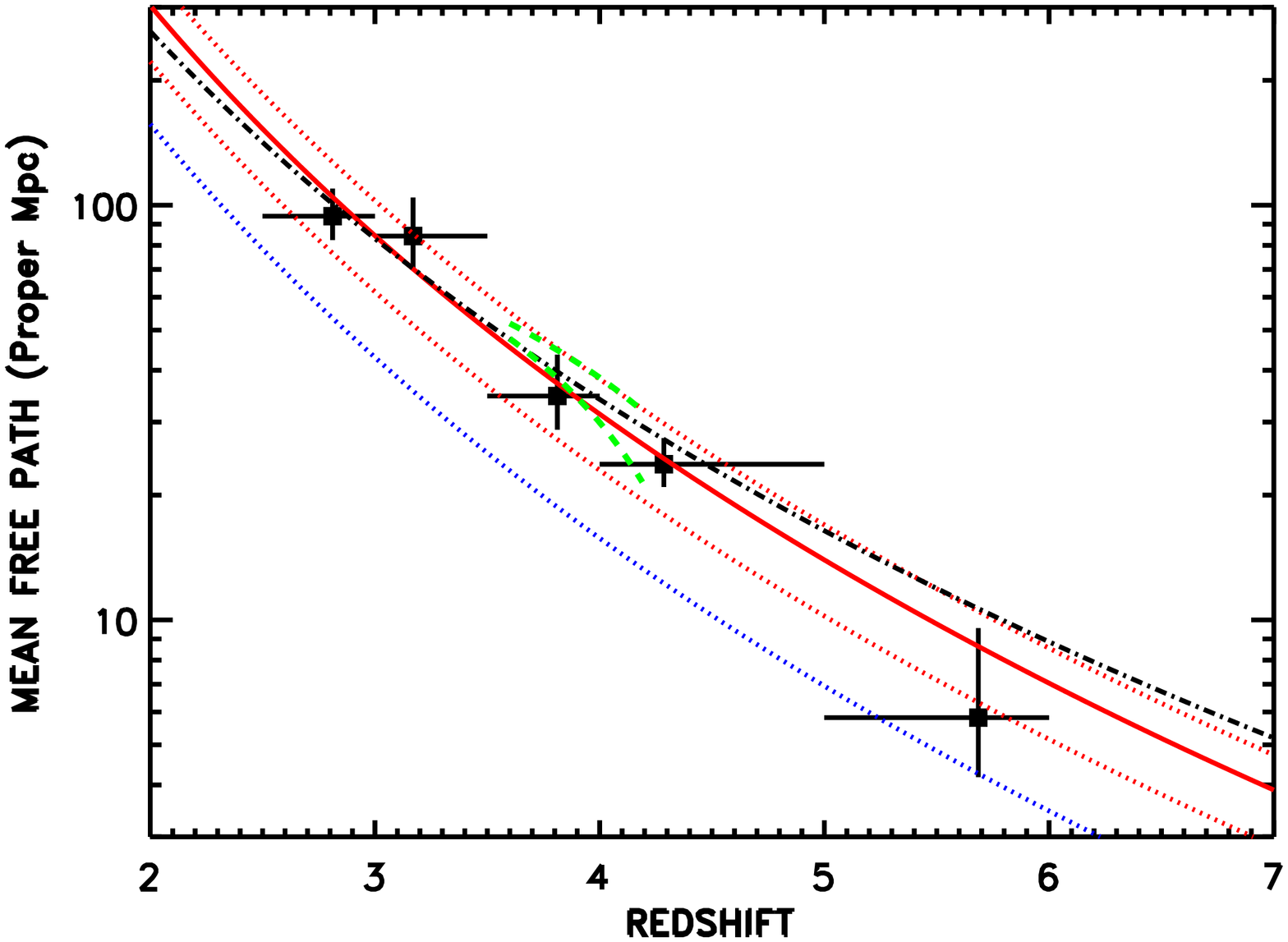}
        \includegraphics[width=3in,angle=90,scale=0.85]{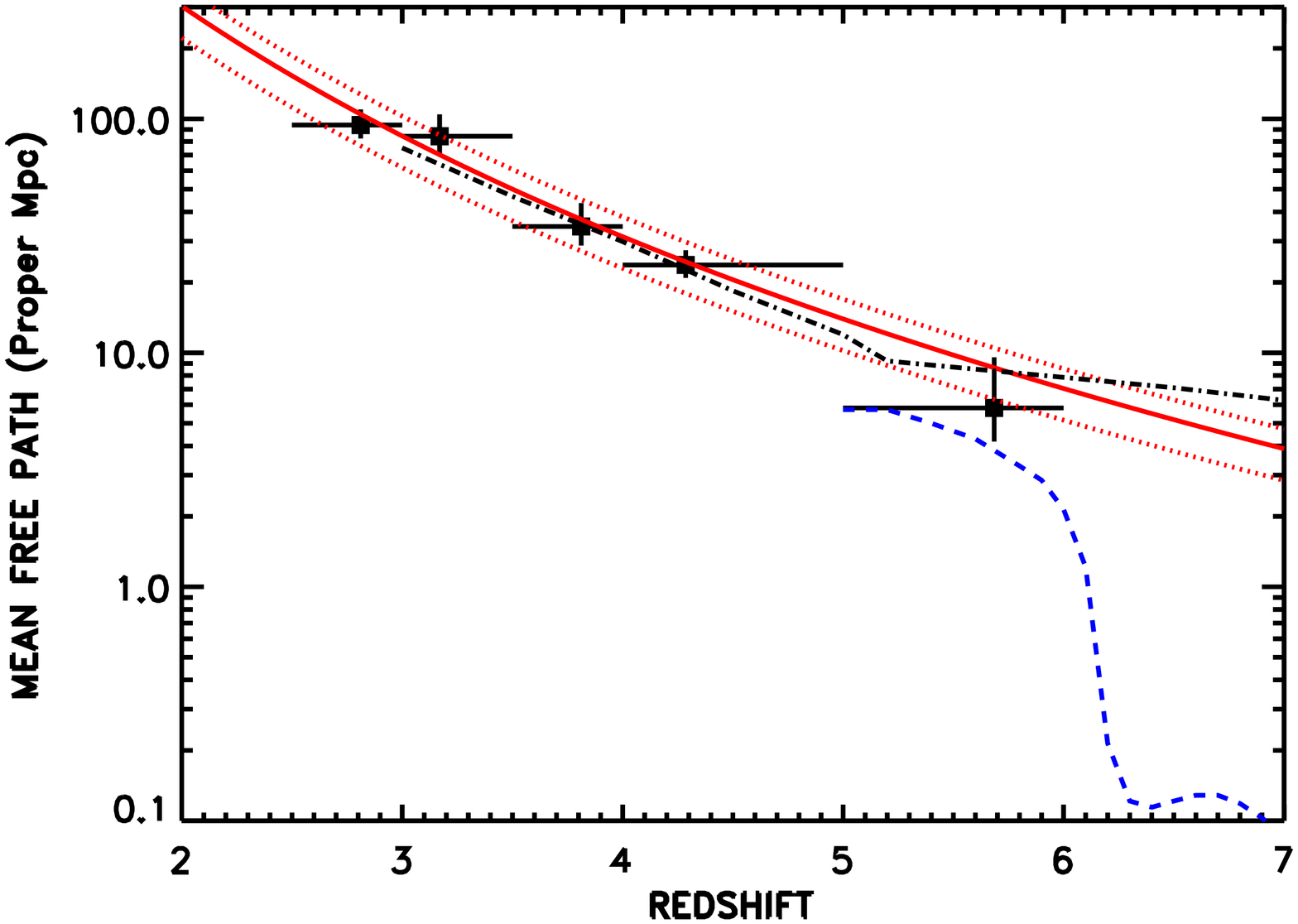}
        
  \caption{(Top)  Mean free path as a function of redshift.  The filled squares show the data points computed from Equation~(\ref{mfp3}) using the maximum likelihood  estimate of $n_{\rm LLS}$\ in the redshift bins indicated by the horizontal lines.  $1~\sigma$\ errors are shown in the y direction.  the points are computed for $\beta = 1.28$\ but may be simply scaled to other choices of $\beta$\ using the functional form of Equation~(\ref{mfp3}).  
  The thick red line is the fit to our combined sample, from Equation~(\ref{mfp3}) with $\gamma = 1.94$, $\beta = 1.28$; the thin red lines show the $1~\sigma$ range in $\beta$\ ($1.1$\ to $1.5$).   
The black dot-dash line is the result of Faucher-Gigu\`ere et al.\ (2008), with $\gamma = 1.5$, $\beta = 1.39$.  We note that, in comparing with MHR (blue dotted line), Faucher-Gigu\`ere et al. did not correct for the cosmology, making the difference between their results and MHR appear larger in their Figure~4 than in the present figure.  The green dashed lines are the power law fit of Prochaska et al. (2009) over the range $3.6<z<4.2$\ based on stacked SDSS quasar spectra.  (Bottom) The blue dashed line is the model of Fig.~7 of Gnedin \& Fan (2006) computed for the present cosmology.  The black  dot-dash line is the mfp computed from the LLS density of Choudhury (2009, Fig.~2(g)) with $\beta = 1.3$.
\label{fig:mfp_fig}
}
\end{figure}

We show the mean free path of equation~(\ref{mfp3}), and also that computed from the measured $n_{\rm LLS}$ at that redshift, in Figure~\ref{fig:mfp_fig},  where we compare with the mean values computed by MHR and Faucher-Gigu\`ere  et al.\ (2008) and with the mfp determination of Prochaska et al.\ (2009a) in the range $3.6<z<4.2$.  Note that the $1~\sigma$\ errors shown are formal statistical errors only and there may be additional systematic errors. 
Rescaled to the present cosmology, the MHR value is $25 ((1+z)/4.5)^{-4.5}\ {\rm Mpc}$\ which, while very similar in shape, is about a factor of two lower than the present results.
The mean free path has a steeper slope than that of Faucher-Gigu\`ere and is reduced by a factor of 1.2 at $z=5$, increasing the emissivity required to ionize the
IGM by the same factor. This increases the well-known discrepancy
between the required and known sources of ionization at
these high redshifts relative to the results of that paper.

Fan et al.\ (2006) have made an estimate of the mfp
based on the measured transmission and an assumed
log normal density distribution function, following Miralda-Escud\'e,
Haehhnelt \& Rees (2000). This procedure is very uncertain
at the high redshifts where the transmission is dominated
by the extreme underdense tail of the distribution function,
which is poorly described by the log normal function (Becker
et al. 2007). Perhaps not surpisingly, given this and
other major uncertainties in the interpretation, the Fan et al.\ (2006) estimates are steeper in shape and about a factor of 2.5 -- 4 lower in normalization than the present measurements in the interval $4.5<z<6$. 

The steep drop in the mfp with redshift
to $z=6$, if extrapolated to higher redshift,
predicts a very small proper length value of around
5~Mpc at $z=6.5$. If this is smaller than the
size of the HII bubbles at the overlap epoch
there could be important consequences for the
evolution of the ionization rate (e.g. Furlanetto
and Mesinger 2009).

However, it is probable that as we move to earlier
periods the evolution of the geometry and ionization
rates will combine to modify the evolution of the LLS,
and simple extrapolations to higher redshifts might
not be justified despite the smooth evolution from
$z=0-6$. Full numerical simulations, capable of the
resolution required to describe the Lyman limit systems,
are still extremely challenging and subject to many
uncertainties. In Figure~\ref{fig:mfp_fig} we compare the present
observations with the simulation of Kohler \& Gnedin (2007)
and Gnedin \& Fan (2006). These models are not a particularly good description of the lower redshift
mfp or $n(z)$,  presumably because of uncertainties in the evolution of the intergalactic ionization parameter.
However, they emphasize that the mfp could evolve smoothly
to $z=6$ and then change rapidly at higher redshifts.
In this case the models correspond to reionization
just above $z=6$ and the mean free path drops rapidly
above this redshift.

However, even the sign of the evolution in the mean free
path is poorly determined at high redshift. Choudhury
\& Ferrara (2005, 2006) and Choudhury (2009) have computed semi-analytic models
based on the formalism of Miralda-Escud\'e (2003)
with ionization histories chosen to match the high redshift
constraints. These models provide a very good match to
the $z=3-6$ LLS density. However, as can be seen in Figure~\ref{fig:mfp_fig},
they predict that the mfp flattens above $z=6$ and the $n(z)$
drops. In these models the bulk of the IGM volume
is still ionized up to $z=8$ and the change in the mfp
corresponds to the transition away from popIII ionization
at redshifts $z<8$.

We conclude that it is probably wise not to extrapolate
the mfp beyond the measured range. However, the shape and
normalization of the mfp evolution at $z<6$ may still provide
useful tests of the higher redshift evolution since any
model must satisfy the $z=6$ boundary condition.

\acknowledgments

This research was
supported by the National Science Foundation under grant AST~06--07871.
The authors acknowledge helpful conversations with Andrea Ferrara and Tirthankar Choudhury.

\newpage
\clearpage

\appendix

\begin{figure*}[h]
\includegraphics[width=6.5in,angle=90,scale=0.8]{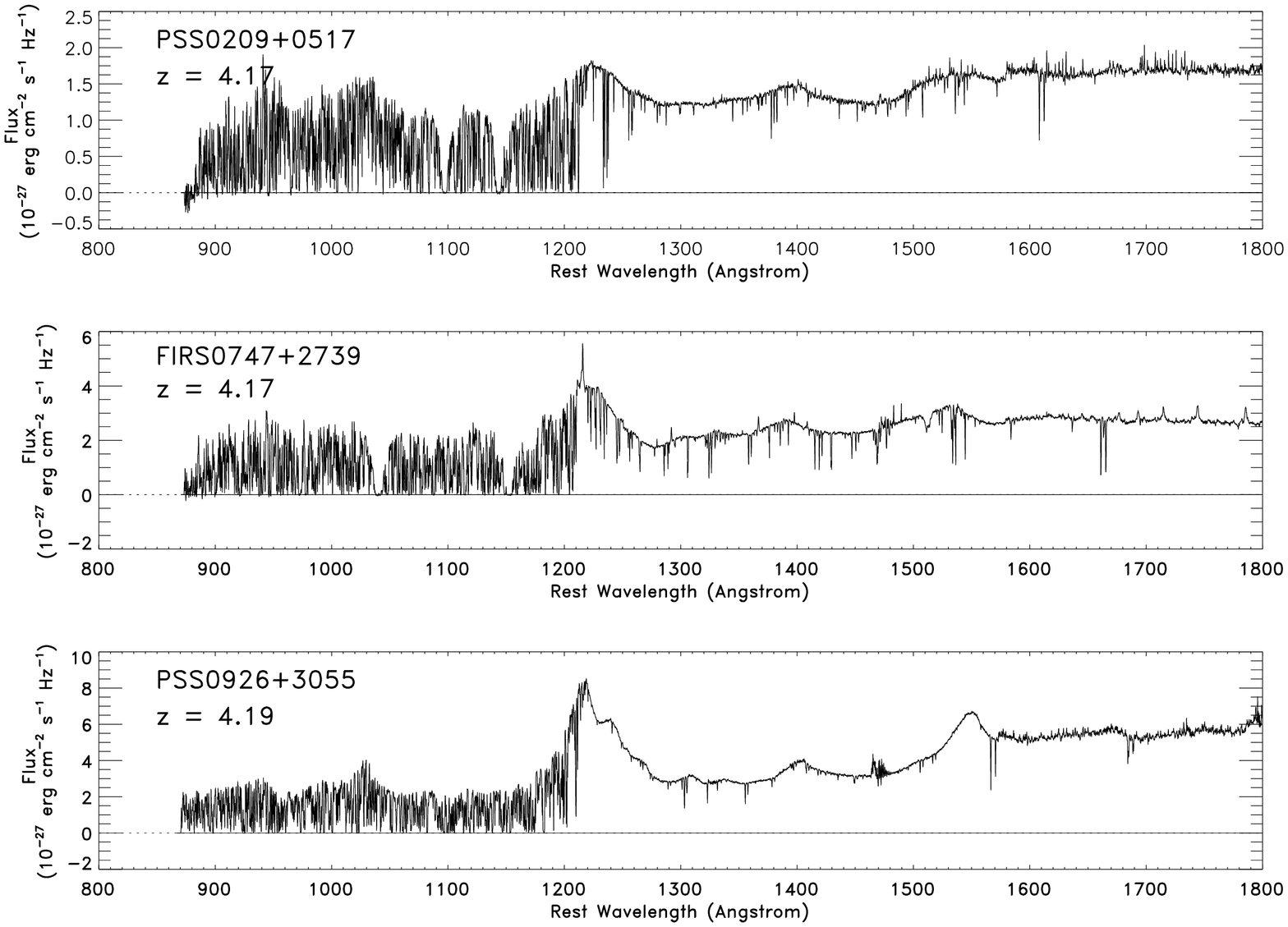}
  \caption{ESI spectra of the QSOs used in this investigation.
\label{fig:spectra1}
}
\figurenum{1}
\end{figure*}

\newpage
\clearpage

\begin{figure*}[h]
  \includegraphics[width=6.5in,angle=90,scale=0.8]{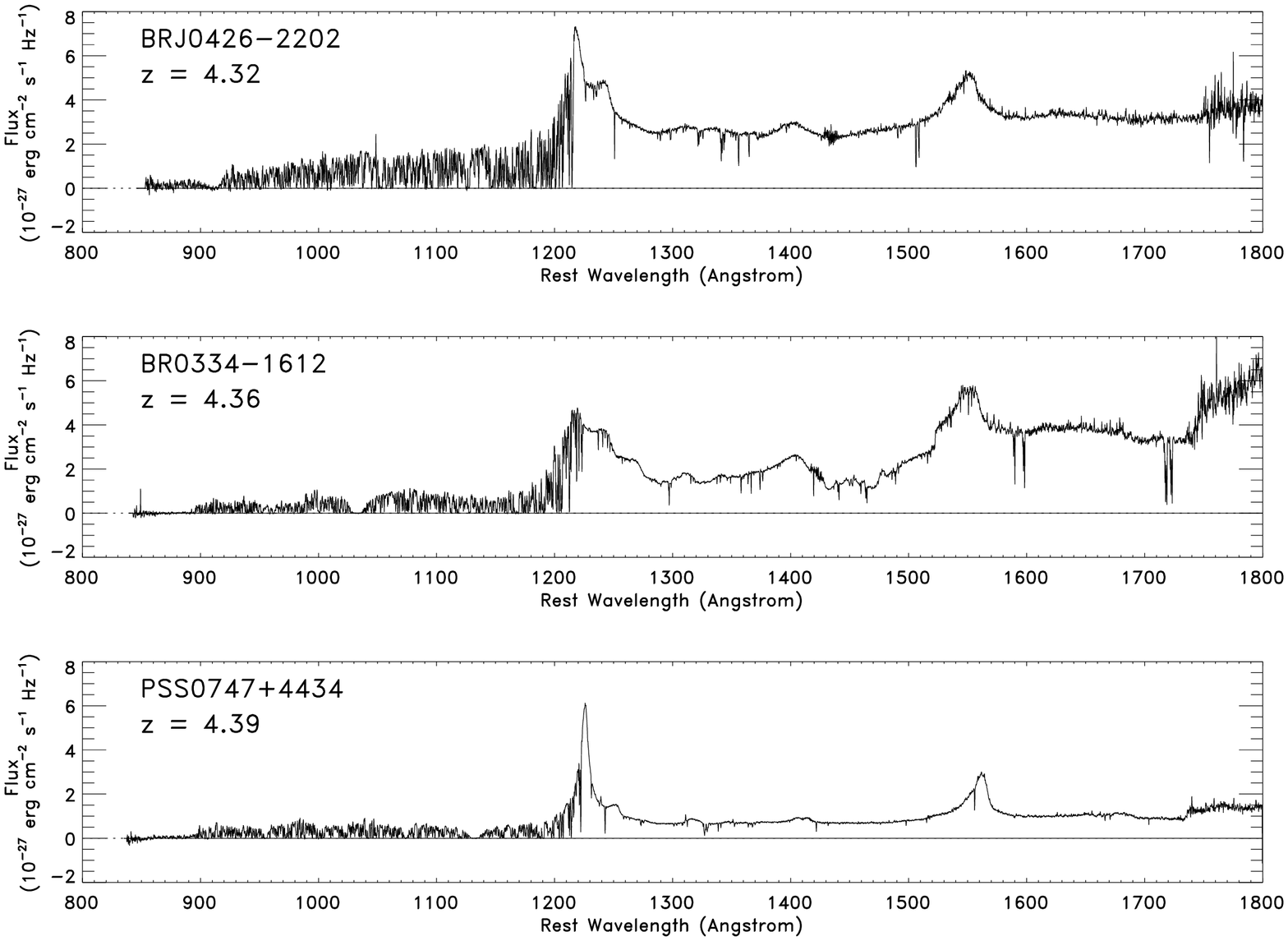}
  \caption{{\it contd.}
\label{fig:spectra2}
}
\figurenum{1}
\end{figure*}

\newpage
\clearpage

\begin{figure*}[h]
  \includegraphics[width=6.5in,angle=90,scale=0.8]{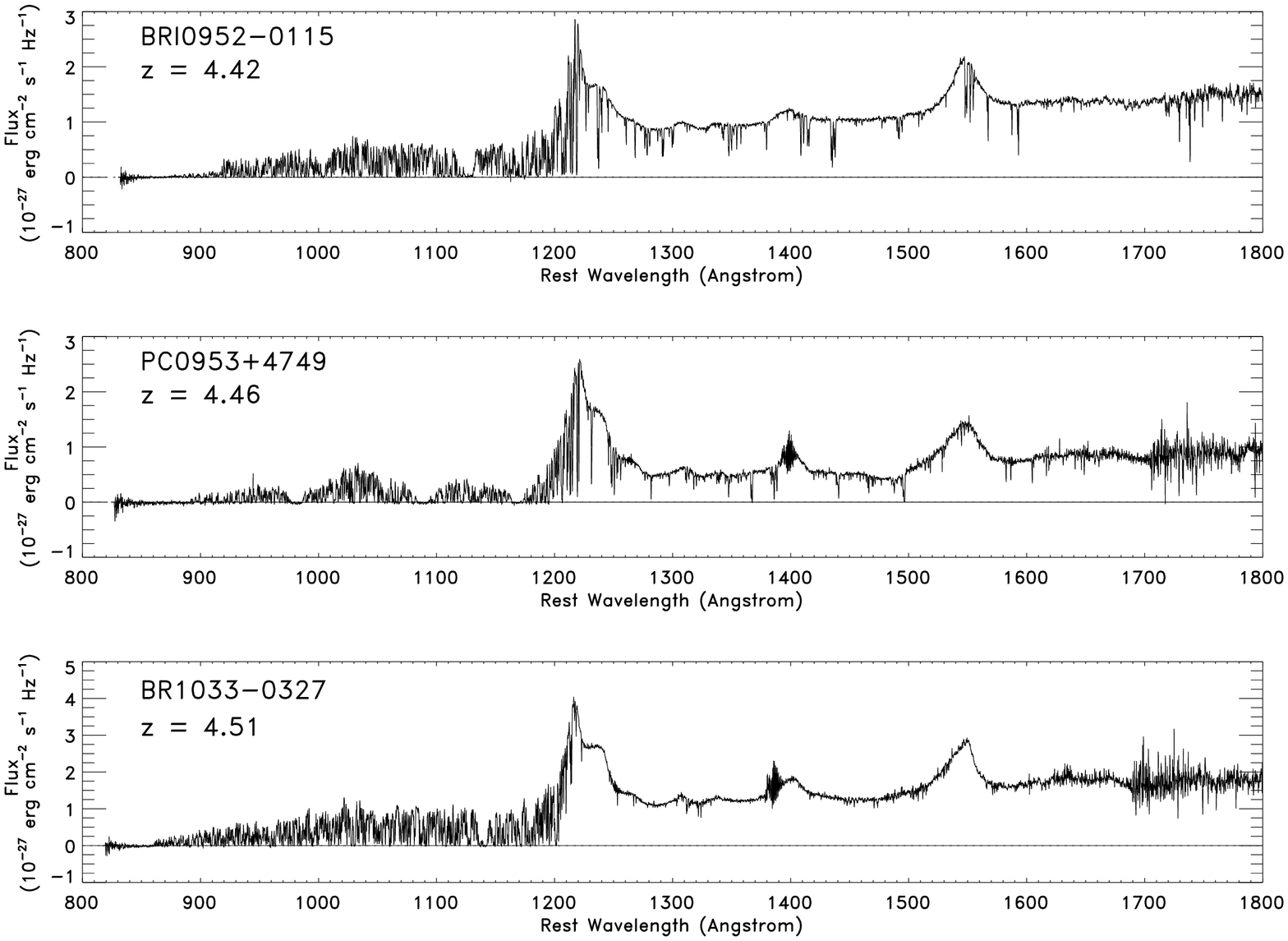}
  \caption{{\it contd.}
\label{fig:spectra3}
}
\figurenum{1}
\end{figure*}

\newpage
\clearpage

\begin{figure*}[h]
  \includegraphics[width=6.5in,angle=90,scale=0.8]{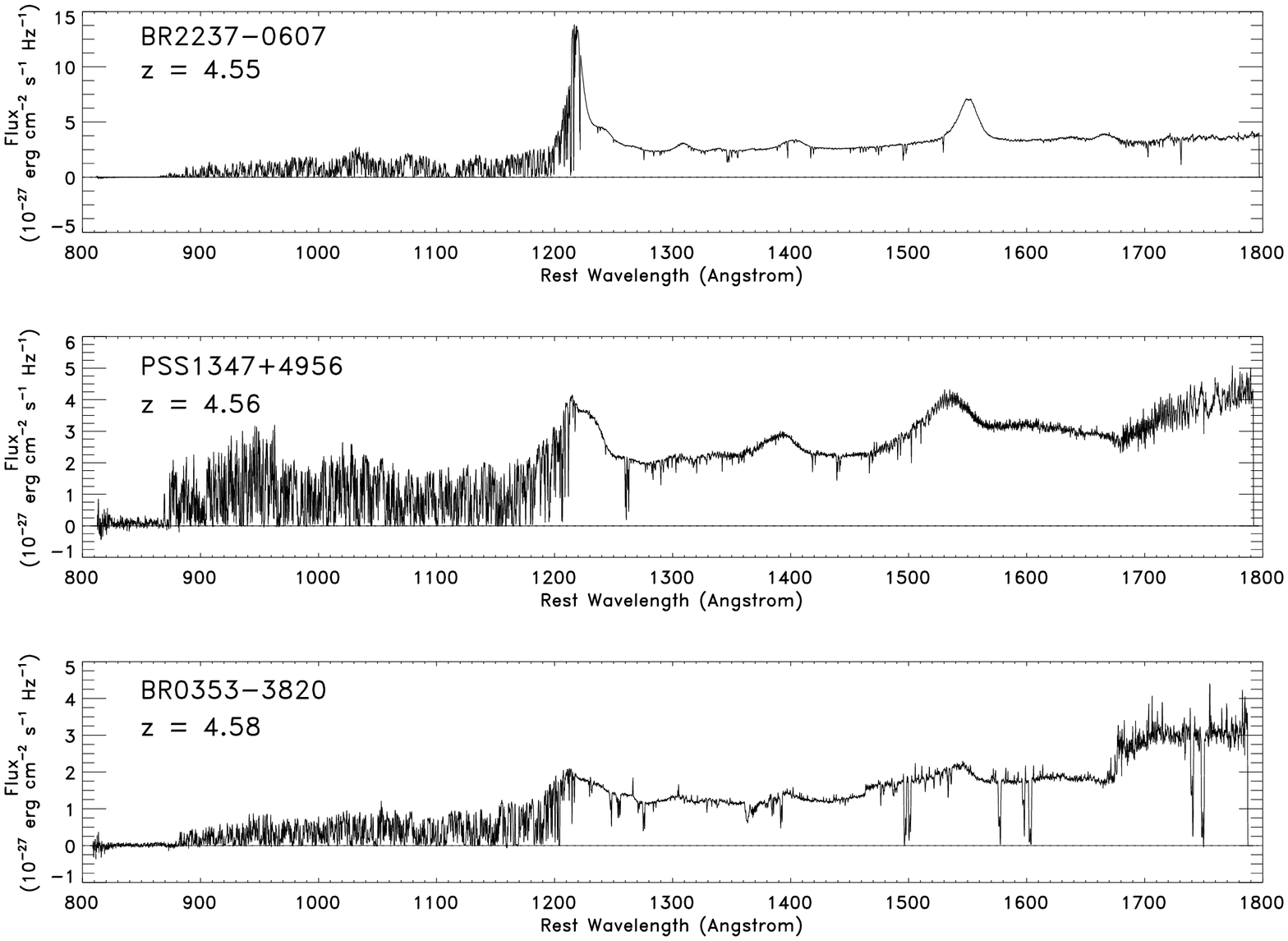}
  \caption{{\it contd.}
\label{fig:spectra4}
}
\figurenum{1}
\end{figure*}

\newpage
\clearpage

\begin{figure*}[h]
  \includegraphics[width=6.5in,angle=90,scale=0.8]{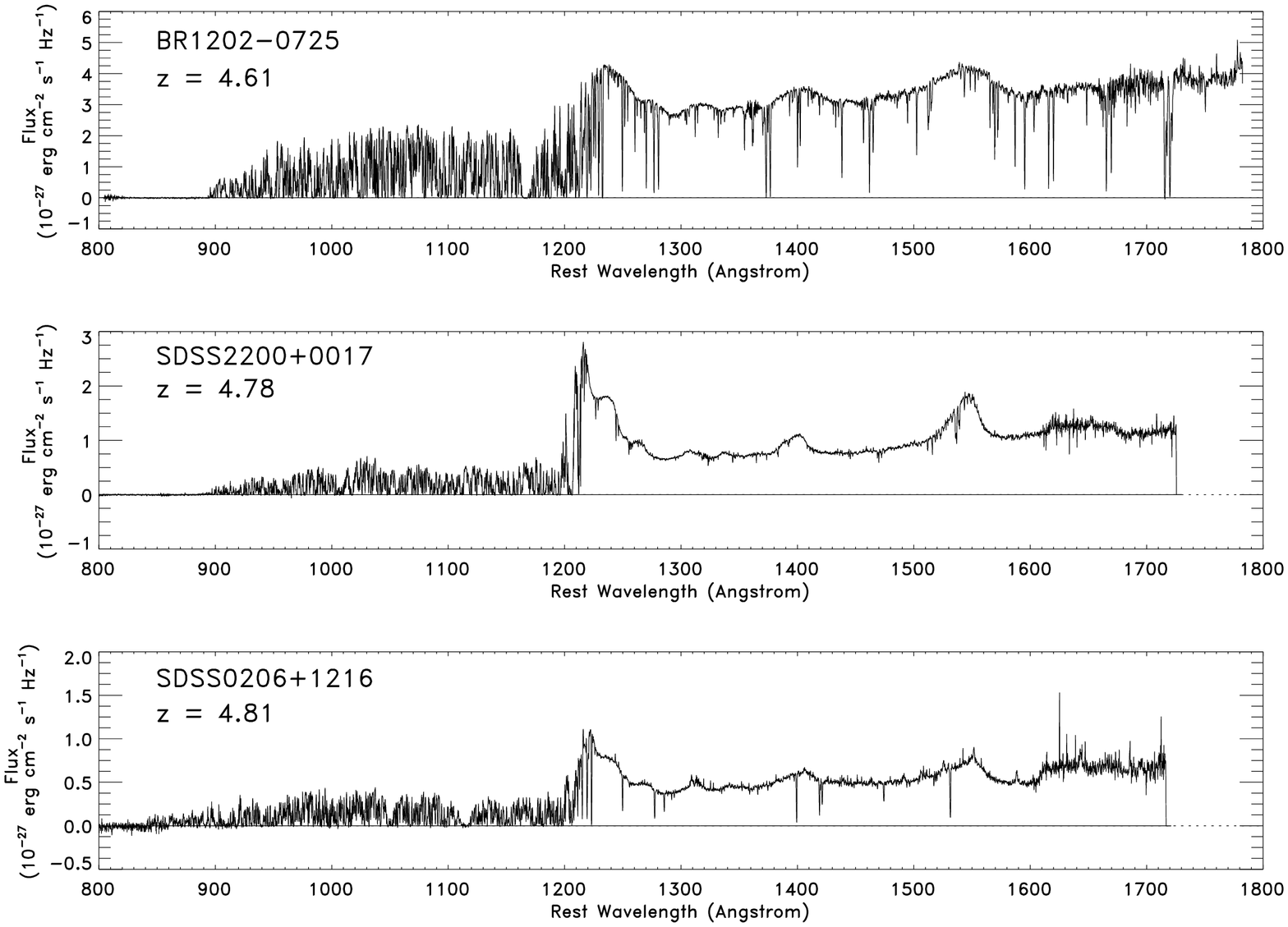}
  \caption{{\it contd.}
\label{fig:spectra5}
}
\figurenum{1}
\end{figure*}

\newpage
\clearpage

\begin{figure*}[h]
  \includegraphics[width=6.5in,angle=90,scale=0.8]{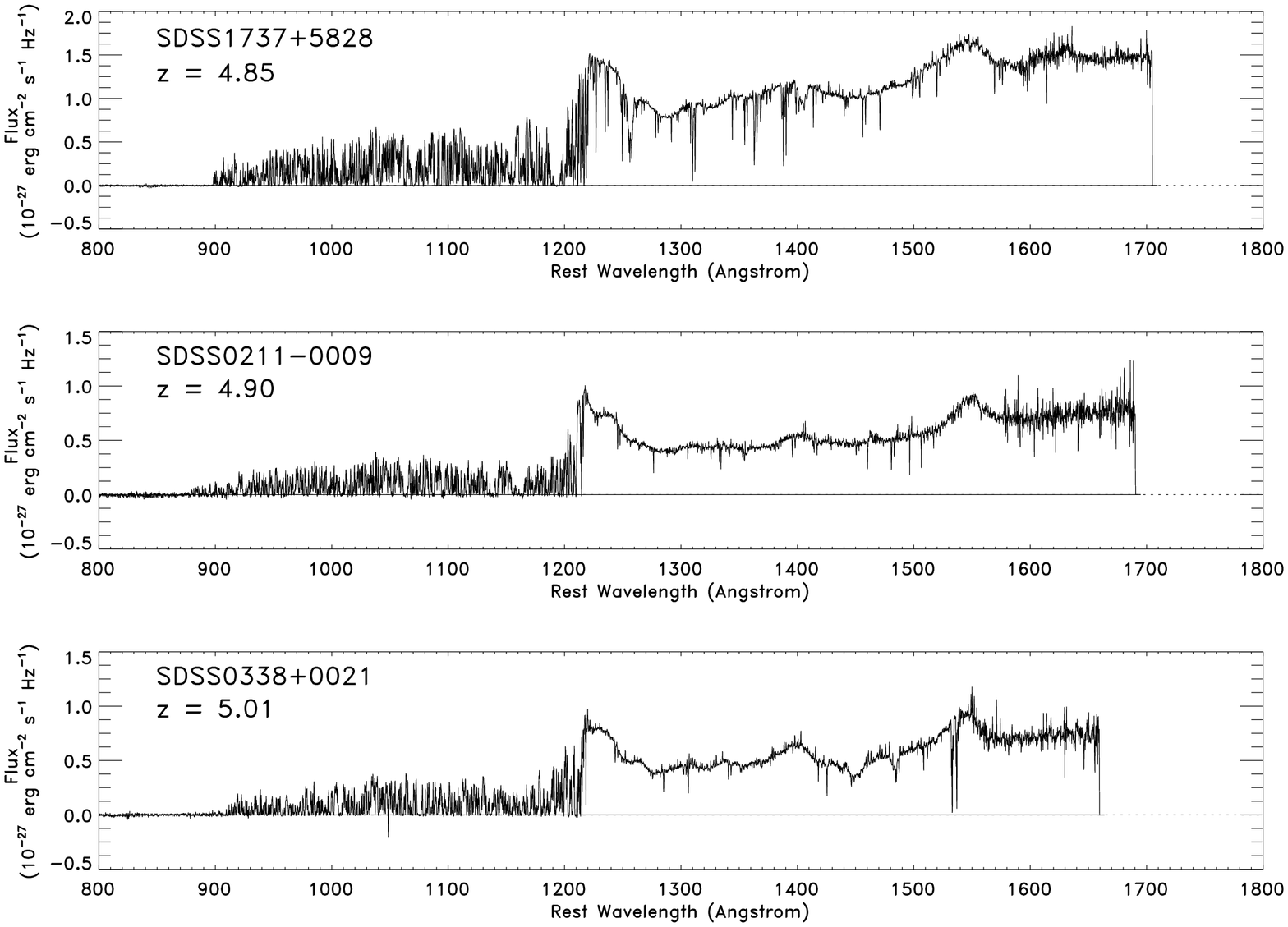}
  \caption{{\it contd.}
\label{fig:spectra6}
}
\figurenum{1}
\end{figure*}

\newpage
\clearpage

\begin{figure*}[h]
  \includegraphics[width=6.5in,angle=90,scale=0.8]{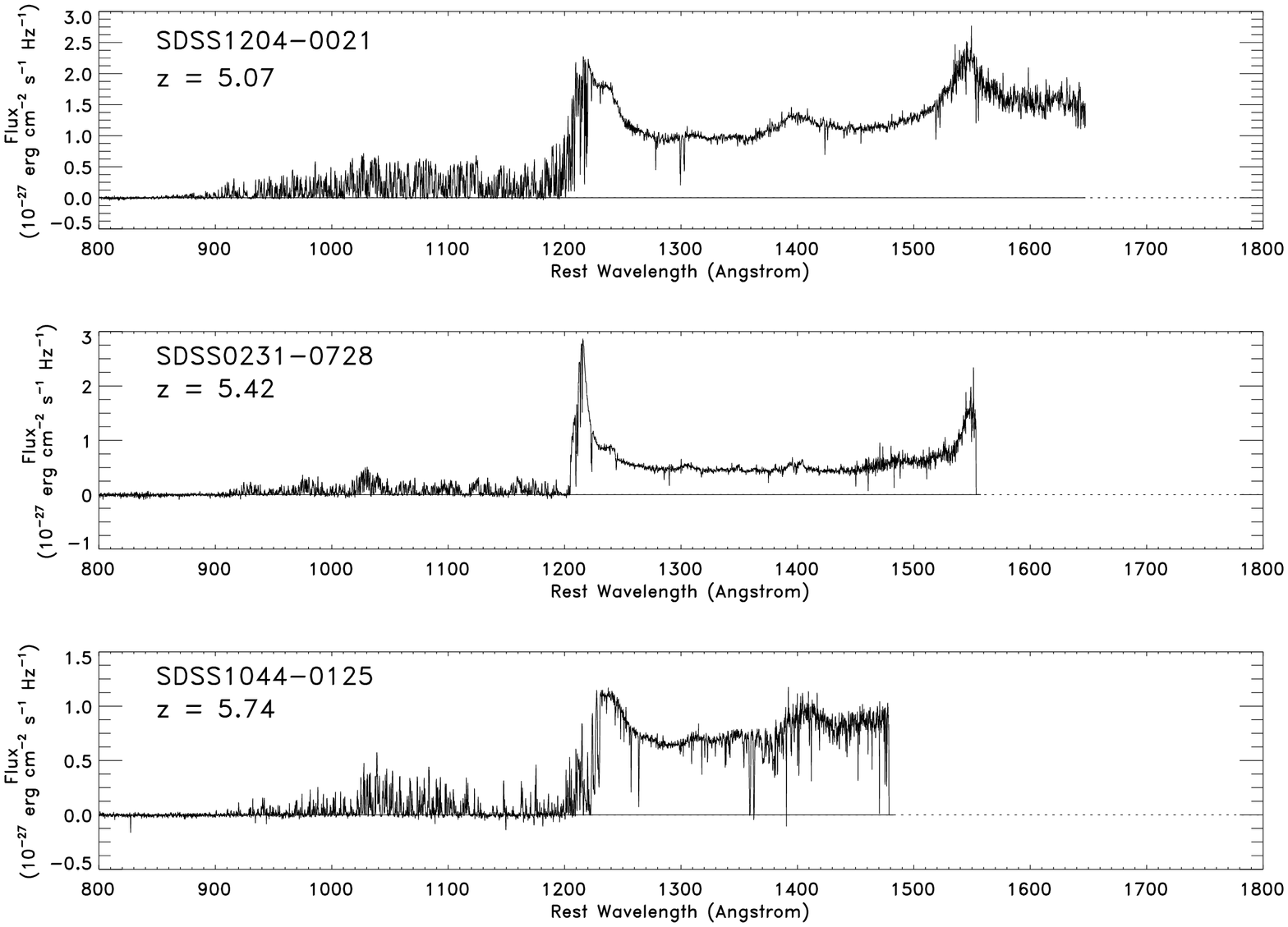}
  \caption{{\it contd.}
\label{fig:spectra7}
}
\figurenum{1}
\end{figure*}

\newpage
\clearpage

\begin{figure*}[h]
  \includegraphics[width=6.5in,angle=90,scale=0.8]{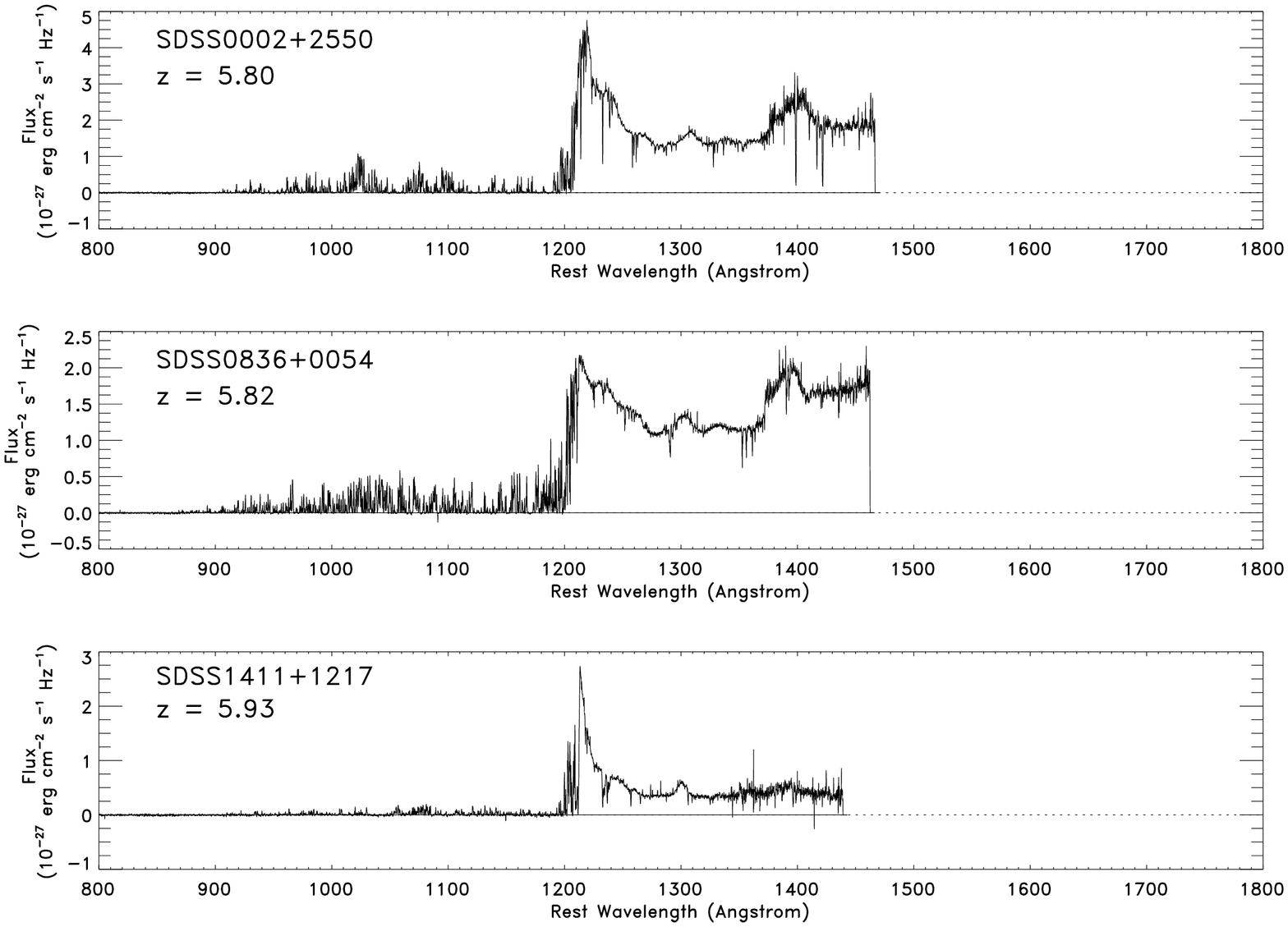}
  \caption{{\it contd.}
\label{fig:spectra8}
}
\figurenum{1}
\end{figure*}

\newpage
\clearpage

\begin{figure*}[h]
  \includegraphics[width=6.5in,angle=90,scale=0.8]{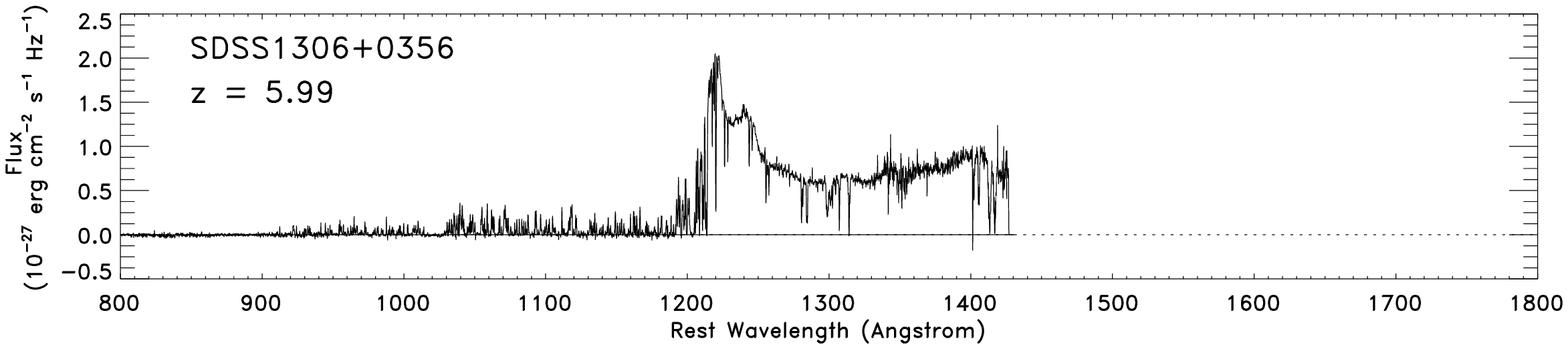}
  \caption{{\it contd.}
\label{fig:spectra9}
}
\figurenum{1}
\end{figure*}

\newpage
\clearpage

\begin{figure}[h]
\figurenum{2}
     \includegraphics[width=3in,angle=90,scale=0.9]{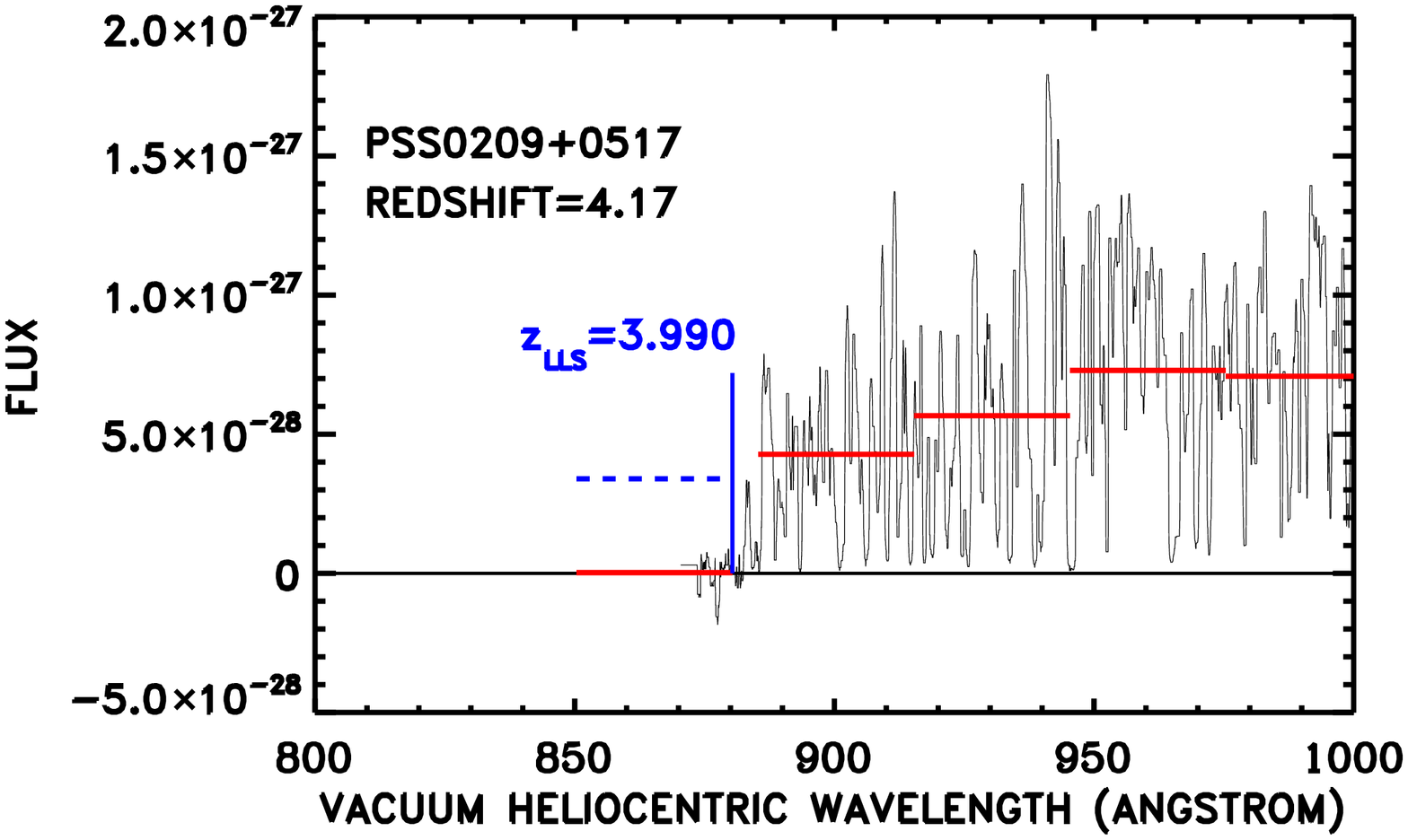}
  \includegraphics[width=3in,angle=90,scale=0.9]{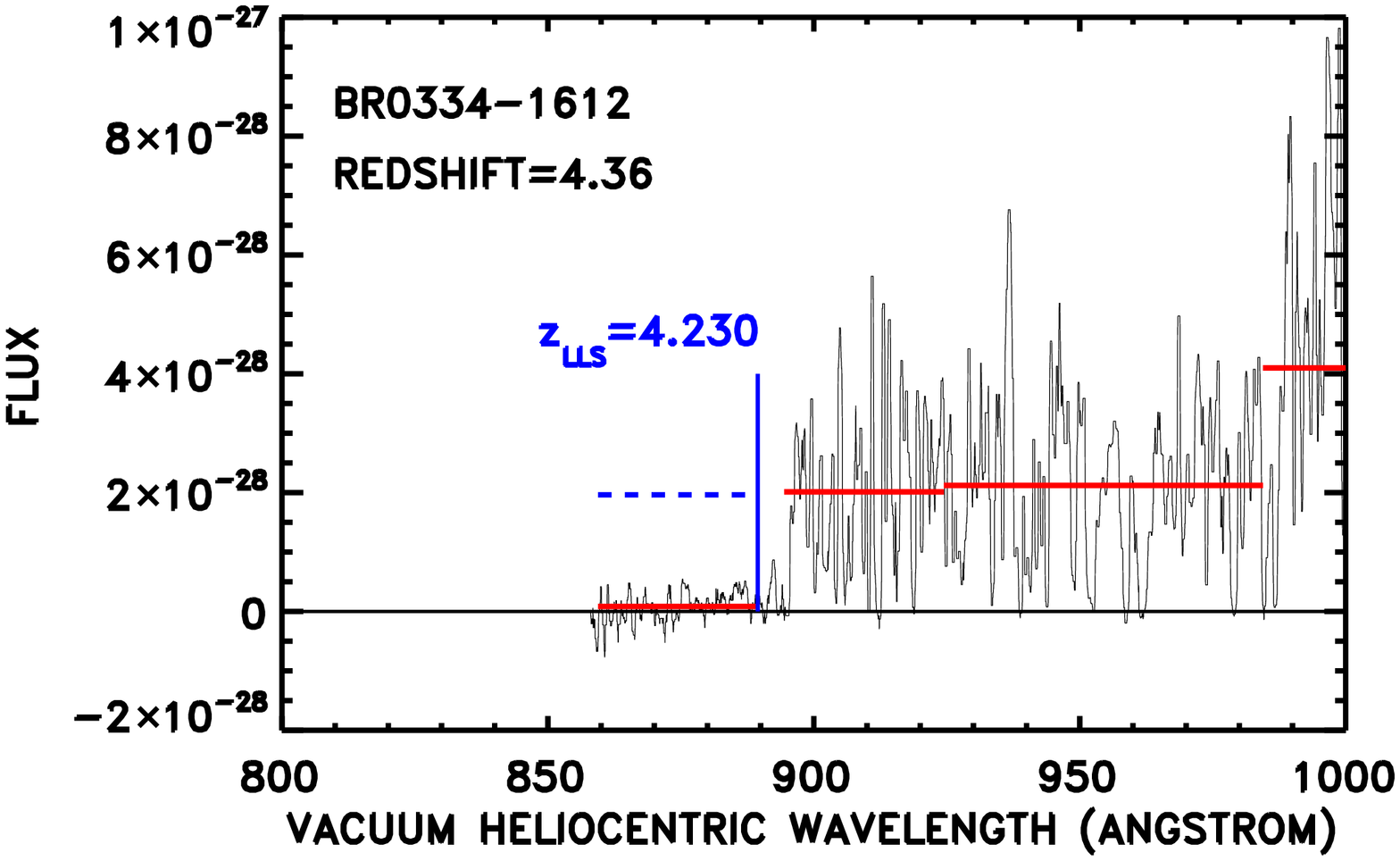}
\caption{Lyman limit systems with $\tau > 1$\ in quasars at intermediate redshift.  The spectra are plotted in the rest frame of the quasar.  The blue vertical line marks the wavelength $\lambda_{\rm LLS}$\ of the LLS and the redshift $z_{\rm LLS}$\ is marked alongside.  The red horizontal bars show the average flux in $30~\rm\AA$\ bins below $\lambda_{\rm LLS}$\ and to redward of $\lambda_{\rm LLS} + 10~{\rm\AA}$.  The blue dashed line shows the predicted continuum level below $\lambda_{\rm LLS}$\ based on a power-law fit to the redward continuum.   
\label{fig:tau_displaya}
}
\end{figure}

\begin{figure}[h]
\figurenum{2}
     \includegraphics[width=3in,angle=90,scale=0.9]{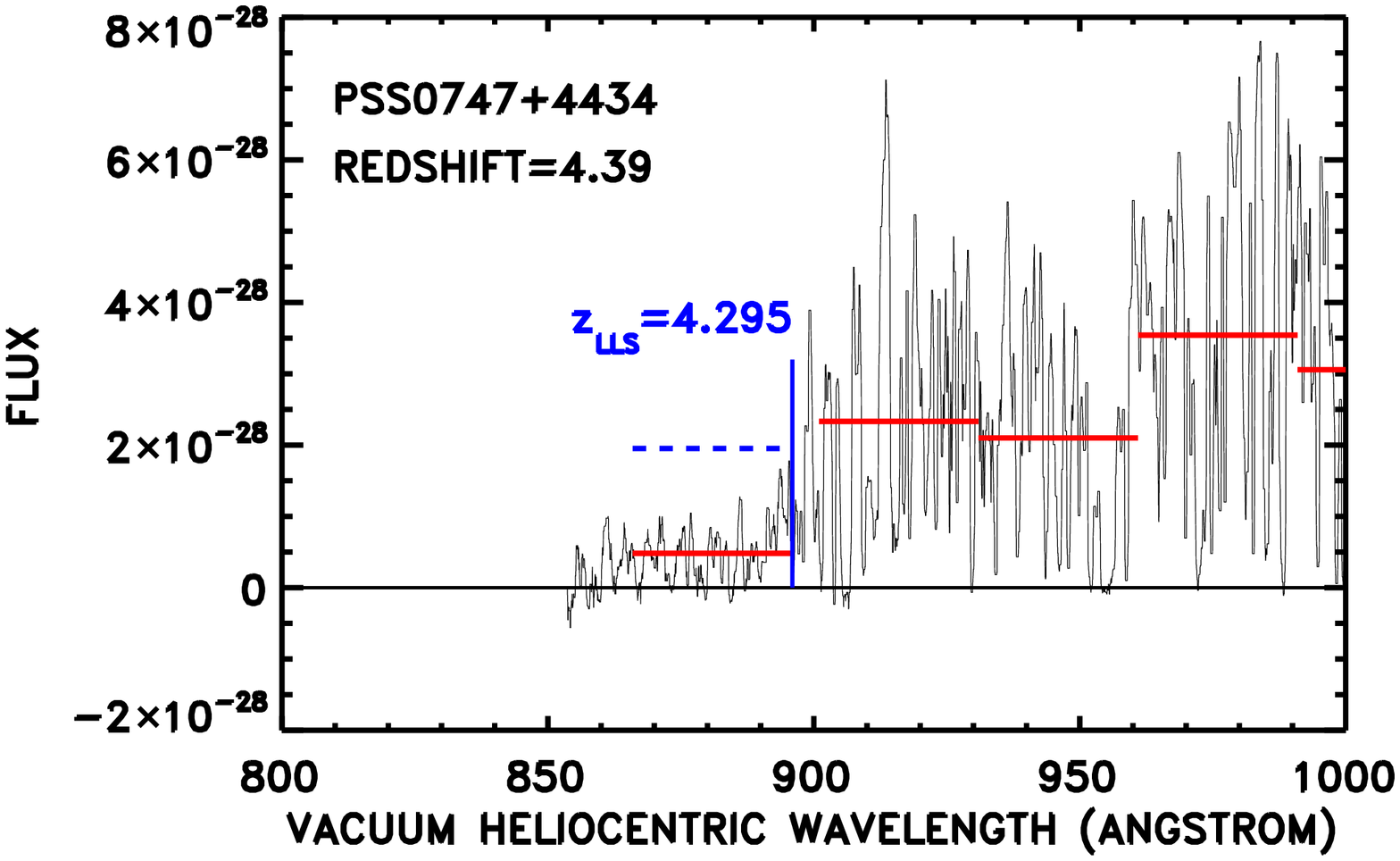}
  \includegraphics[width=3in,angle=90,scale=0.9]{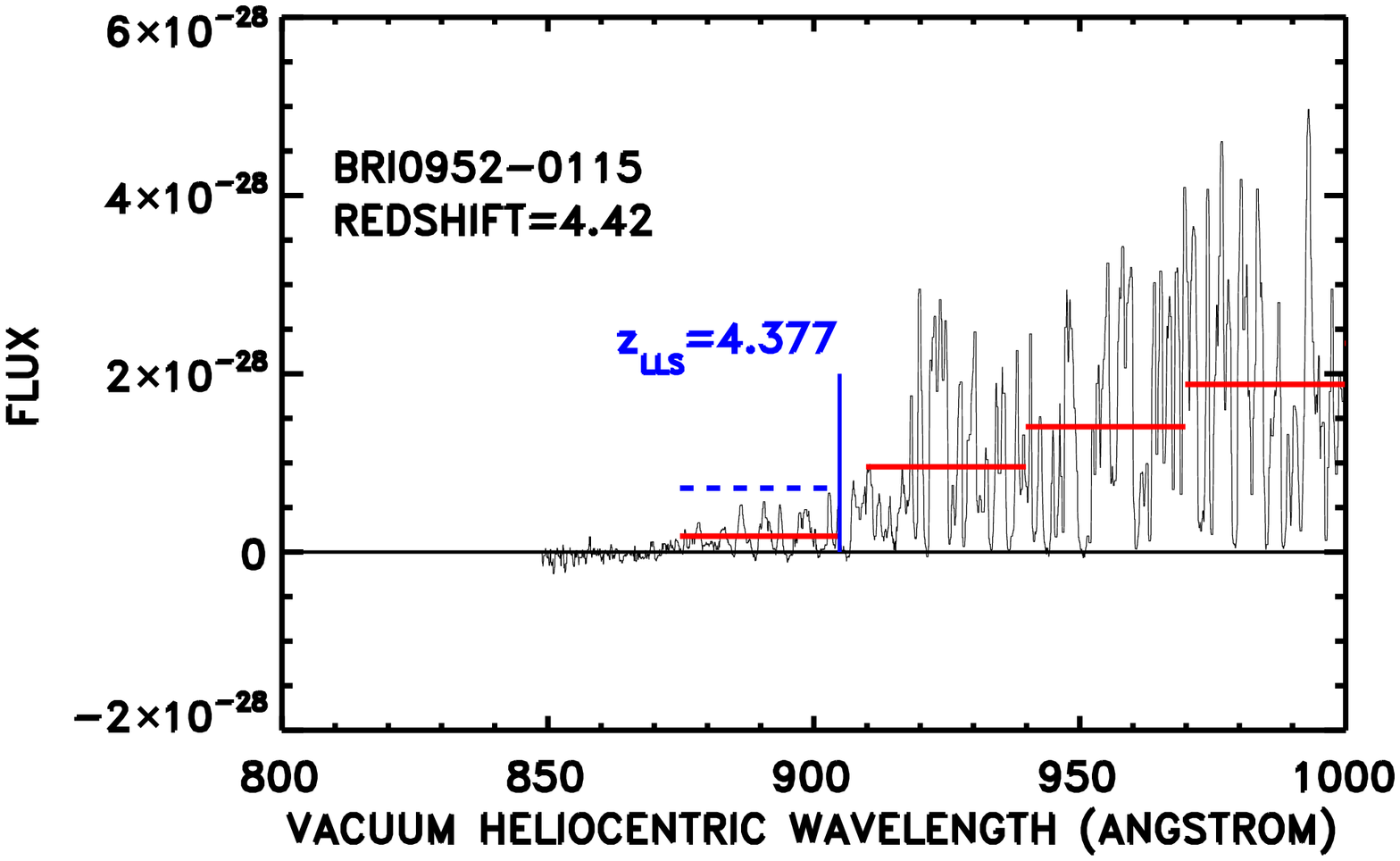}
\caption{{\it contd.}   
\label{fig:tau_displaya}
}
\end{figure}

\begin{figure}[h]
\figurenum{2}
    \includegraphics[width=3in,angle=90,scale=0.9]{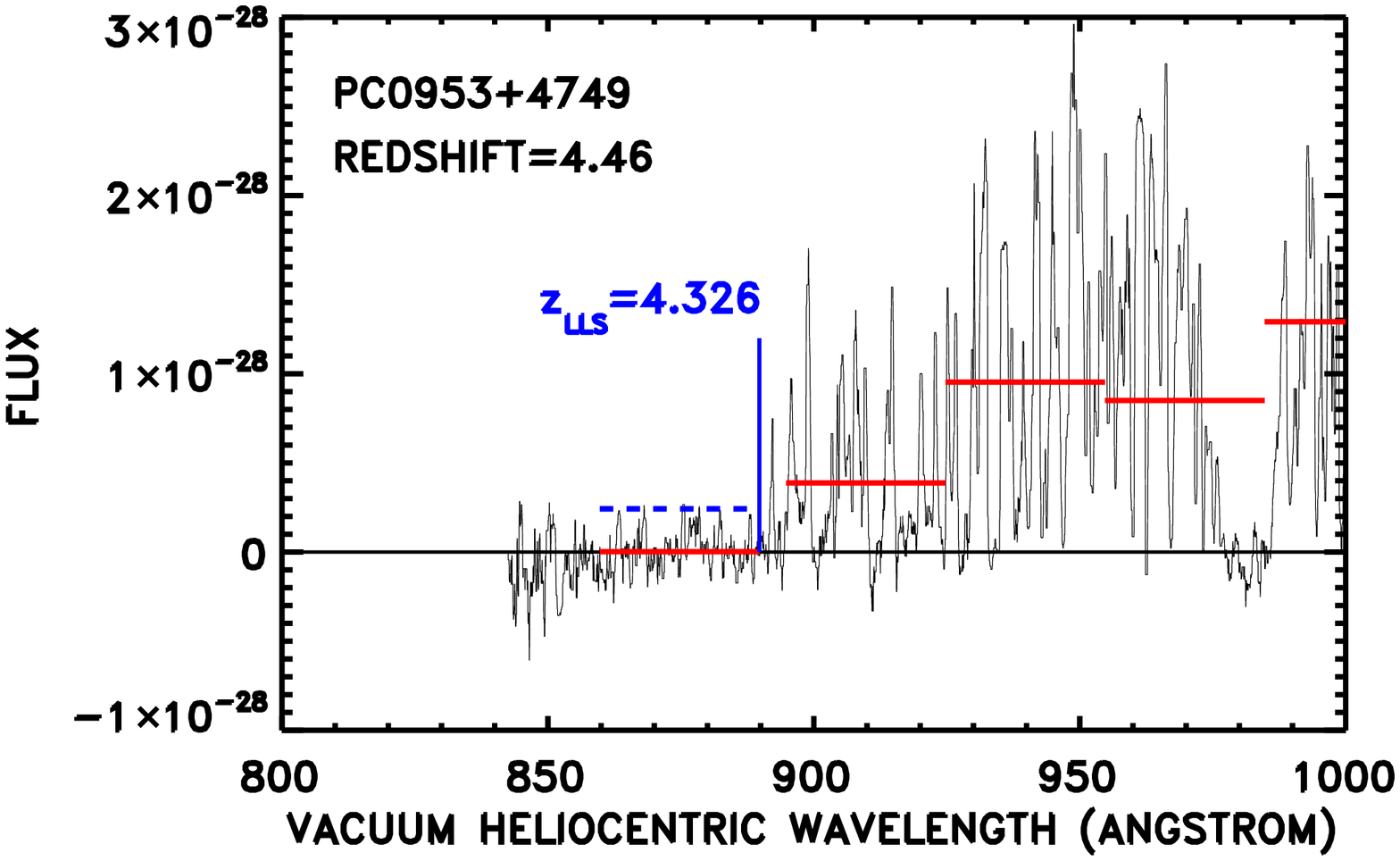}
    \includegraphics[width=3in,angle=90,scale=0.9]{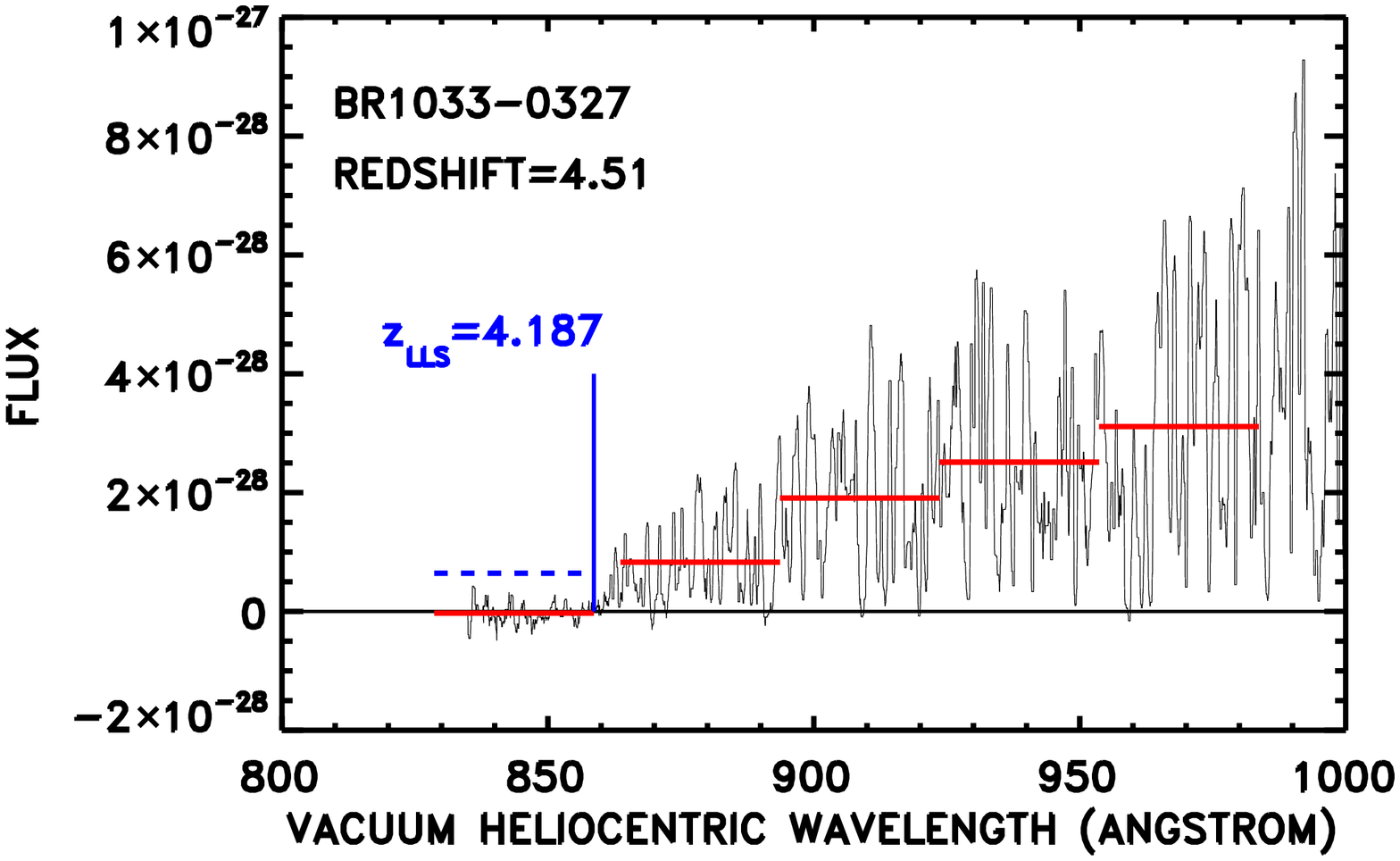}

 \caption{{\it contd.}
\label{fig:tau_displayb}
}
\end{figure}

\begin{figure}[h]
\figurenum{2}
   \includegraphics[width=3in,angle=90,scale=0.9]{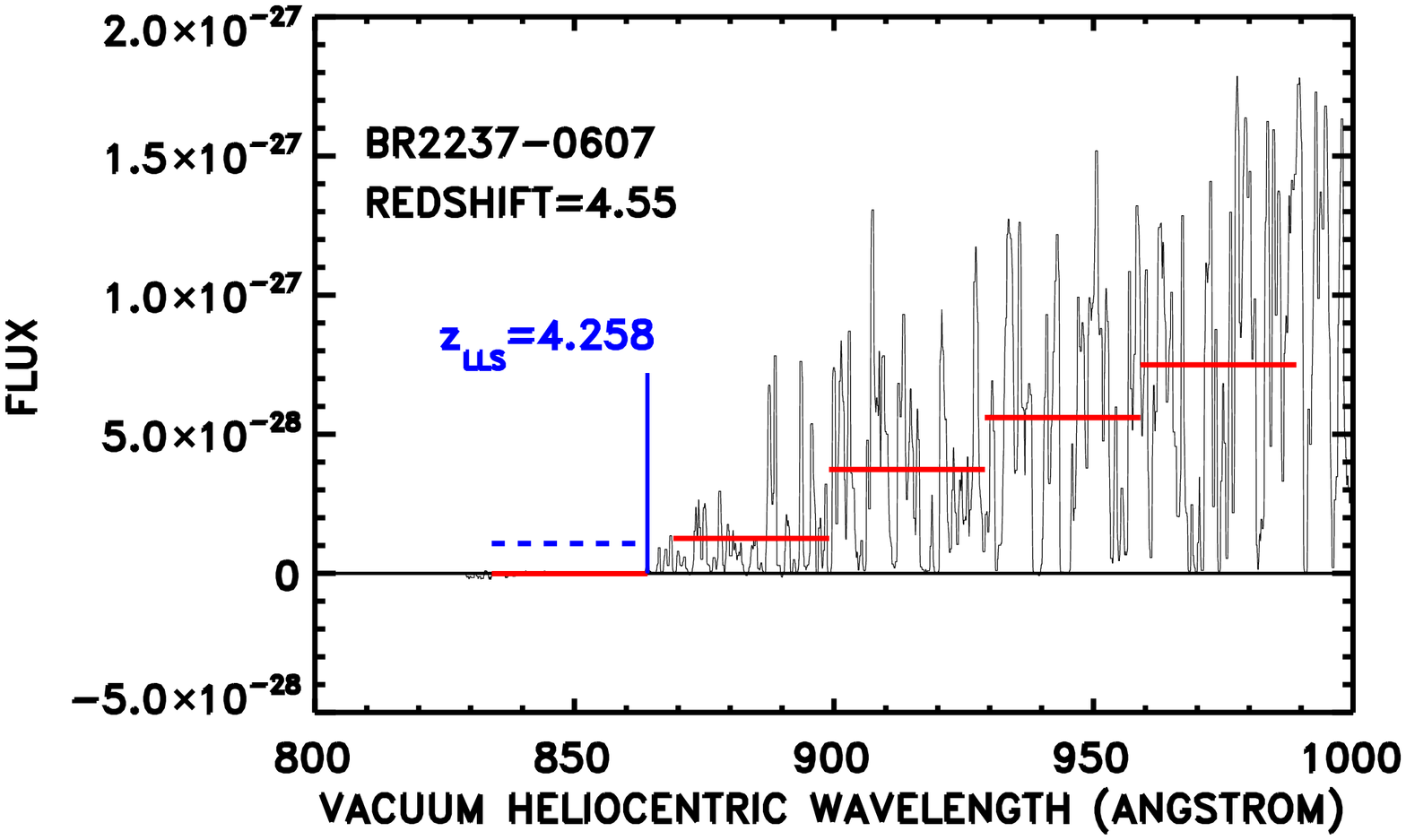}
   \includegraphics[width=3in,angle=90,scale=0.9]{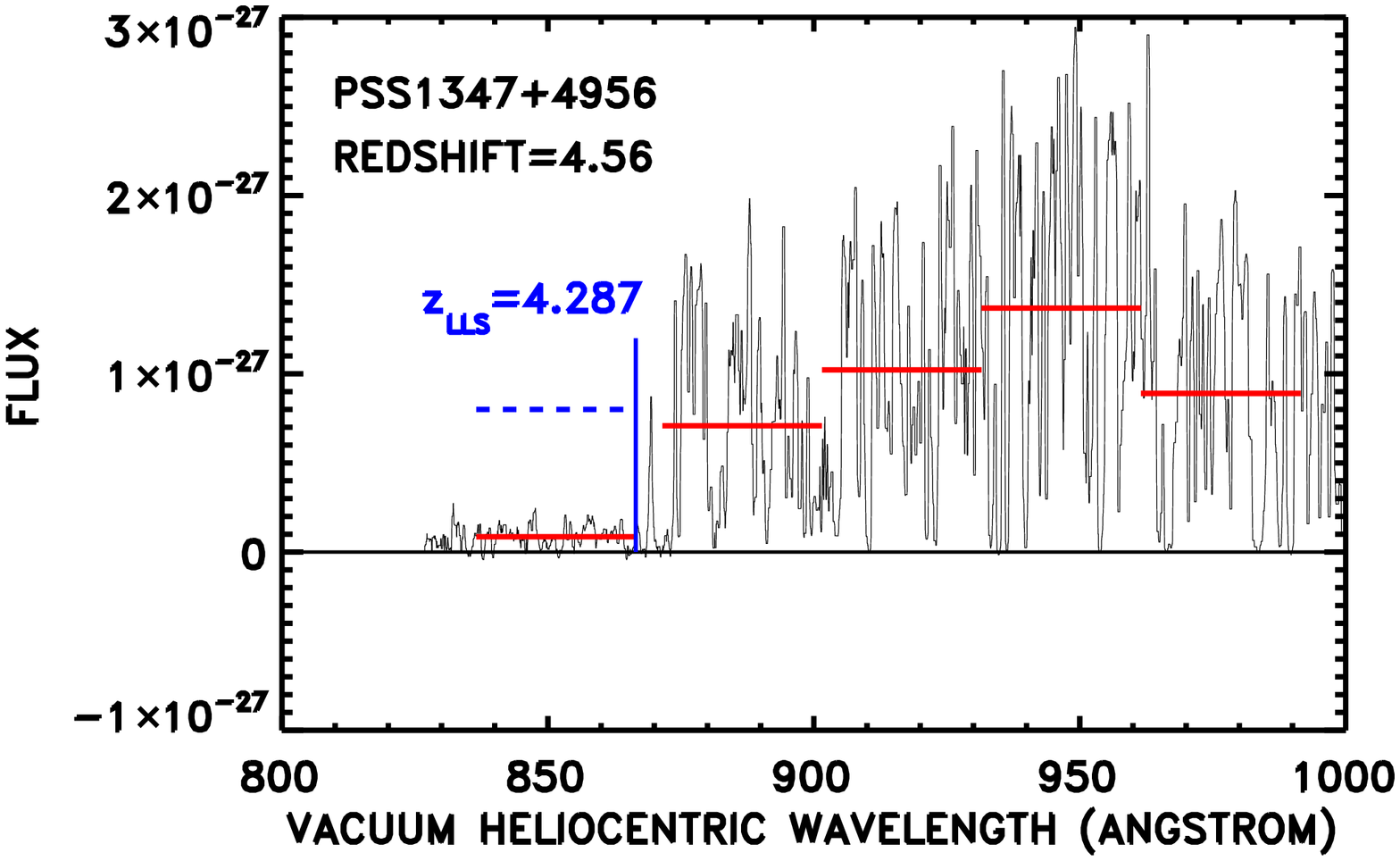}
 \caption{{\it contd.}
\label{fig:tau_displayb}
}
\end{figure}

\begin{figure}[h]
\figurenum{2}
   \includegraphics[width=3in,angle=90,scale=0.9]{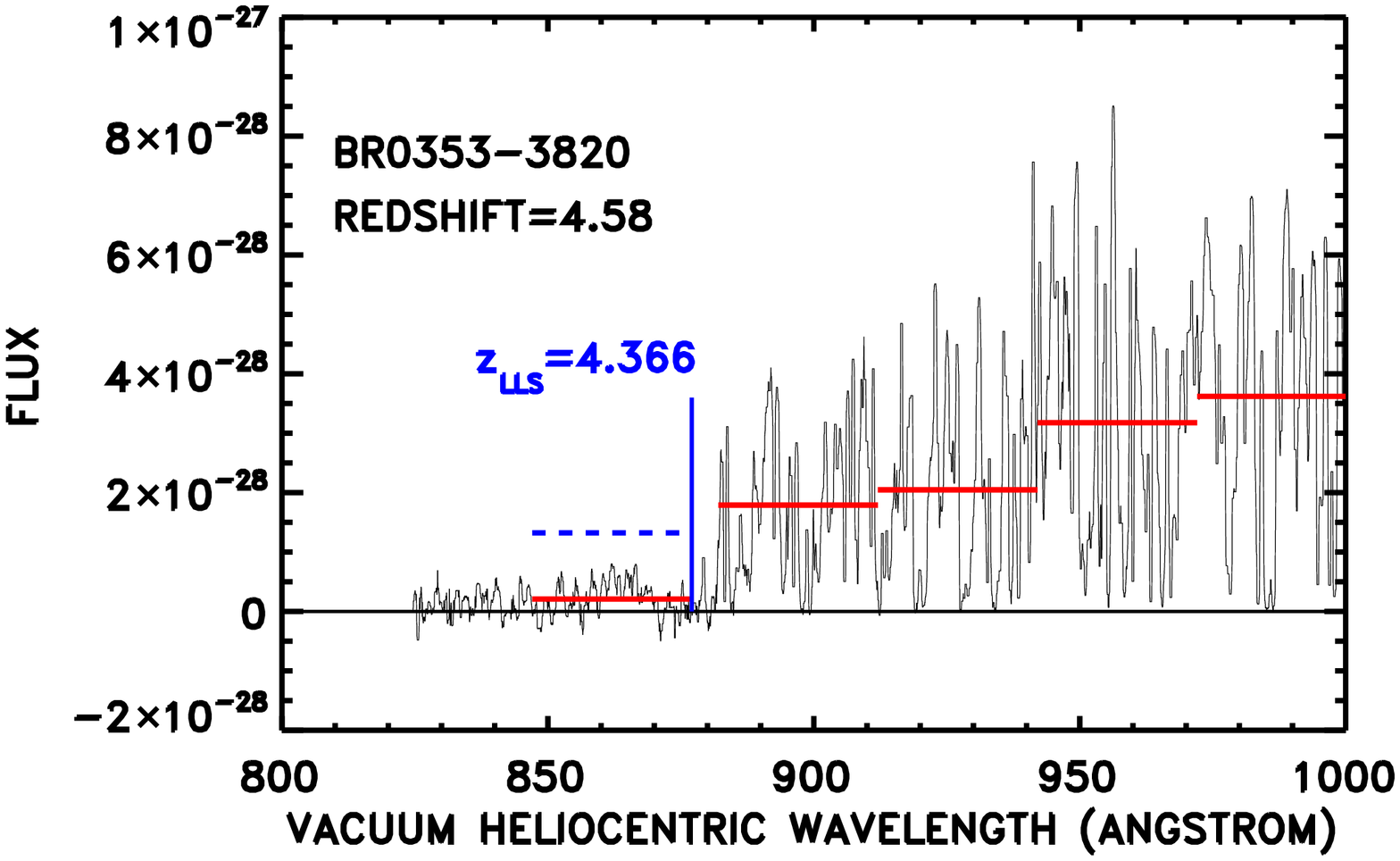}
   \includegraphics[width=3in,angle=90,scale=0.9]{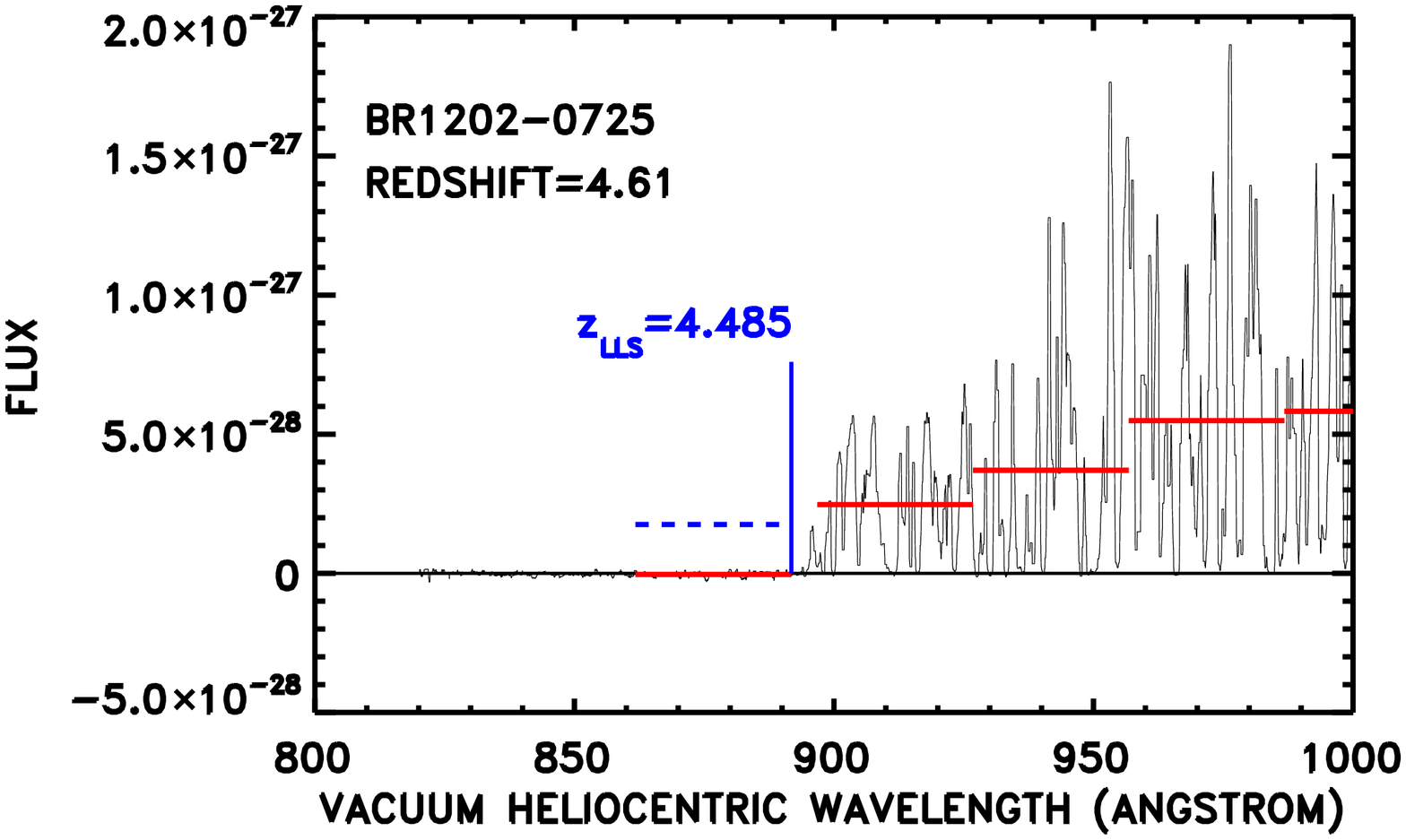}
 \caption{{\it contd.}
\label{fig:tau_displayb}
}
\end{figure}

\begin{figure}[h]
\figurenum{2}
   \includegraphics[width=3in,angle=90,scale=0.9]{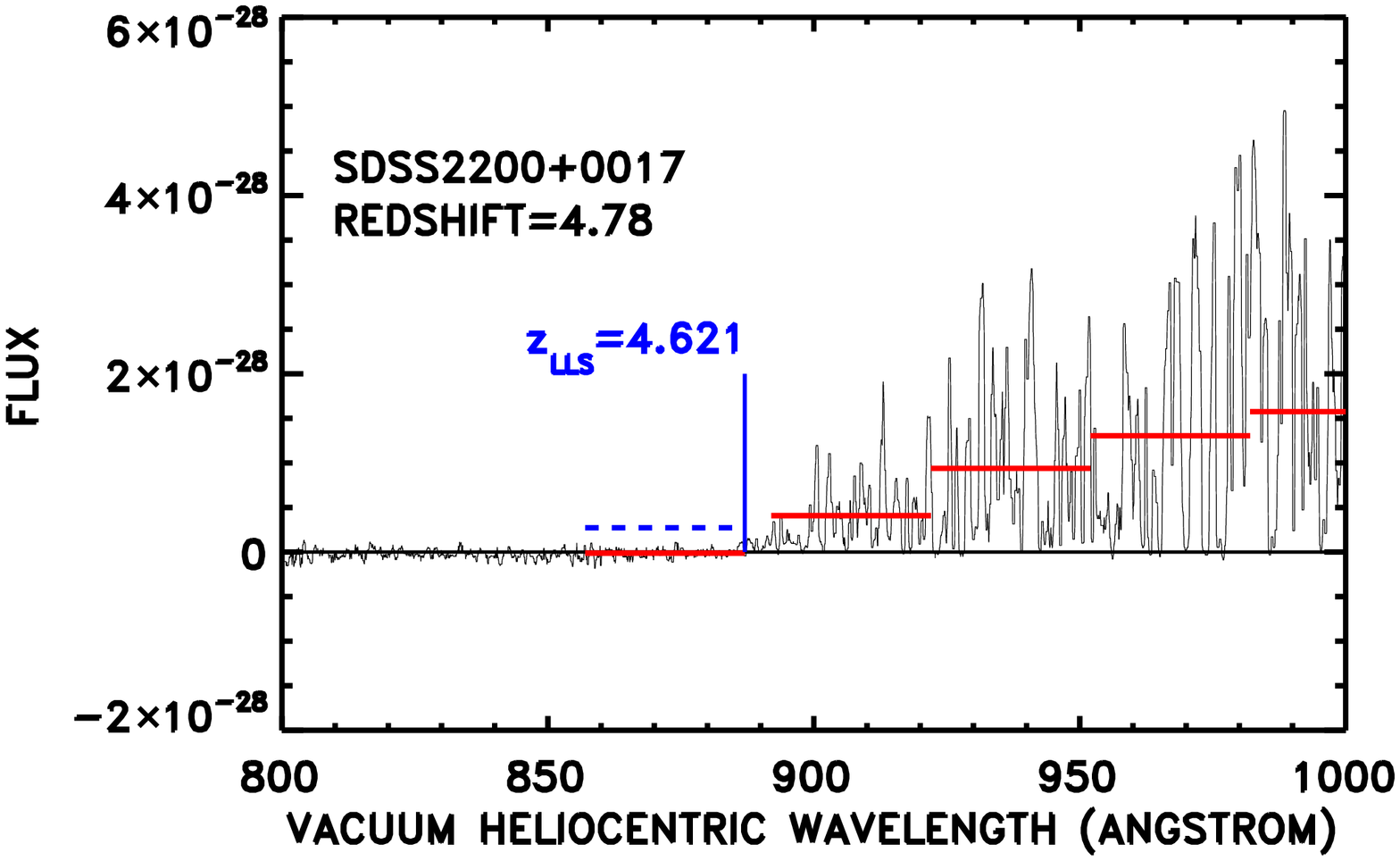}
   \includegraphics[width=3in,angle=90,scale=0.9]{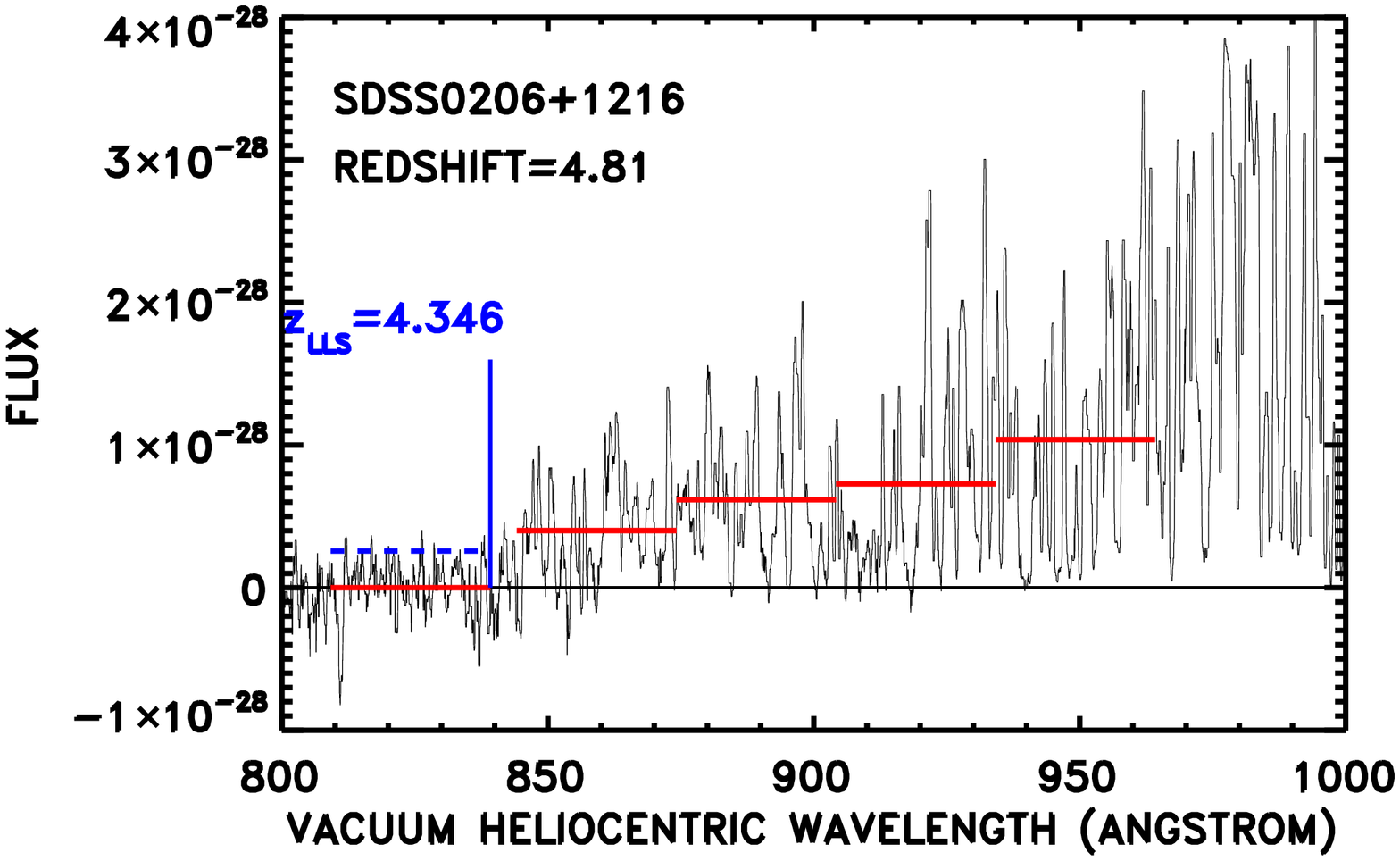}
 \caption{{\it contd.}
\label{fig:tau_displayb}
}
\end{figure}

\begin{figure}[h]
\figurenum{2}
   \includegraphics[width=3in,angle=90,scale=0.9]{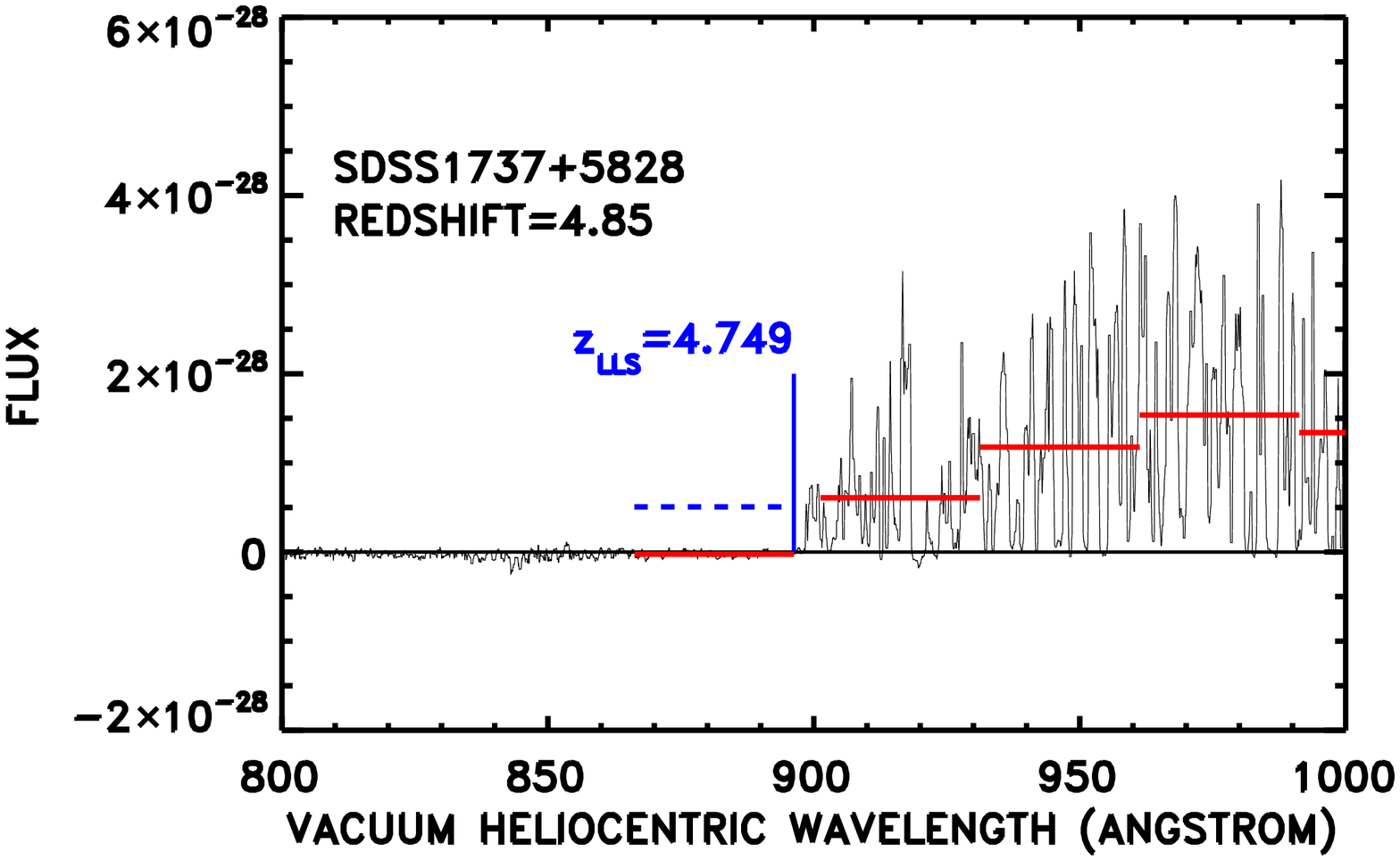}
   \includegraphics[width=3in,angle=90,scale=0.9]{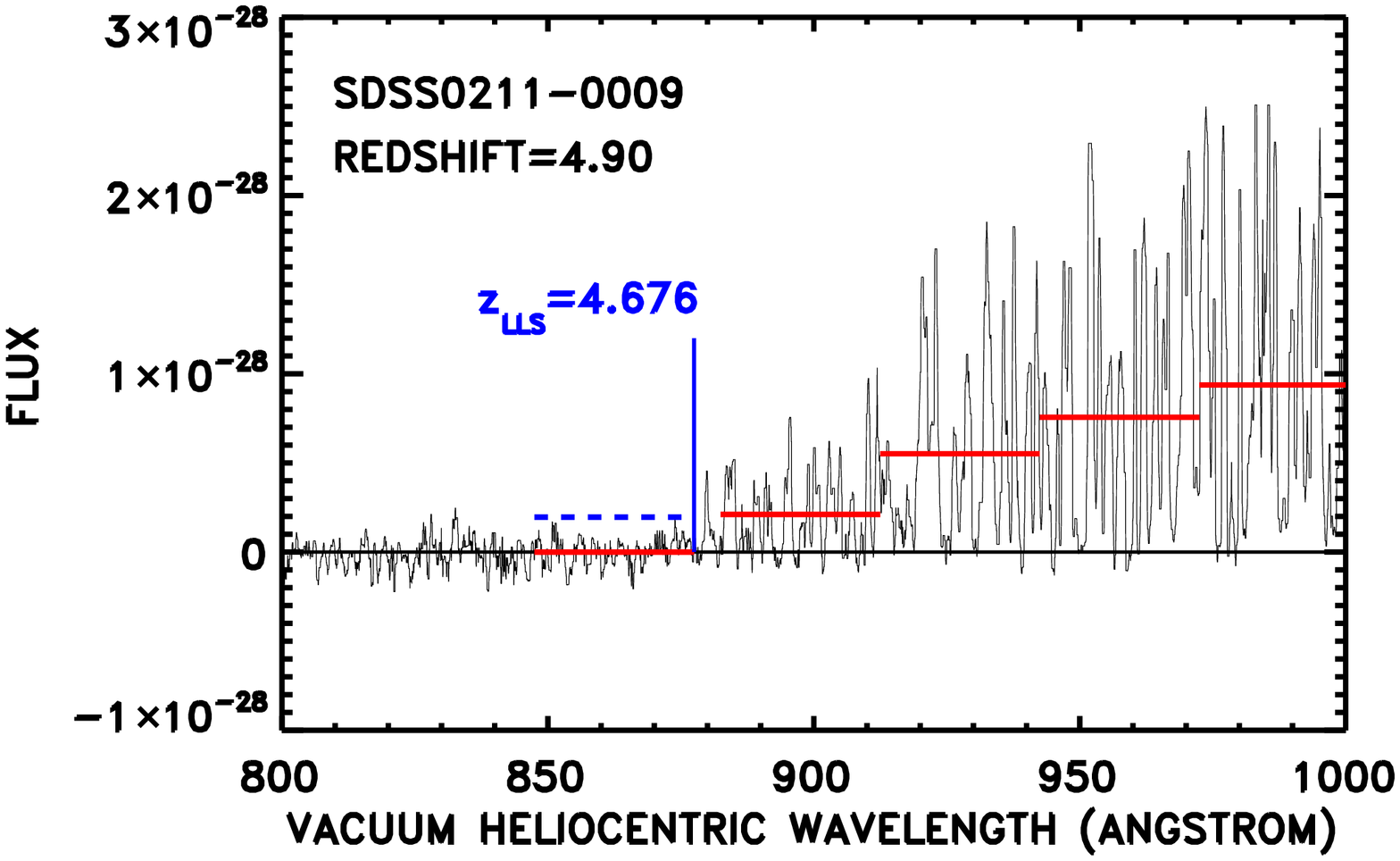}
 \caption{{\it contd.}
\label{fig:tau_displayb}
}
\end{figure}

\begin{figure}[h]
\figurenum{2}
   \includegraphics[width=3in,angle=90,scale=0.9]{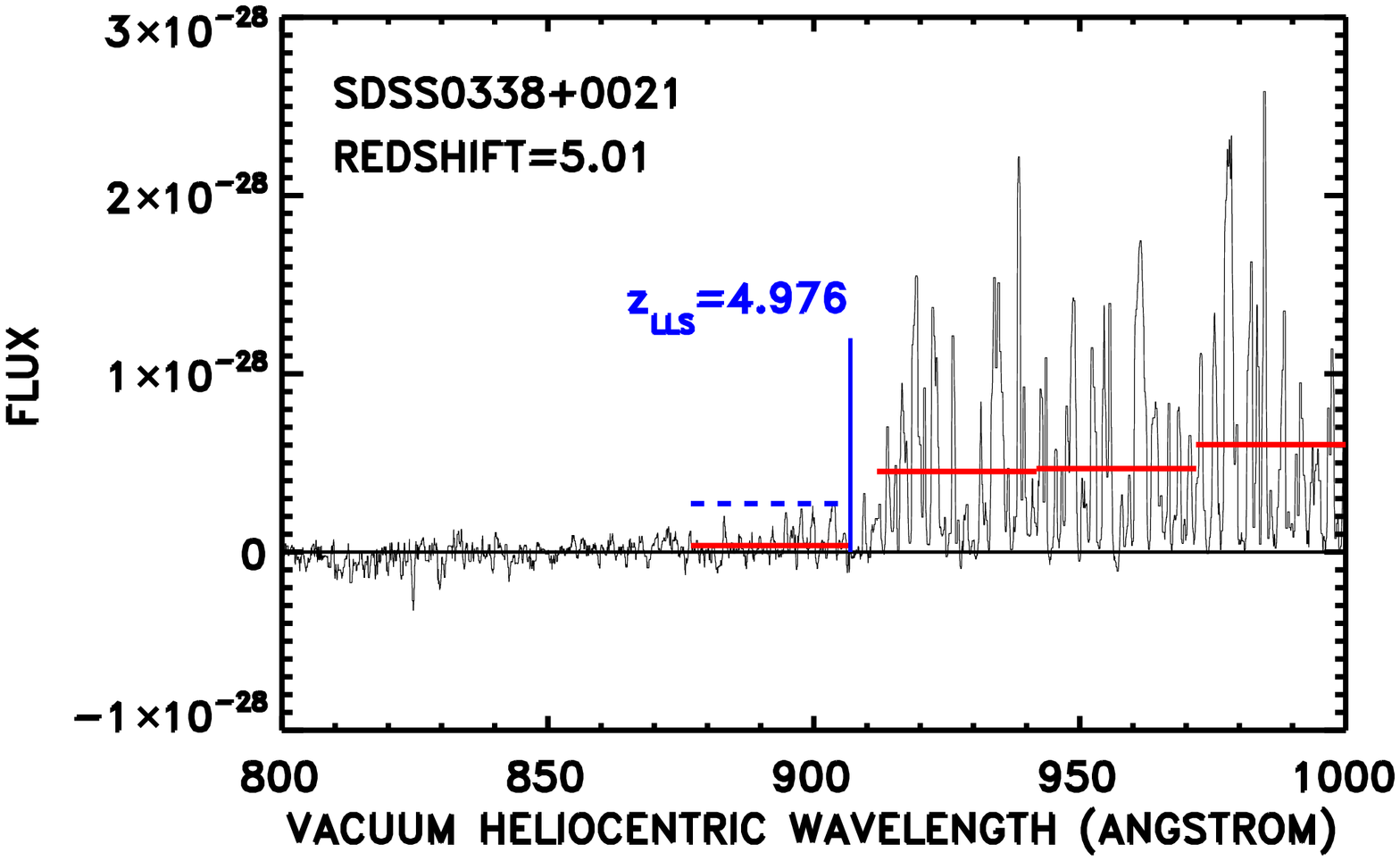}
   \includegraphics[width=3in,angle=90,scale=0.9]{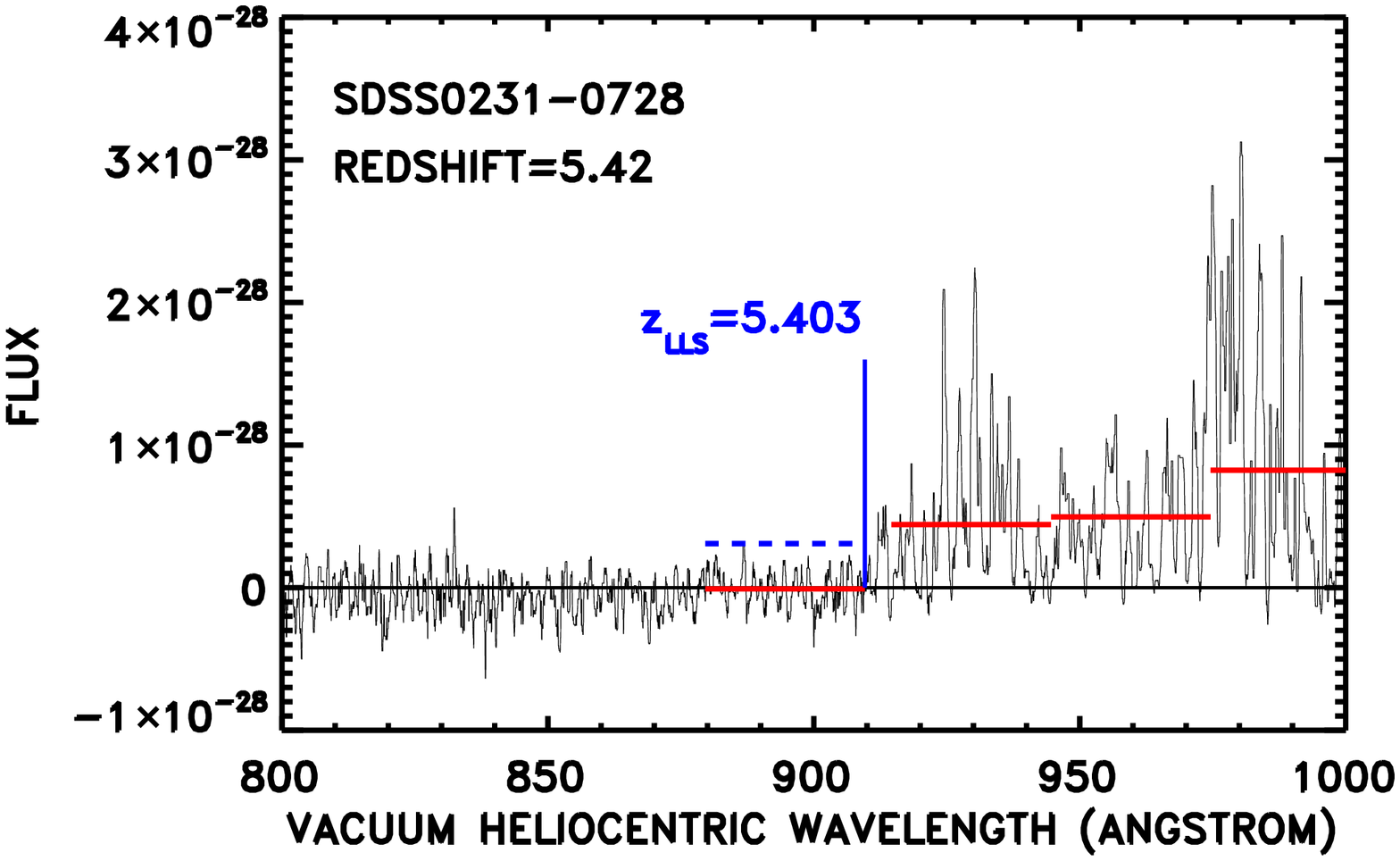}
 \caption{{\it contd.}
\label{fig:tau_displayb}
}
\end{figure}

\begin{figure}[h]
\figurenum{2}
   \includegraphics[width=3in,angle=90,scale=0.9]{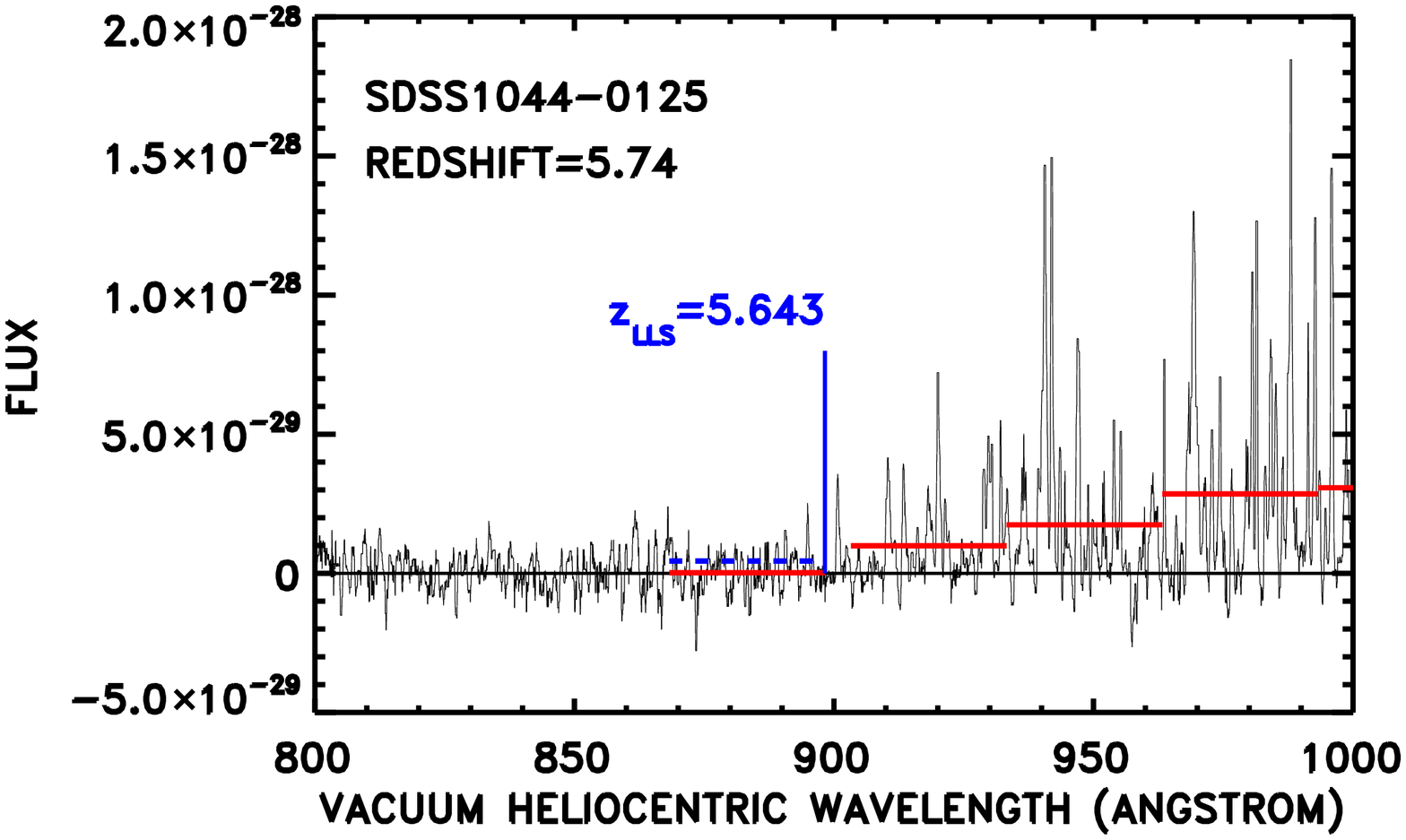}
   \includegraphics[width=3in,angle=90,scale=0.9]{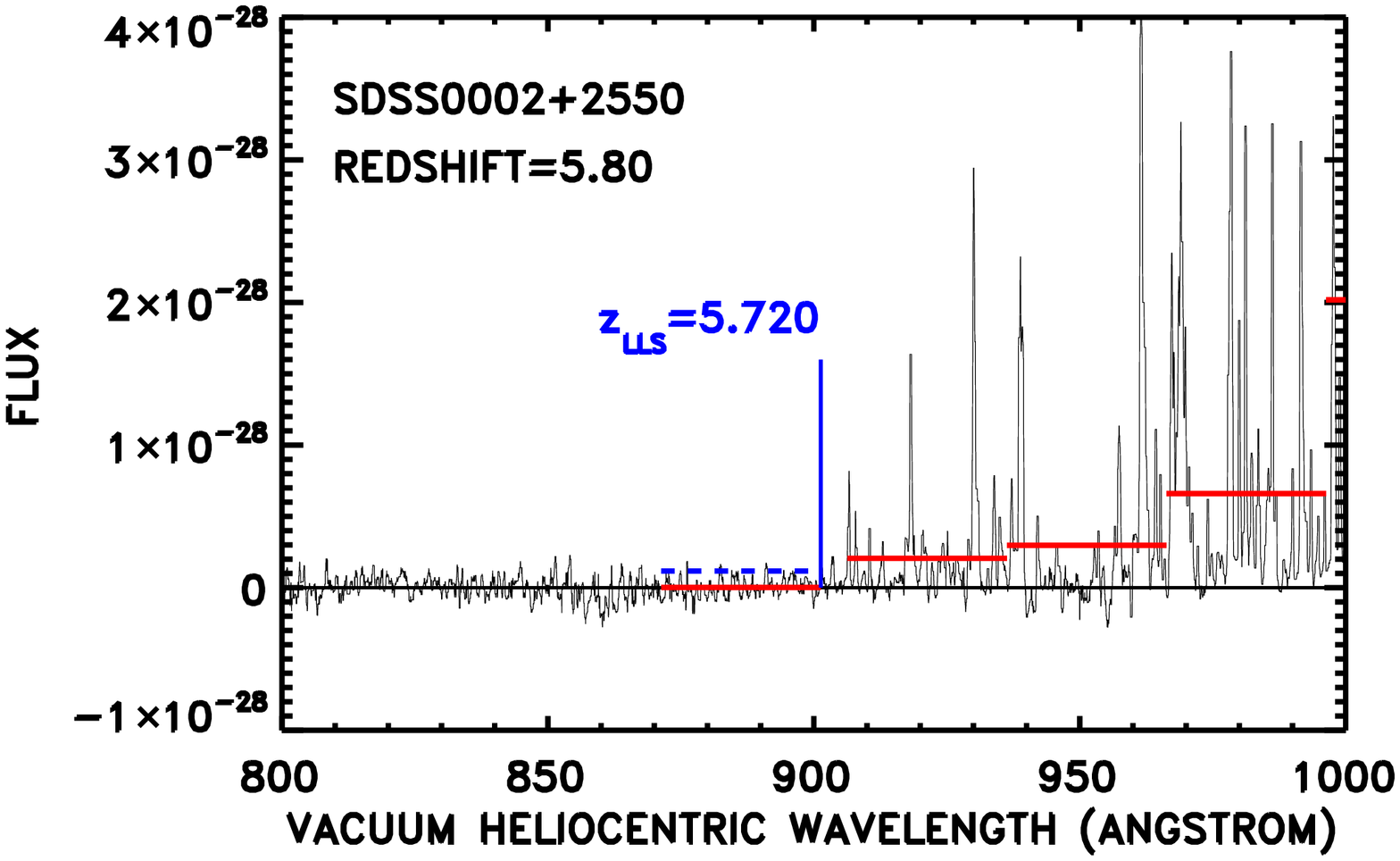}
 \caption{{\it contd.}
\label{fig:tau_displayb}
}
\end{figure}

\begin{figure}[h]
\figurenum{2}
   \includegraphics[width=3in,angle=90,scale=0.9]{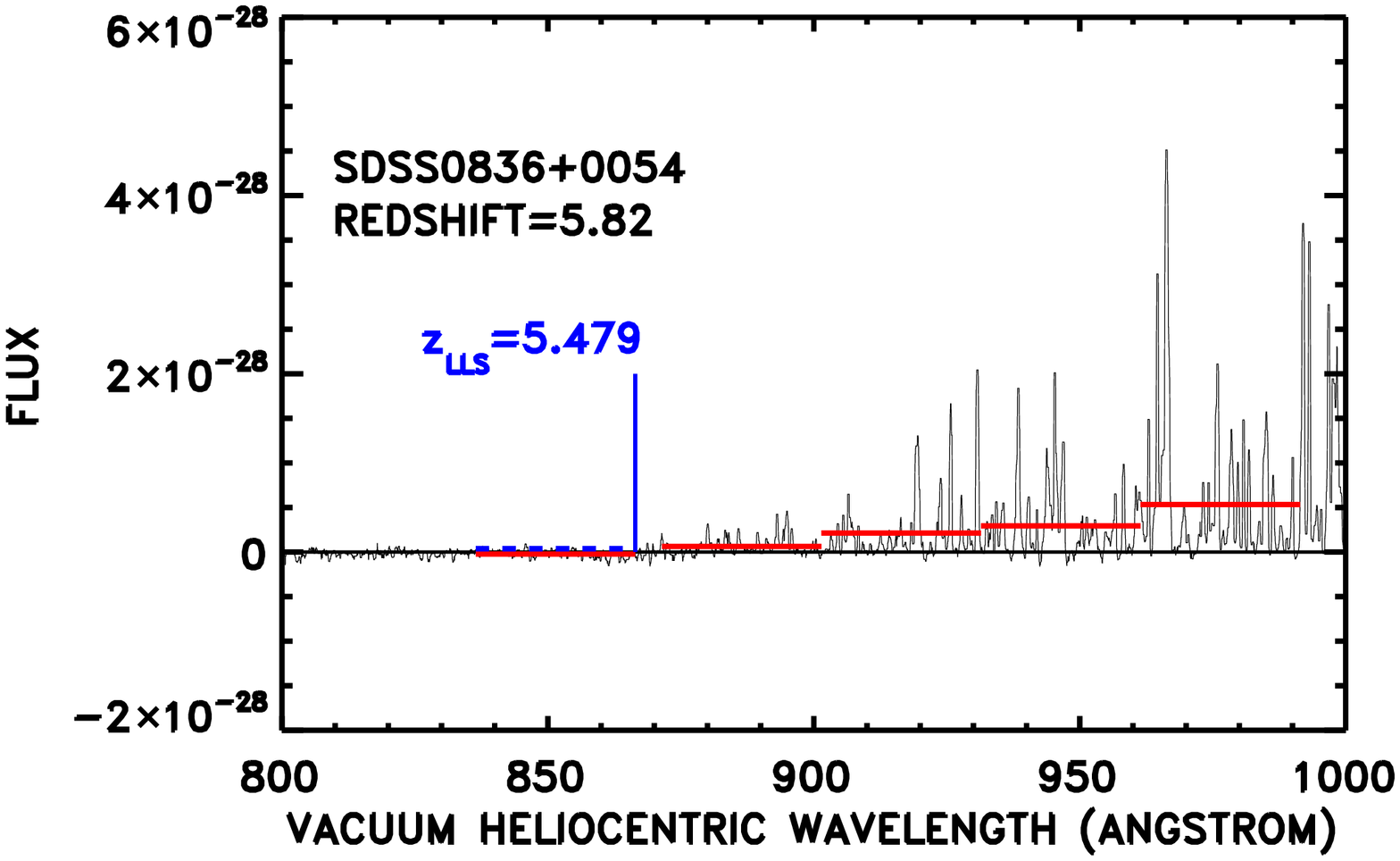}
   \includegraphics[width=3in,angle=90,scale=0.9]{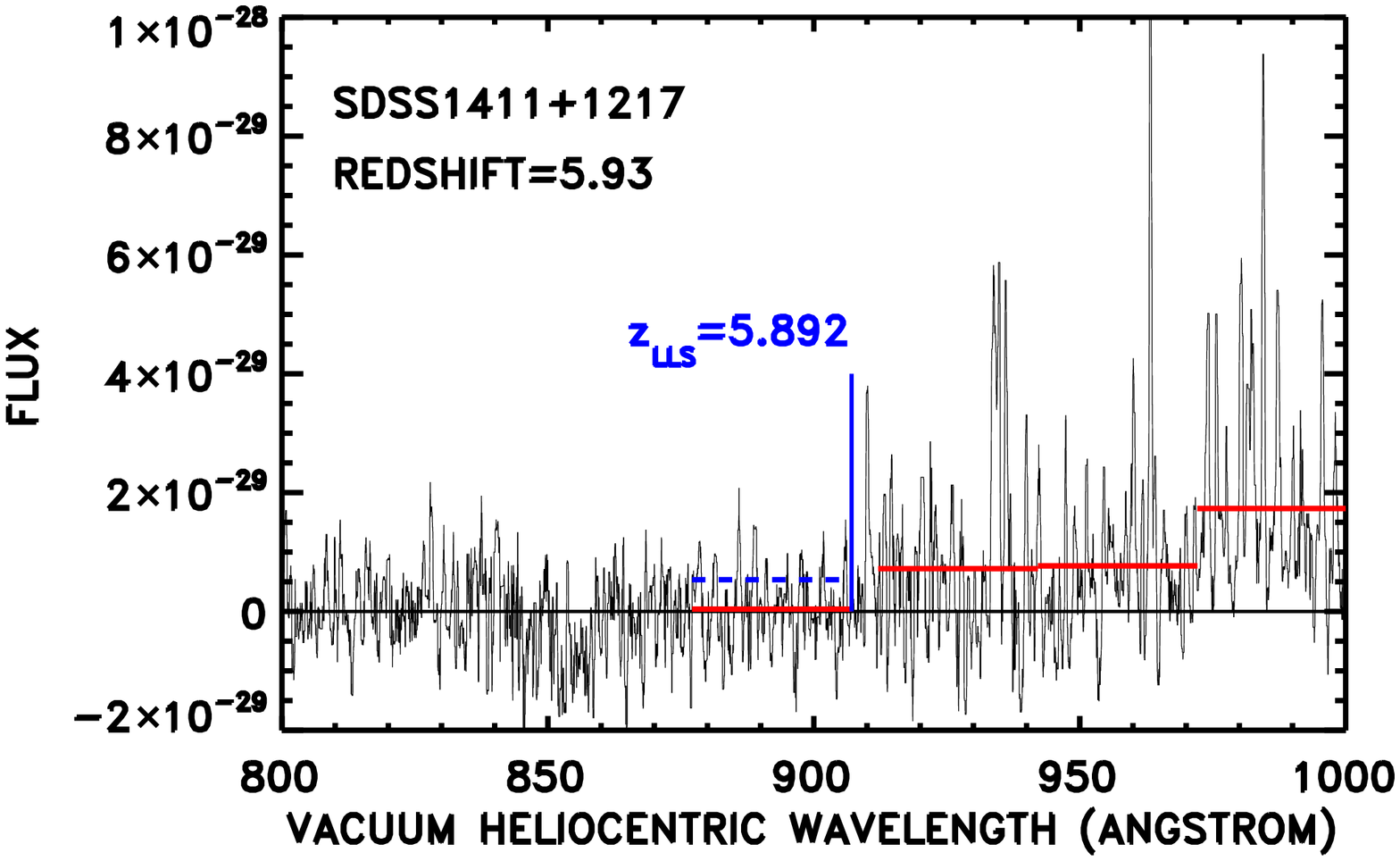}
 \caption{{\it contd.}
\label{fig:tau_displayb}
}
\end{figure}

\begin{figure}[h]
\figurenum{2}
   \includegraphics[width=3in,angle=90,scale=0.9]{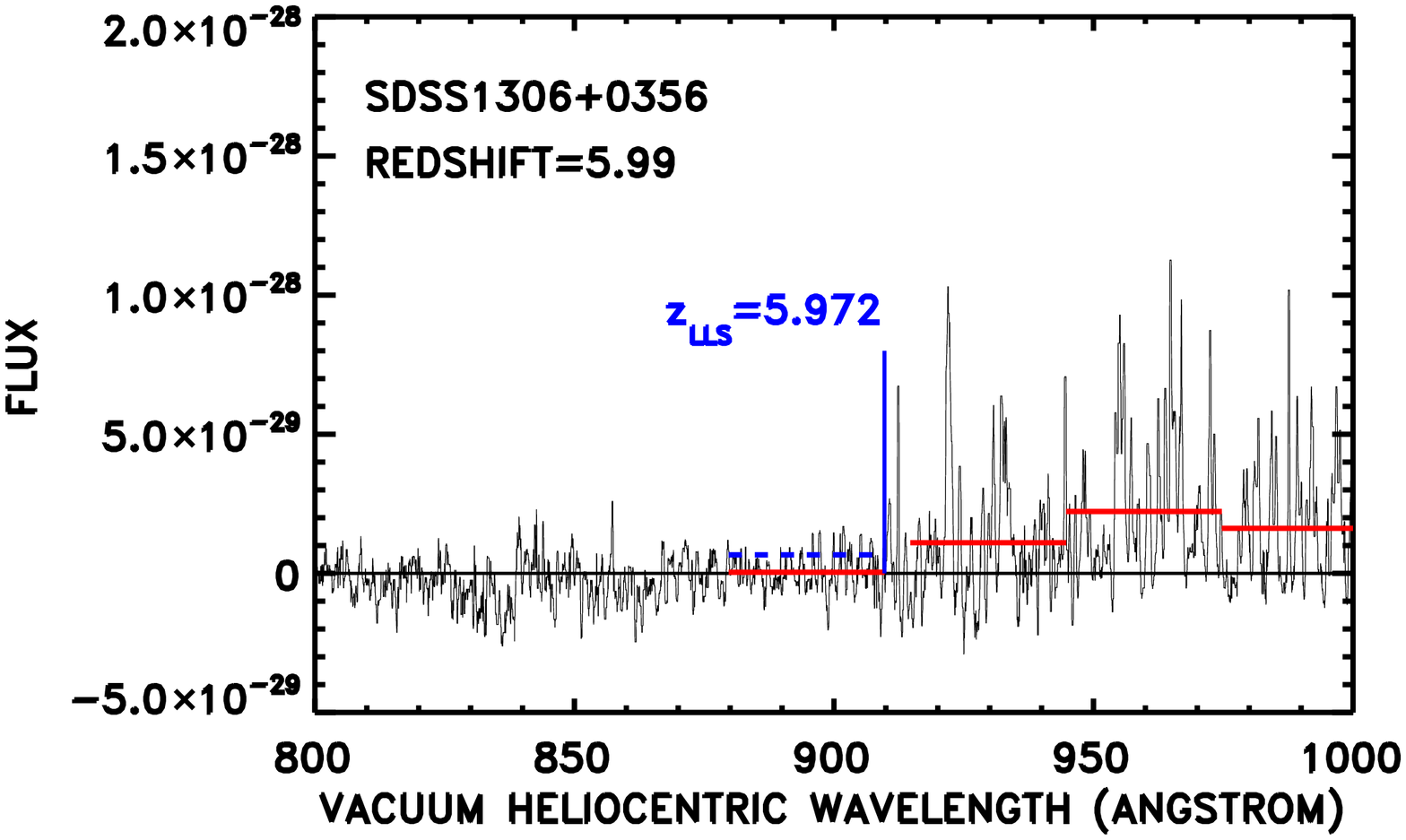}
 \caption{{\it contd.}
\label{fig:tau_displayb}
}
\end{figure}

\newpage
\clearpage

\begin{figure}[h]
\figurenum{4}
 \includegraphics[width=3in,angle=0,scale=1]{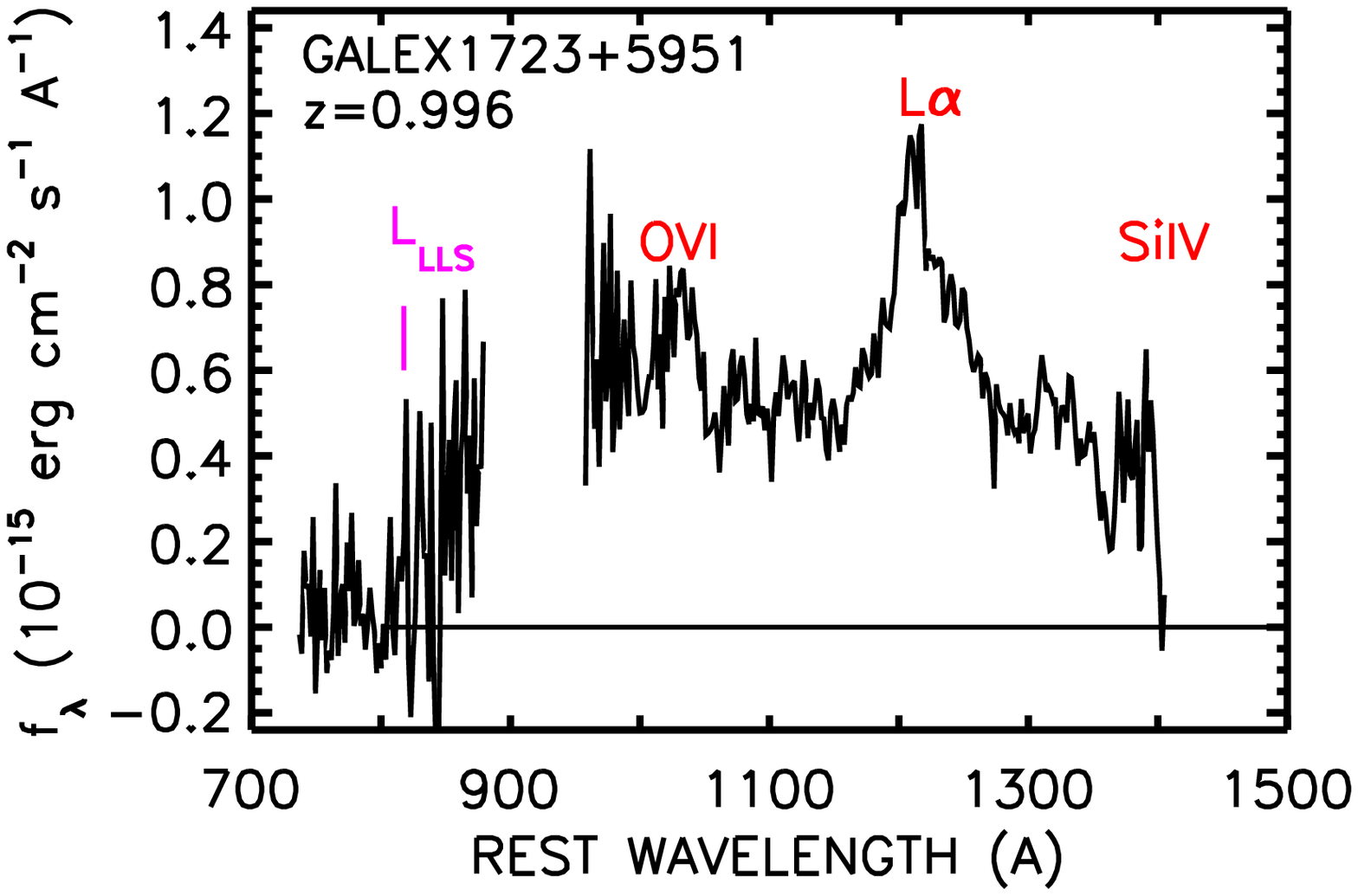}
 \includegraphics[width=3in,angle=0,scale=1]{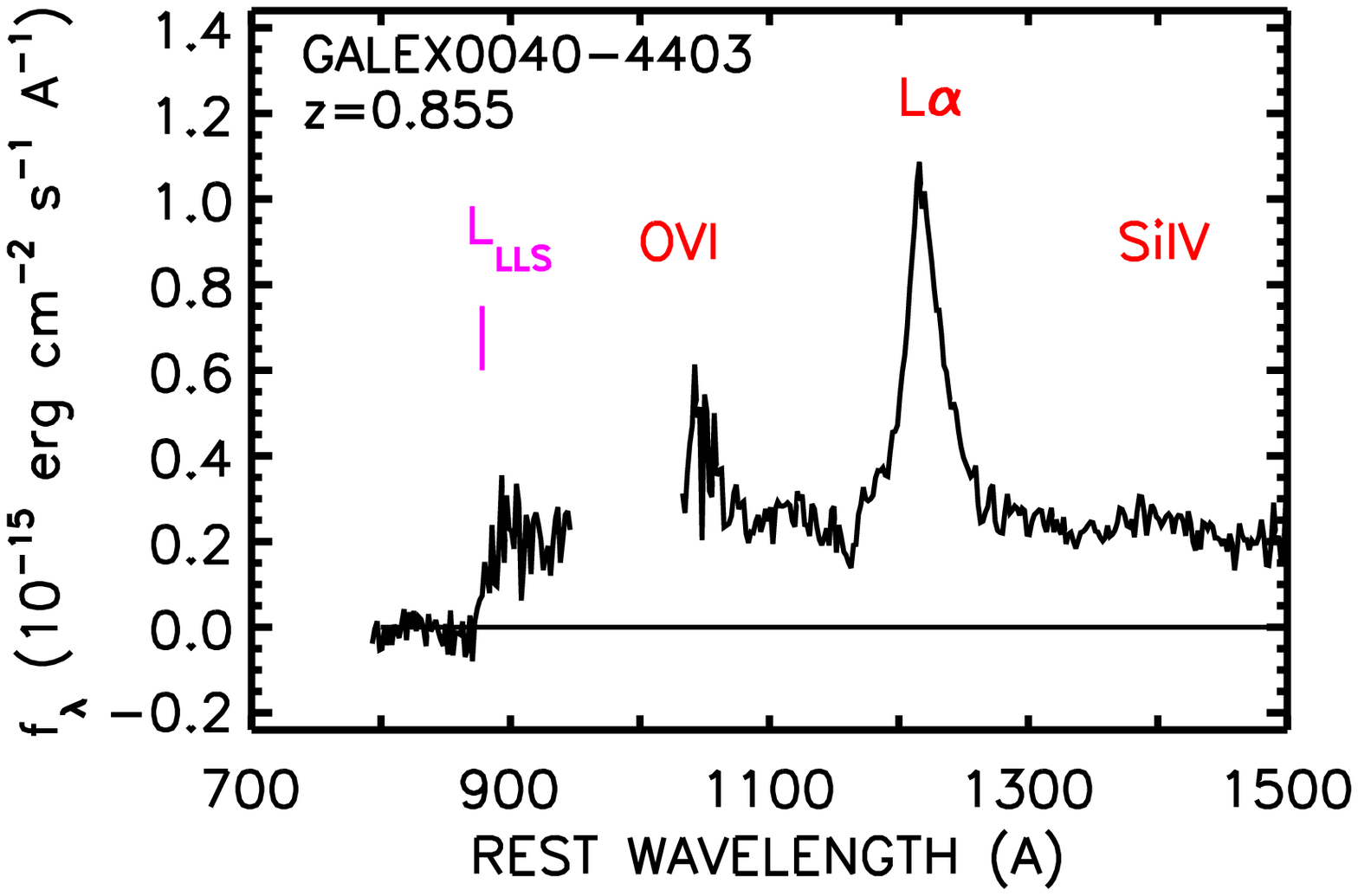}
  \caption{GALEX quasars with Lyman limit systems. The spectrum of the quasar 
   is shown in the rest frame
   with the position of the quasar emission lines marked in red.
   The short wavelength portion of the spectrum is from the GALEX
   FUV grism and the long wavelength portion from the GALEX
   NUV grism. The position of the LLS is marked in purple.  The figures are ordered by the red continuum flux above the Ly$\alpha$\  line.
\label{fig:galex_lls_sample}
}
\end{figure}

\begin{figure}[h]
\figurenum{4}
 \includegraphics[width=3in,angle=0,scale=1]{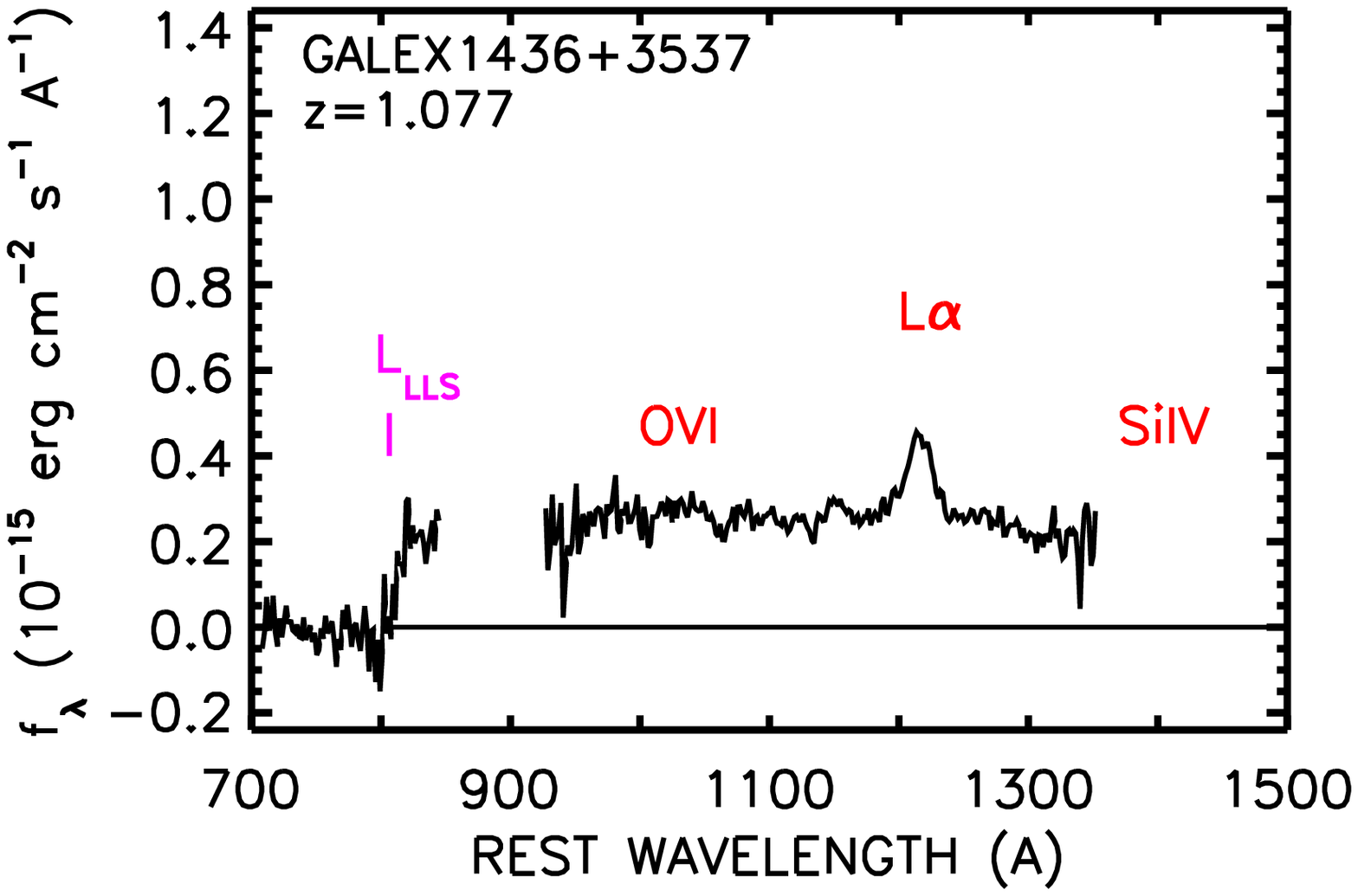}
 \includegraphics[width=3in,angle=0,scale=1]{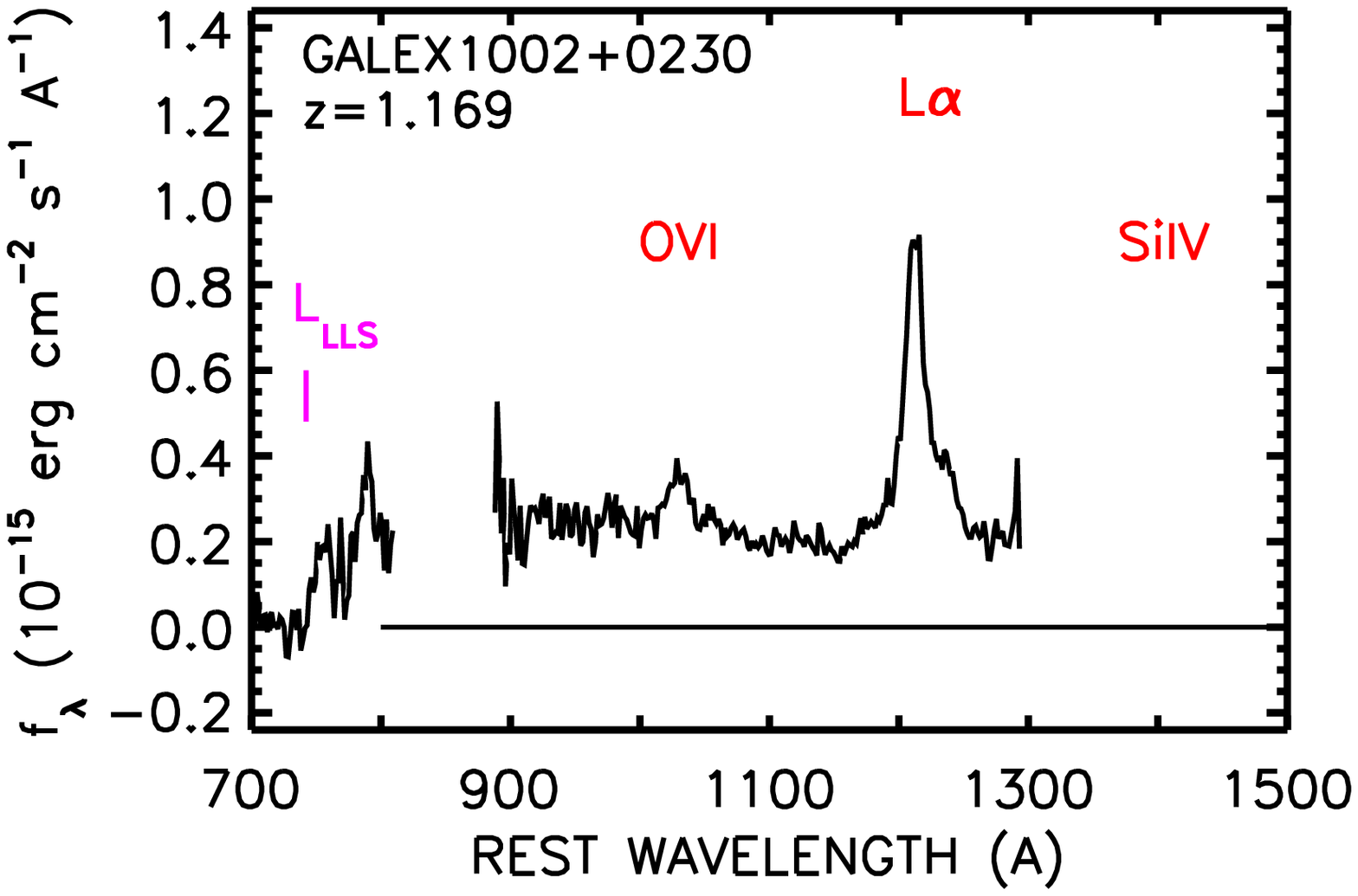}
  \caption{{\it contd.}
\label{fig:galex_lls_sample2}
}
\end{figure}

\begin{figure}[h]
\figurenum{4}
 \includegraphics[width=3in,angle=0,scale=1]{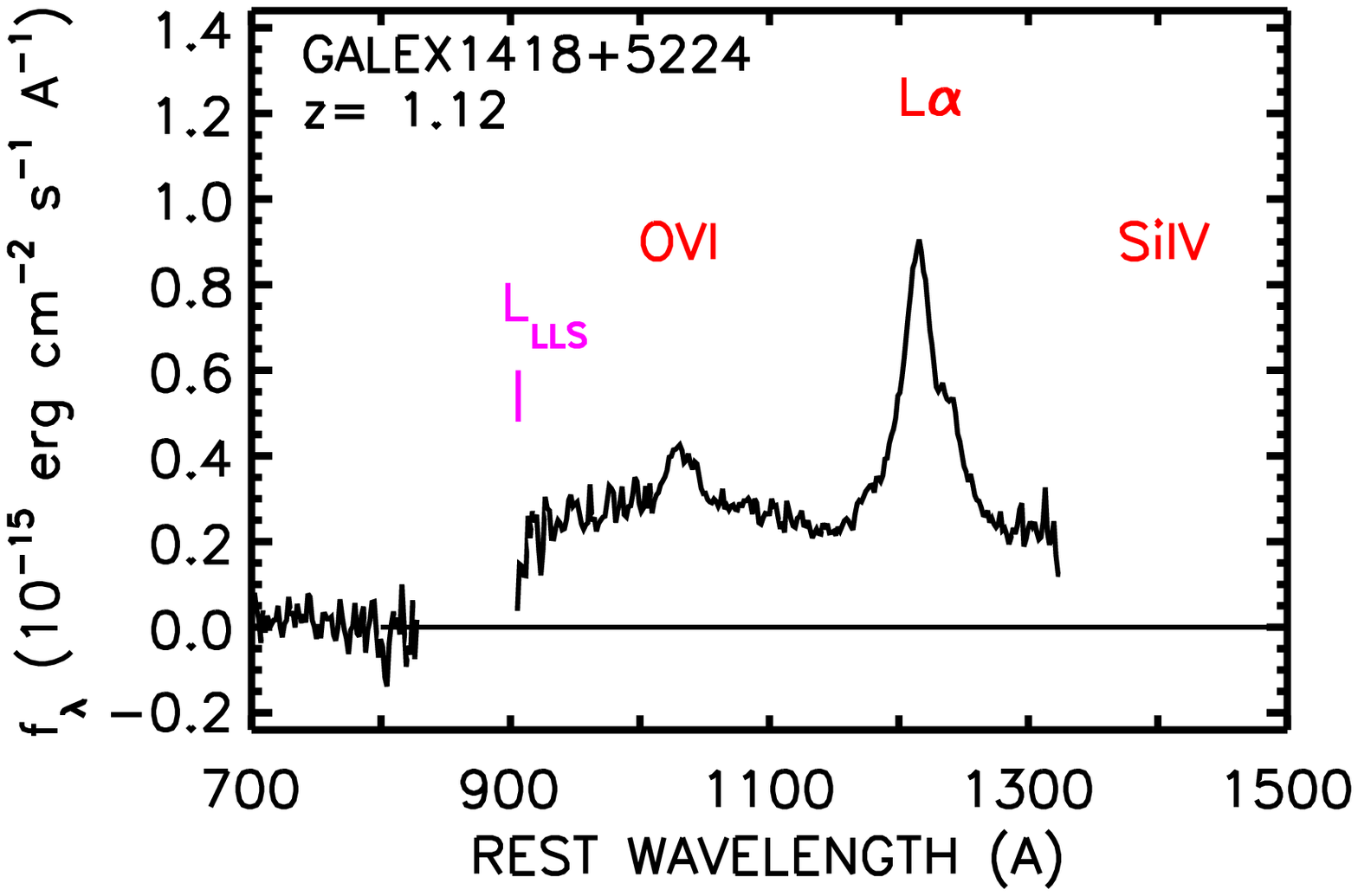}
 \includegraphics[width=3in,angle=0,scale=1]{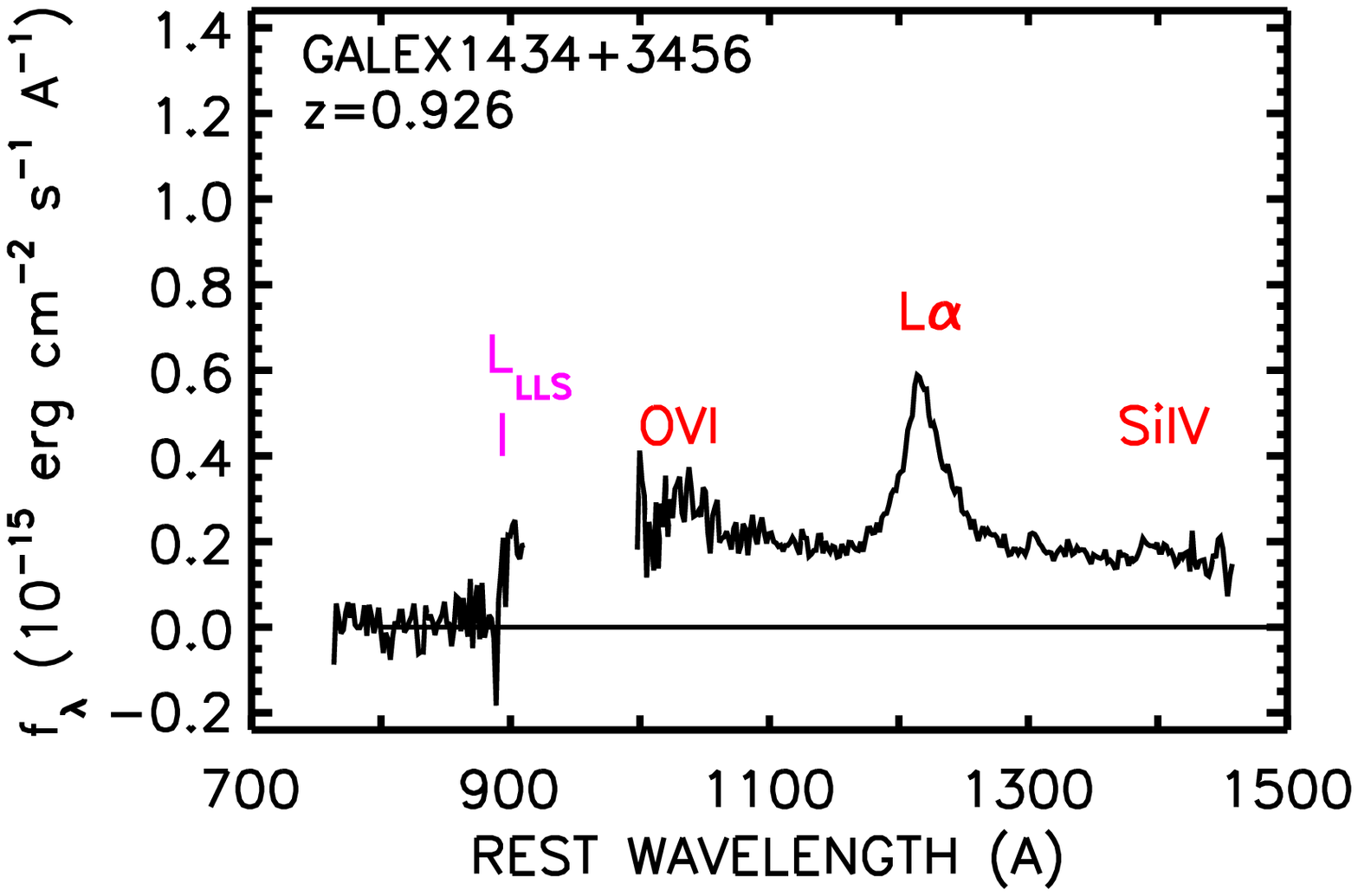}
  \caption{{\it contd.}
\label{fig:galex_lls_sample3}
}
\end{figure}

\begin{figure}[h]
\figurenum{4}
 \includegraphics[width=3in,angle=0,scale=1]{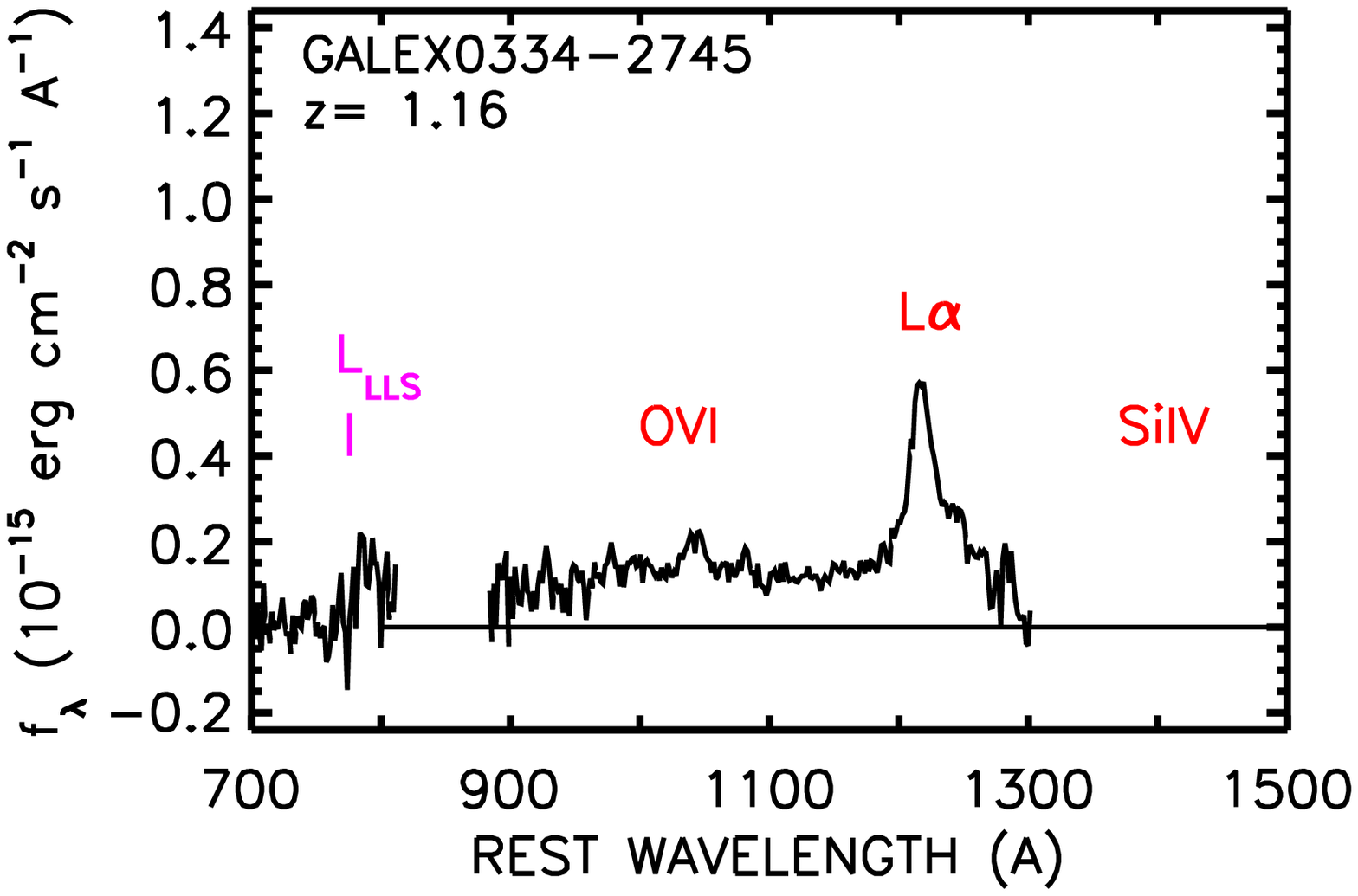}
 \includegraphics[width=3in,angle=0,scale=1]{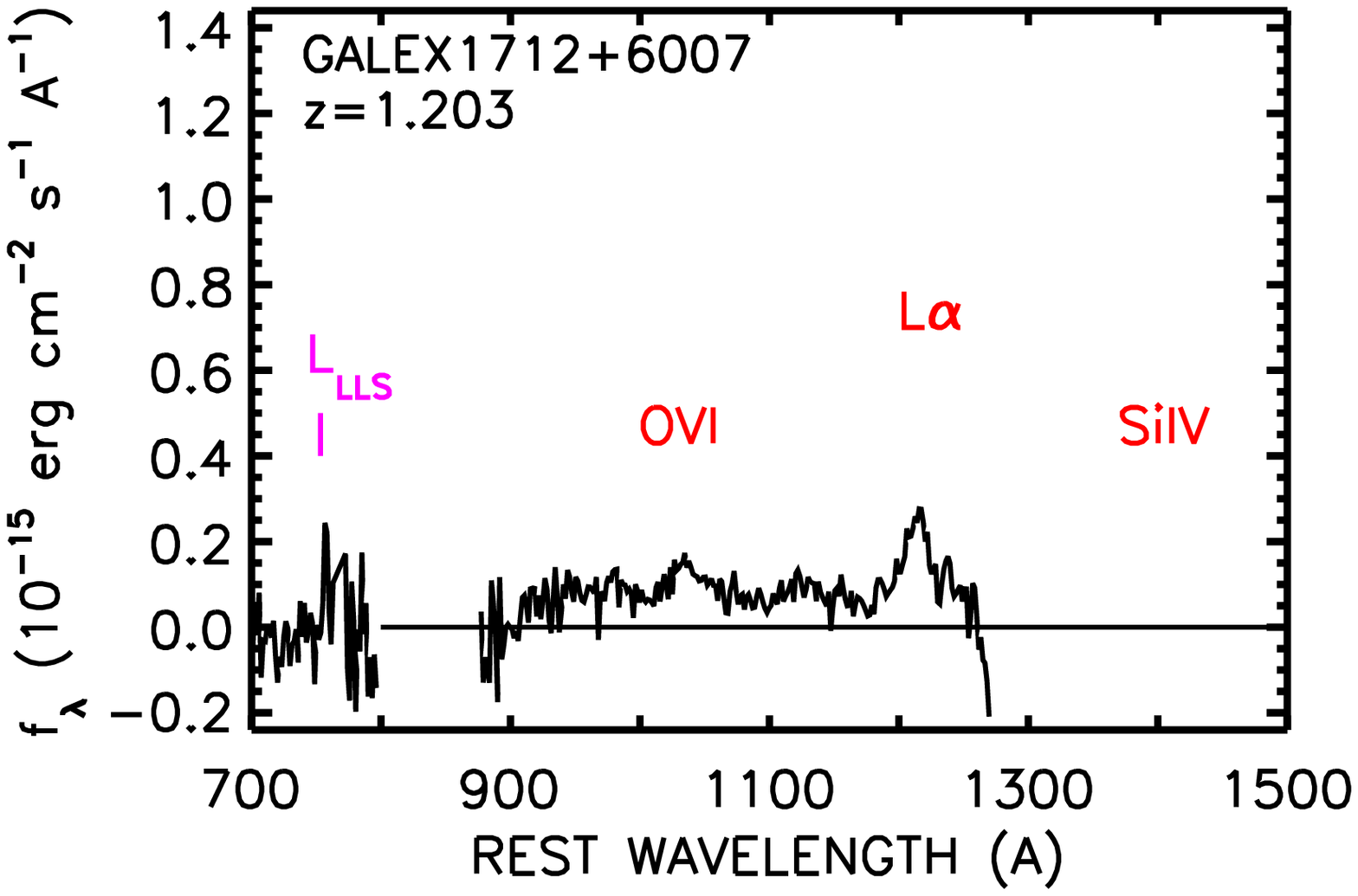}
  \caption{{\it contd.}
\label{fig:galex_lls_sample4}
}
\end{figure}

\begin{figure}[h]
\figurenum{4}
 \includegraphics[width=3in,angle=0,scale=1]{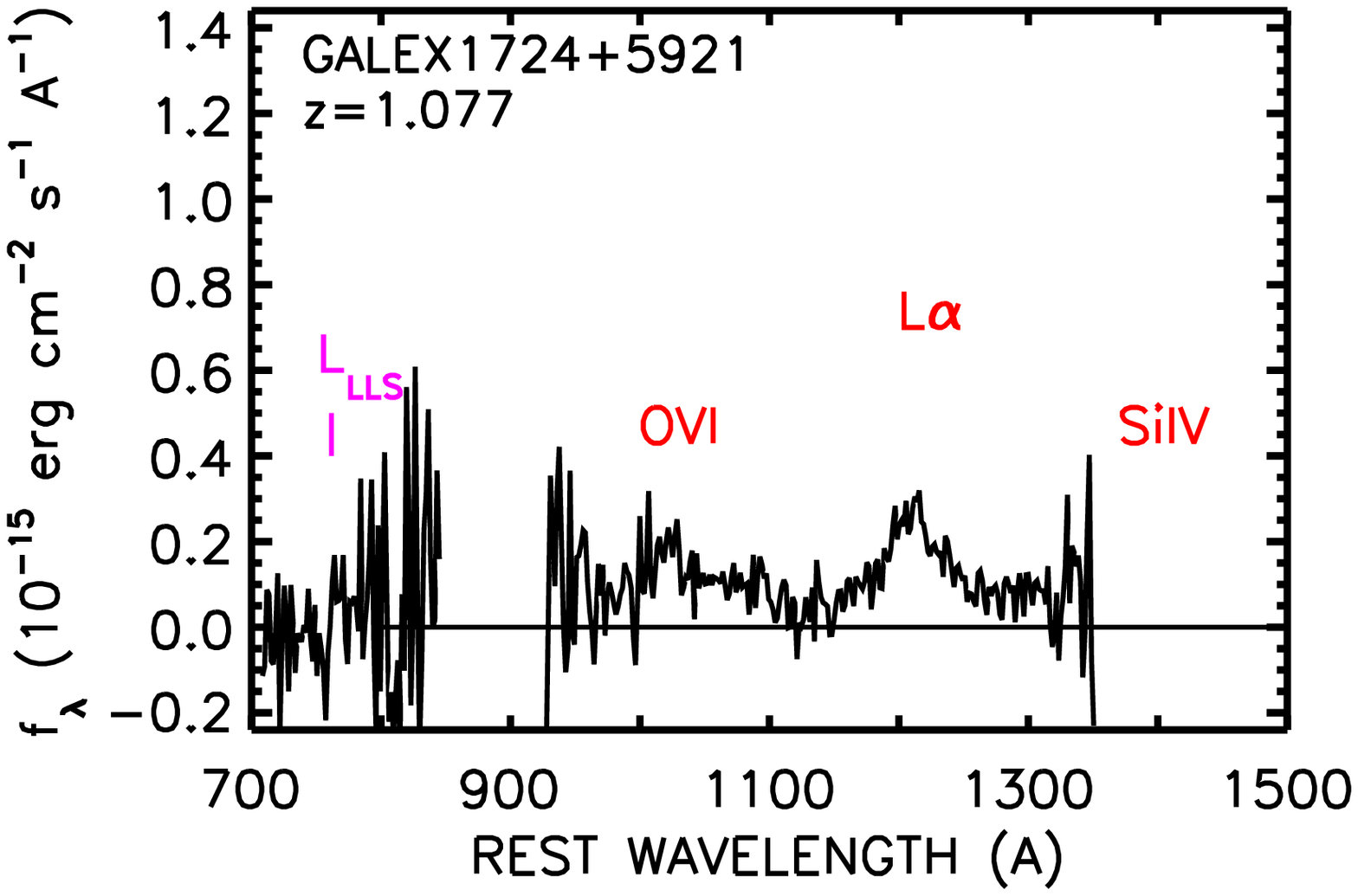}
  \caption{{\it contd.}
\label{fig:galex_lls_sample5}
}
\end{figure}

\newpage
\clearpage

\begin{figure*}[h]
\figurenum{7}
   \includegraphics[width=8in,angle=90,scale=0.3]{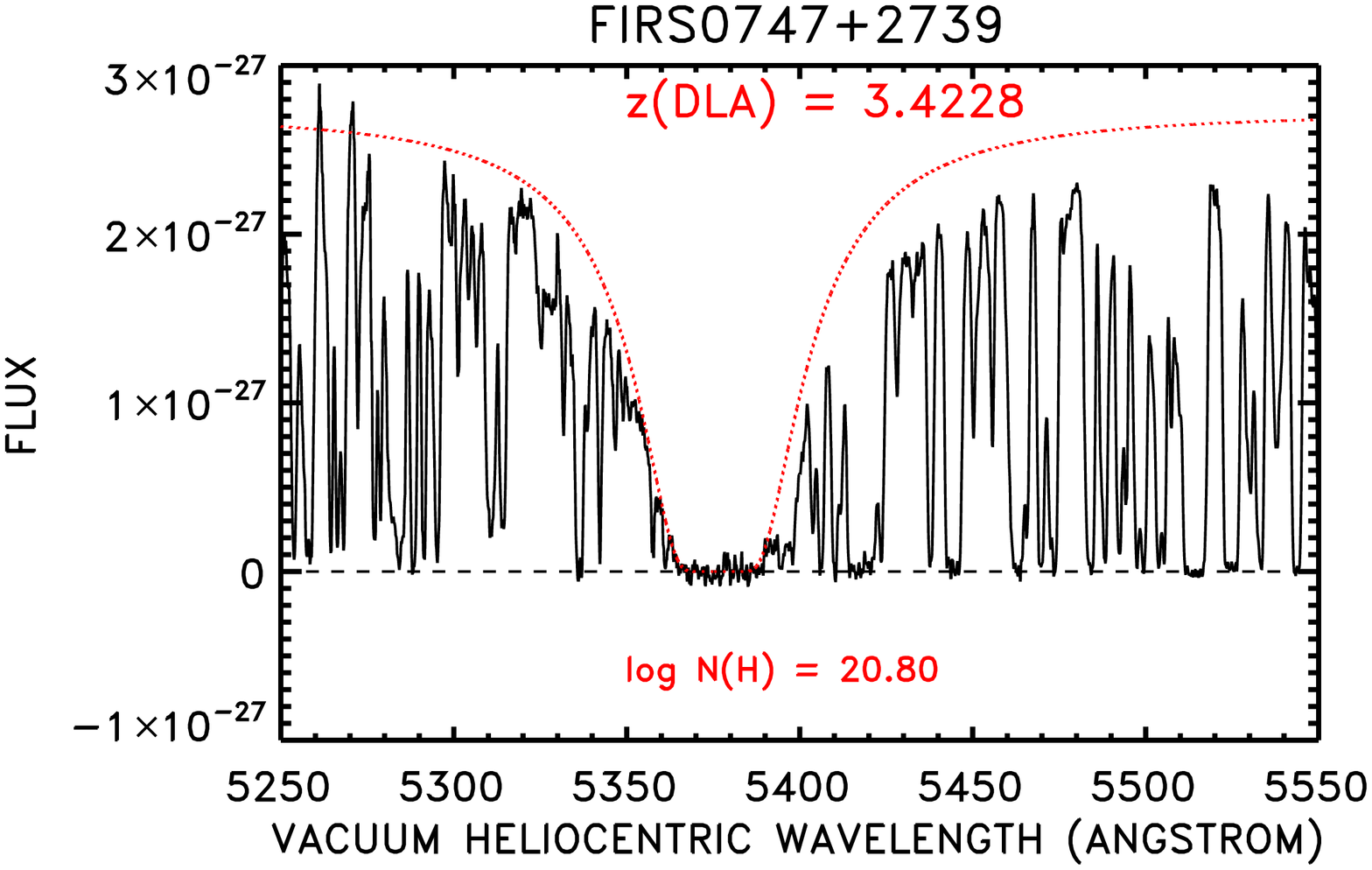}
   \includegraphics[width=8in,angle=90,scale=0.3]{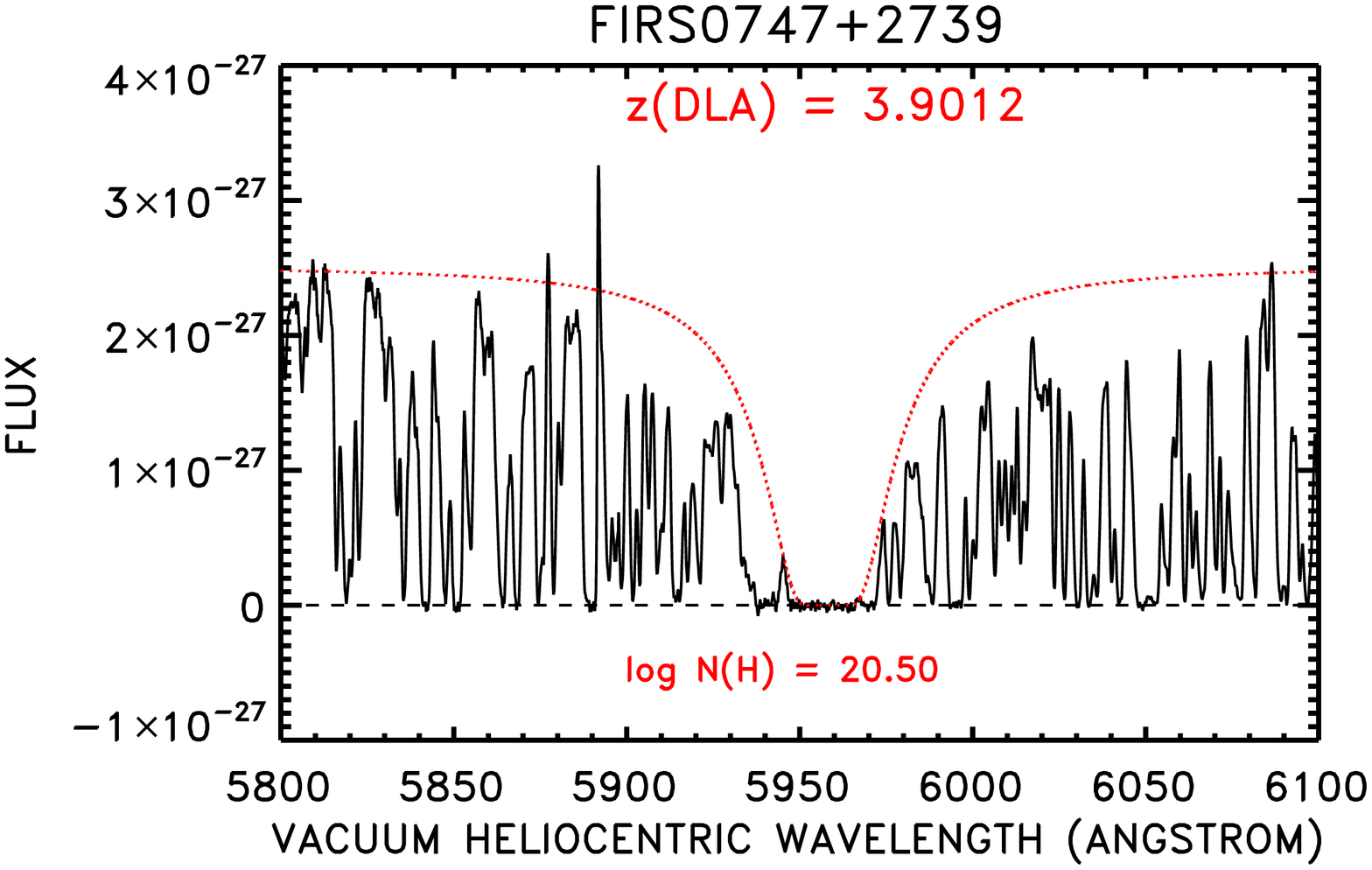}
 \includegraphics[width=8in,angle=90,scale=0.3]{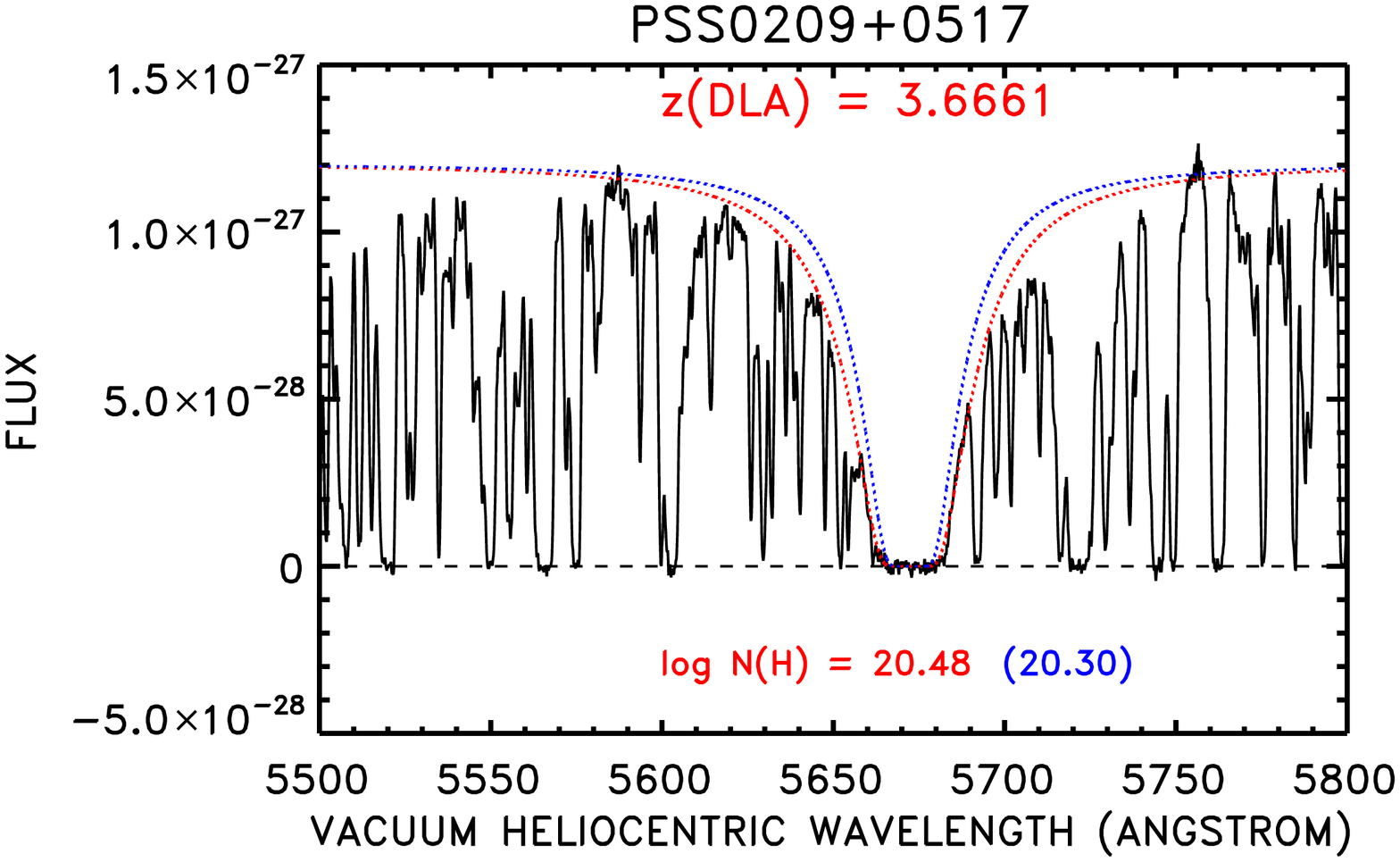}
 \includegraphics[width=8in,angle=90,scale=0.3]{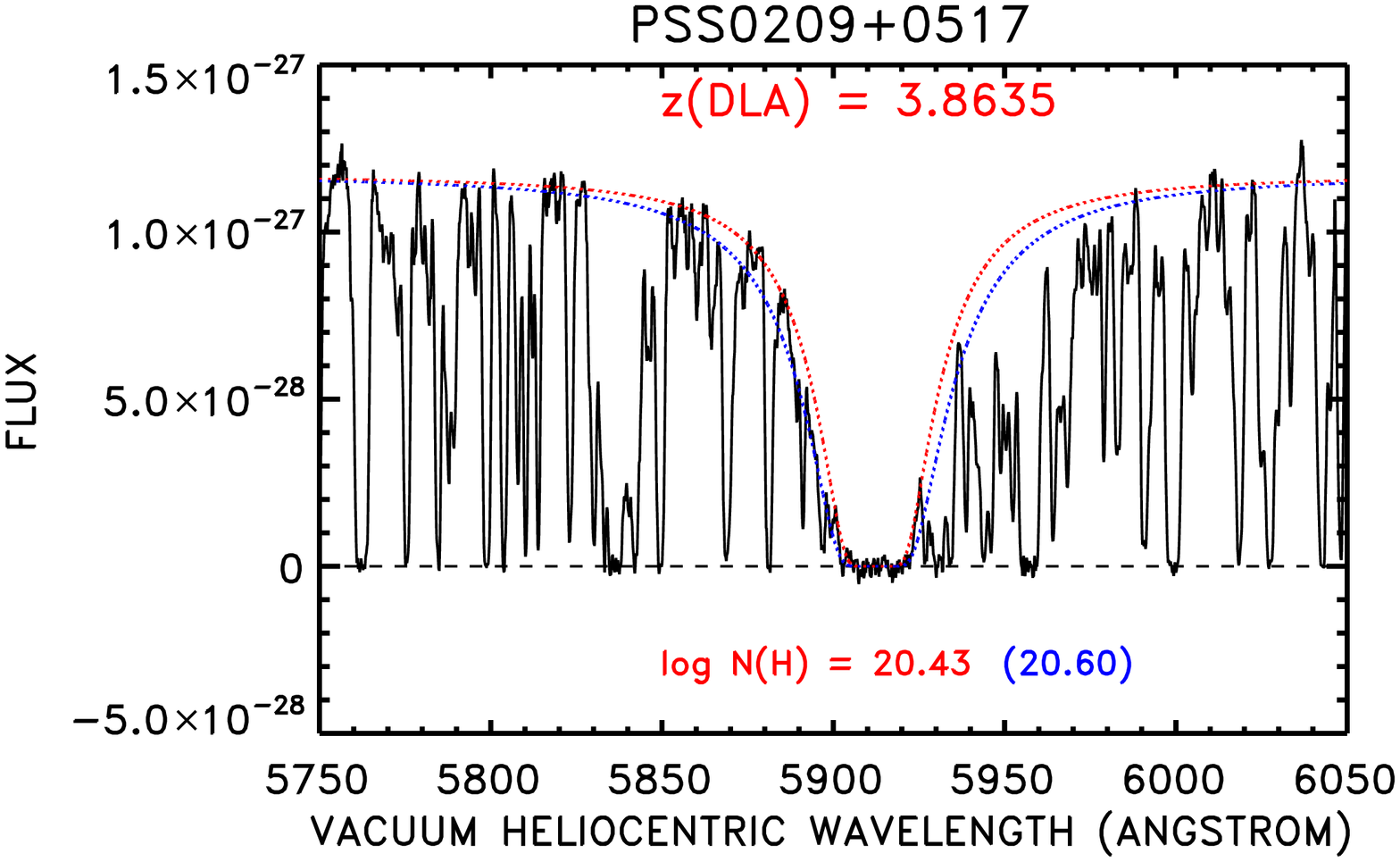}
 \includegraphics[width=8in,angle=90,scale=0.3]{Fig7.5.eps}
 \includegraphics[width=8in,angle=90,scale=0.3]{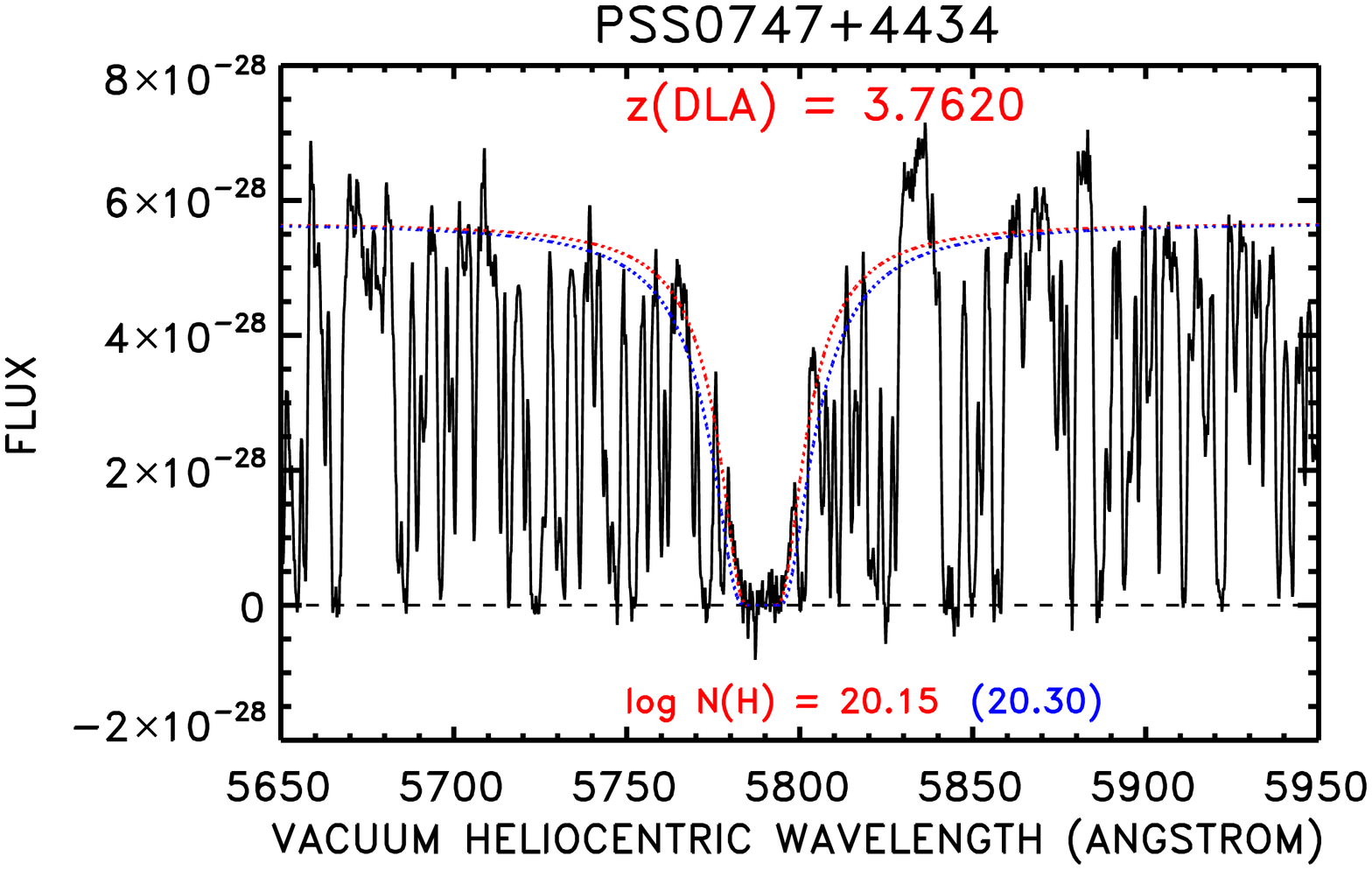}
  \caption{Damped Lyman alpha systems in the high redshift sample.  The red dotted line shows the fit corresponding to $\log N_{\rm H}$\ from Table 1.  Where present, the blue dotted line shows the fit from P03 (see Table 5).
\label{fig:dlas_a}
}
\end{figure*}

\newpage
\clearpage

\begin{figure*}[h]
\figurenum{7}
   \includegraphics[width=8in,angle=90,scale=0.3]{Fig7.7.eps}
  \includegraphics[width=8in,angle=90,scale=0.3]{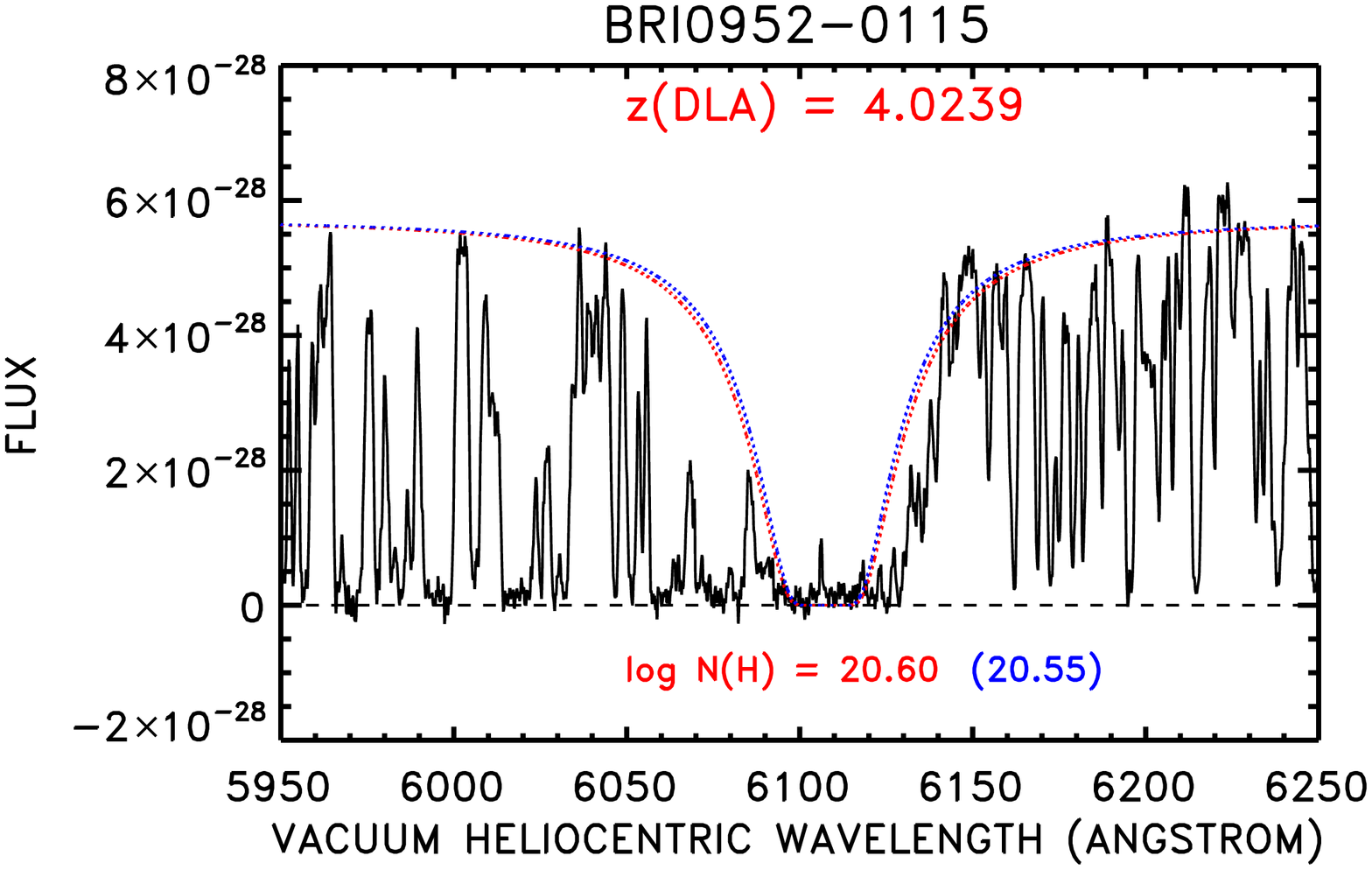}
 \includegraphics[width=8in,angle=90,scale=0.3]{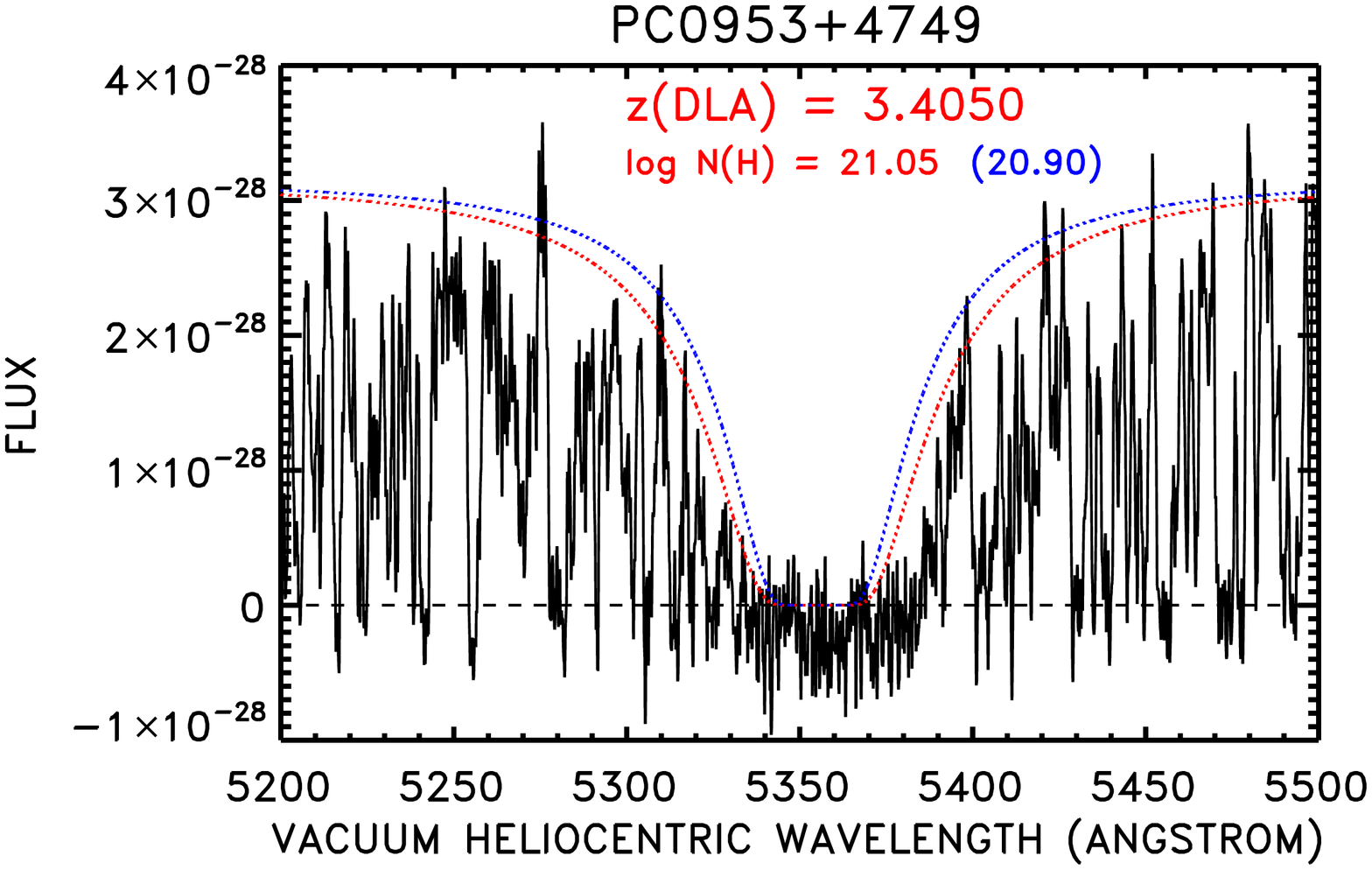}
 \includegraphics[width=8in,angle=90,scale=0.3]{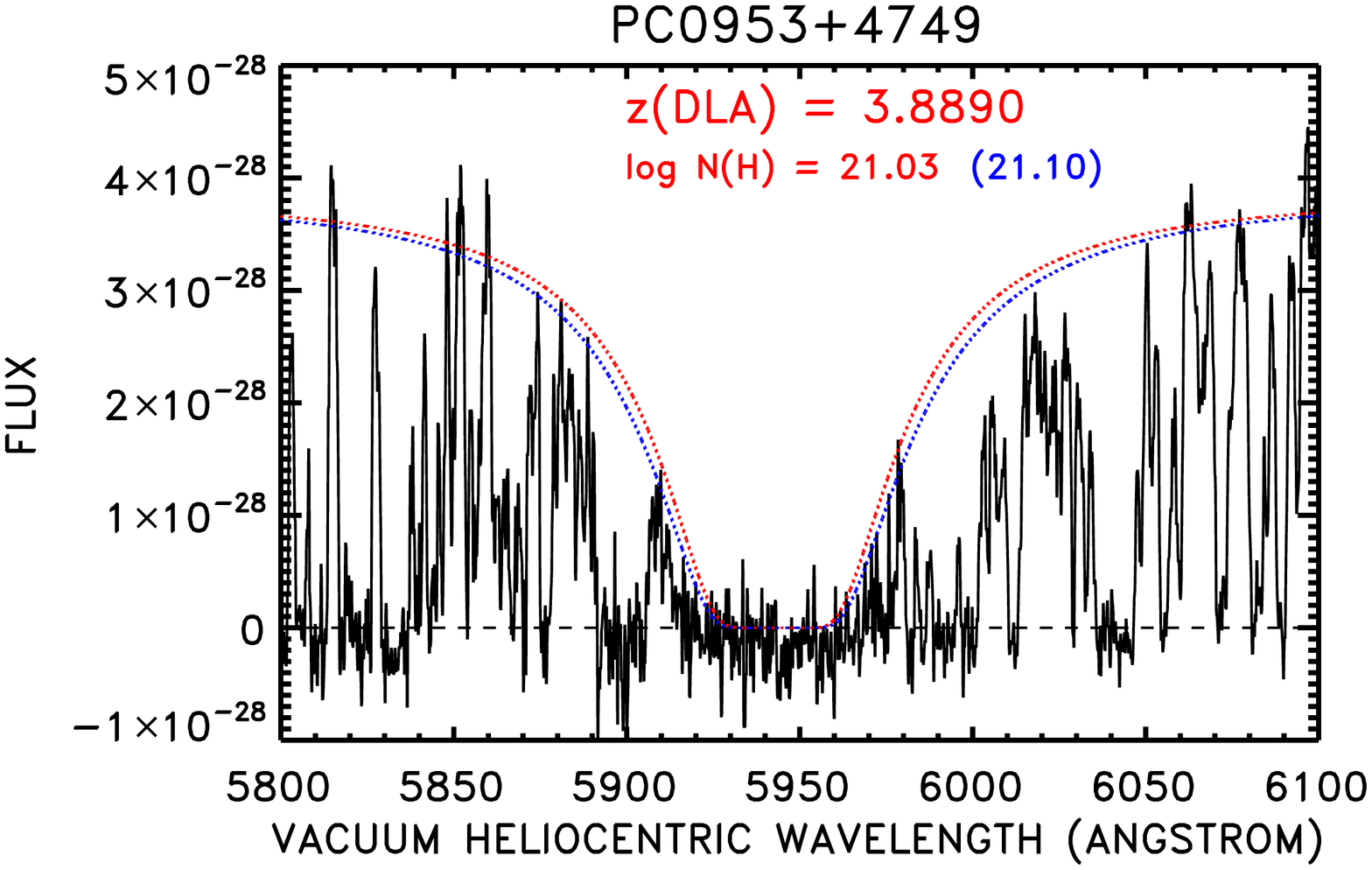}
 \includegraphics[width=8in,angle=90,scale=0.3]{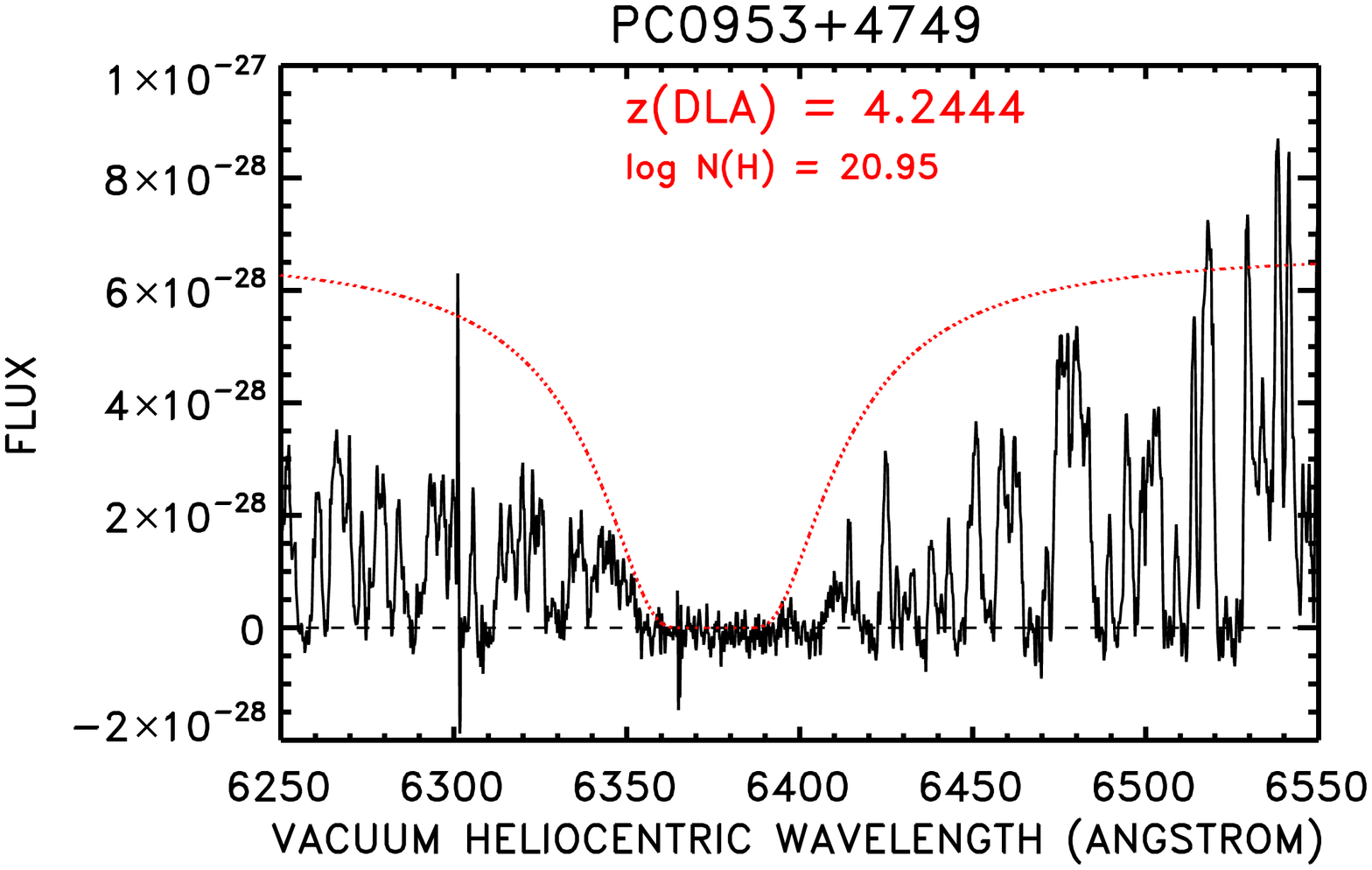}
 \includegraphics[width=8in,angle=90,scale=0.3]{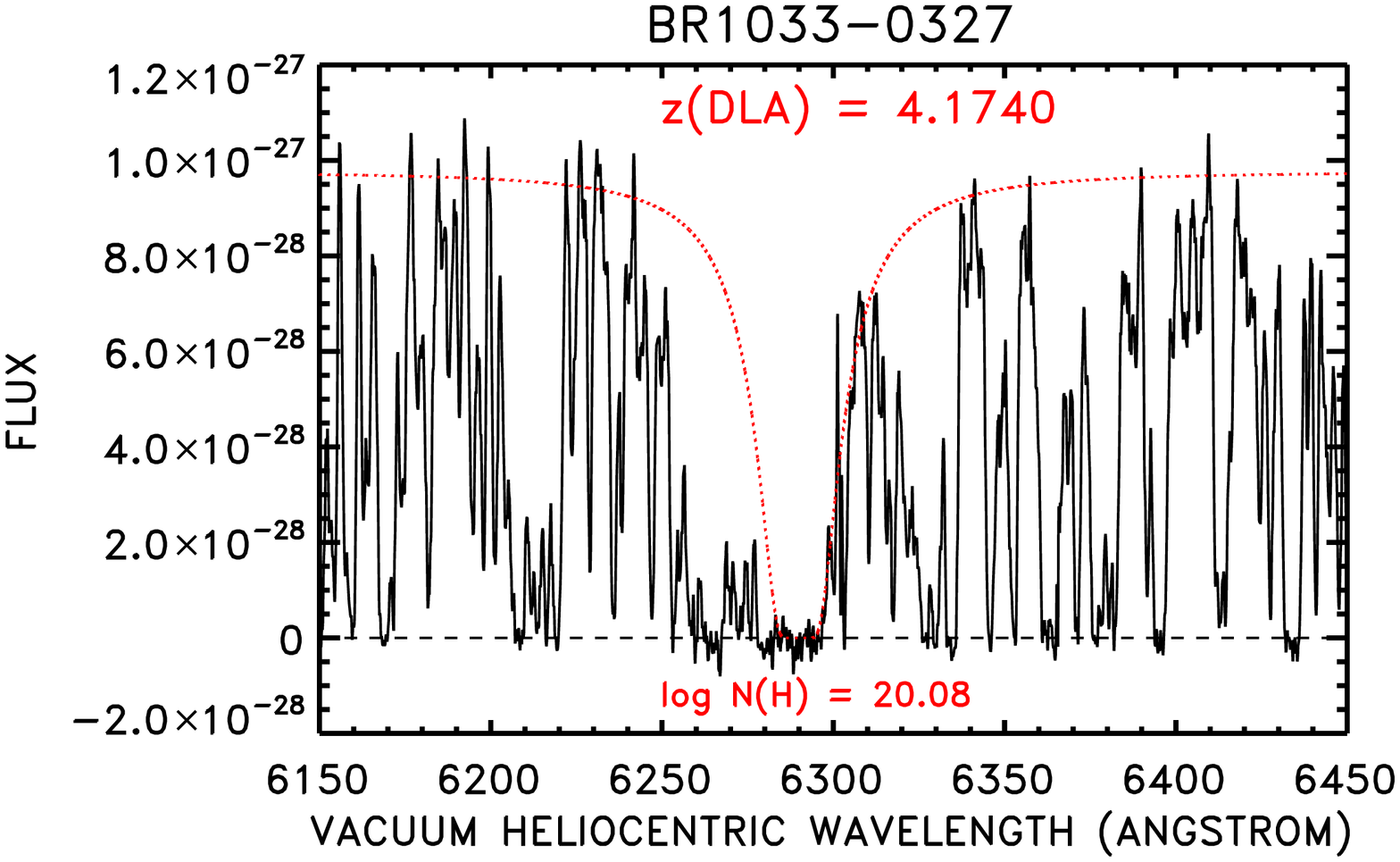}
  \caption{{\it Contd.}
\label{fig:dlas_b}
}
\end{figure*}

\newpage
\clearpage

\begin{figure*}[h]
\figurenum{7}
   \includegraphics[width=8in,angle=90,scale=0.3]{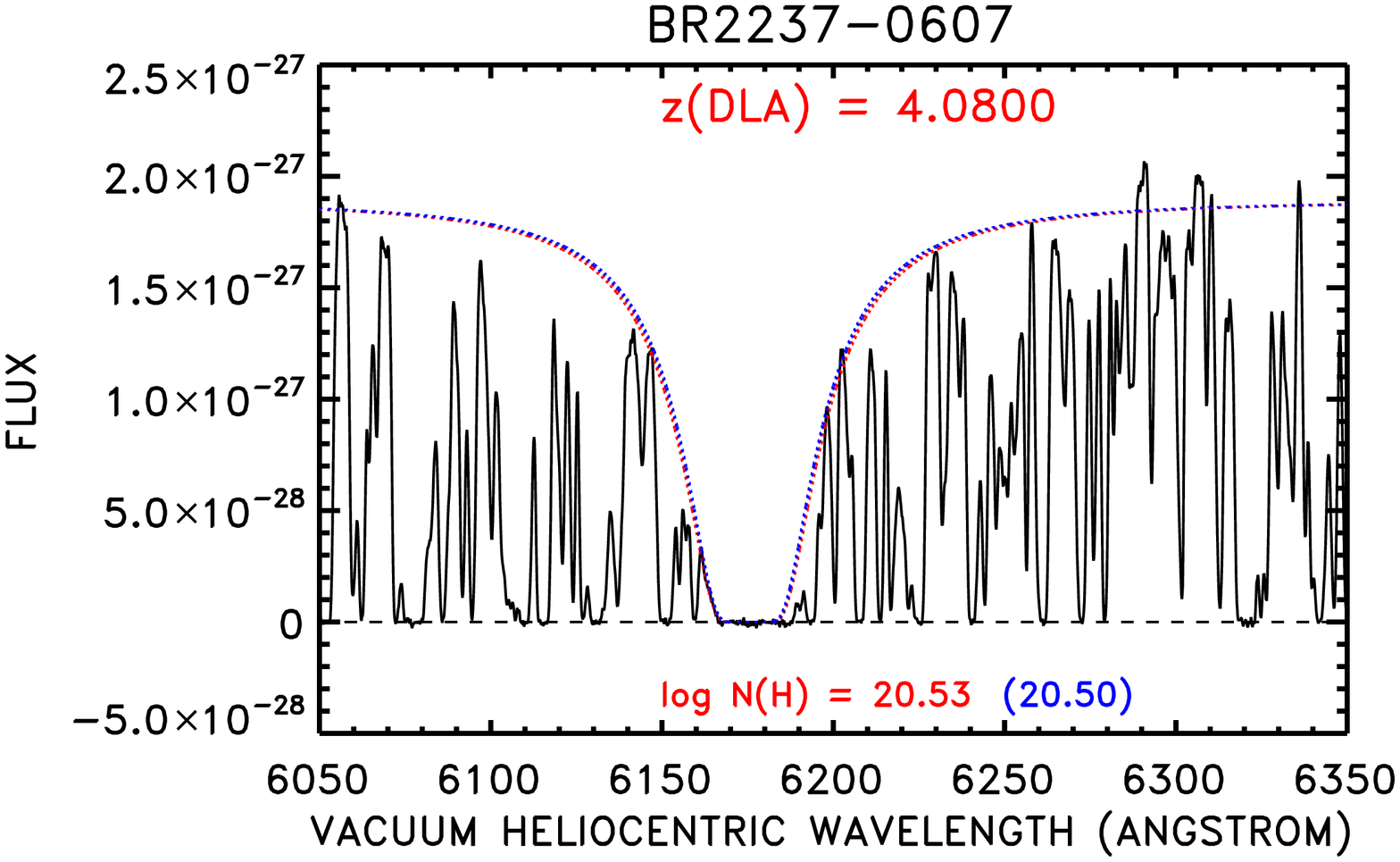}
  \includegraphics[width=8in,angle=90,scale=0.3]{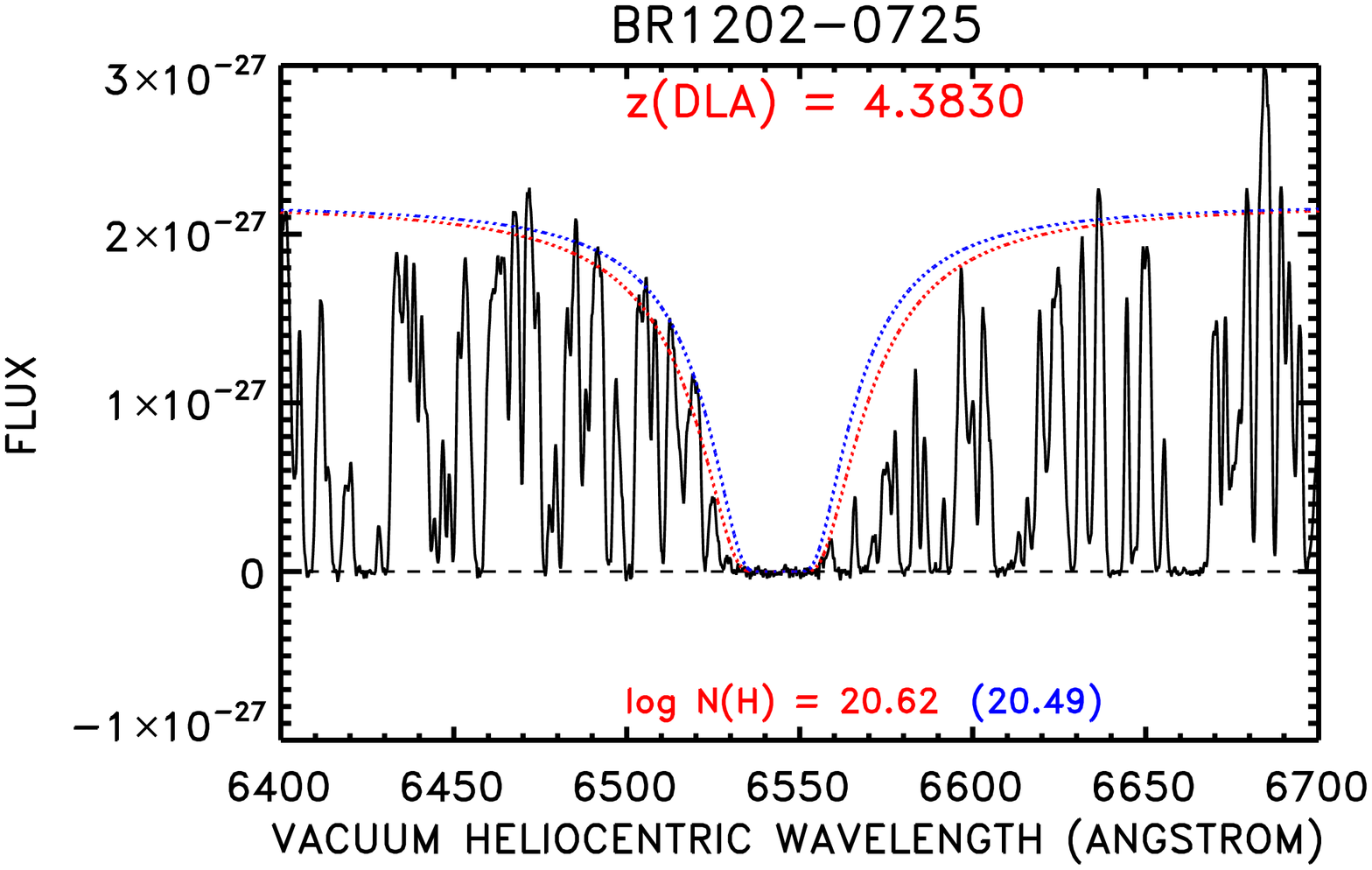}
 \includegraphics[width=8in,angle=90,scale=0.3]{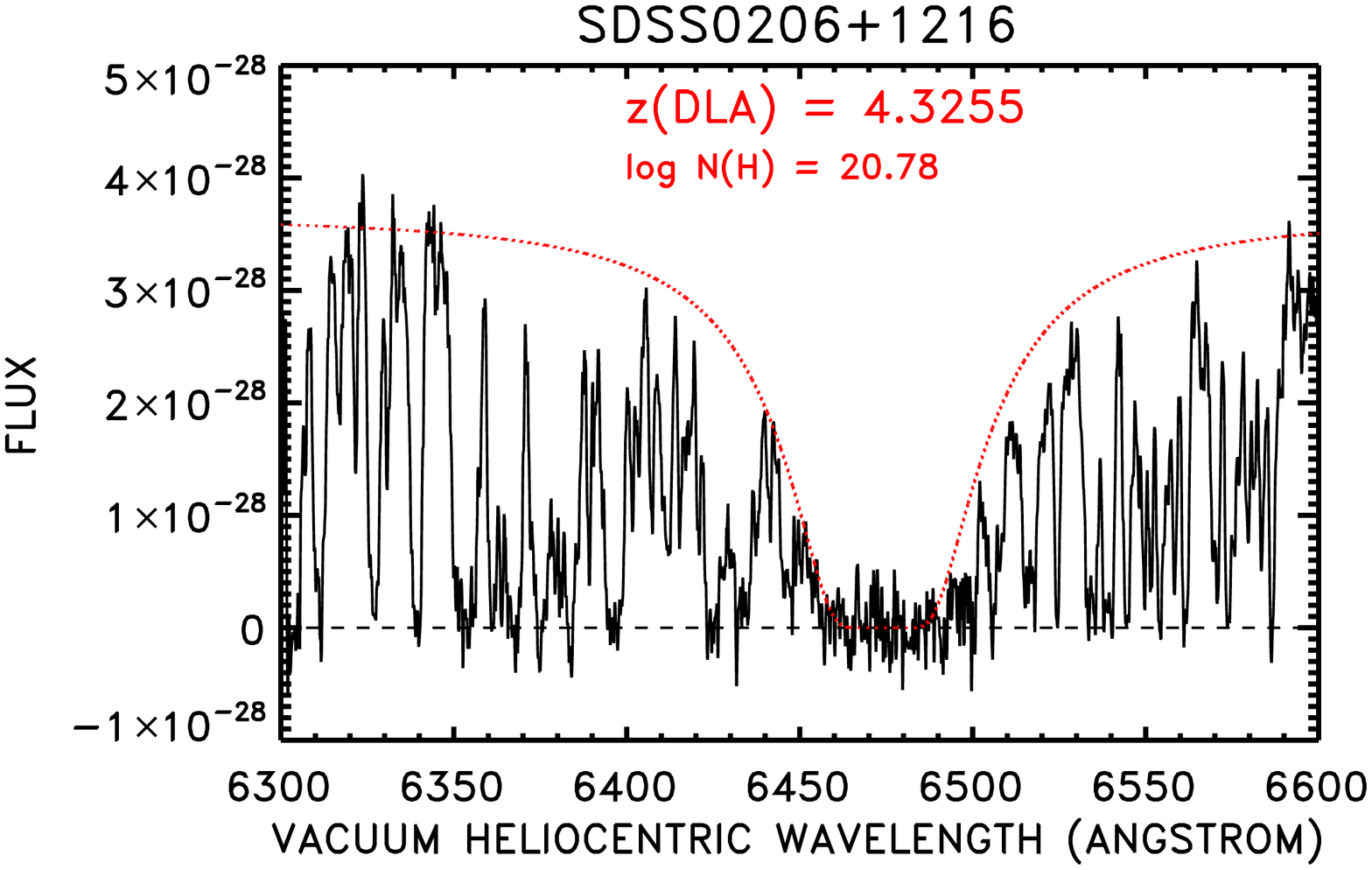}
 \includegraphics[width=8in,angle=90,scale=0.3]{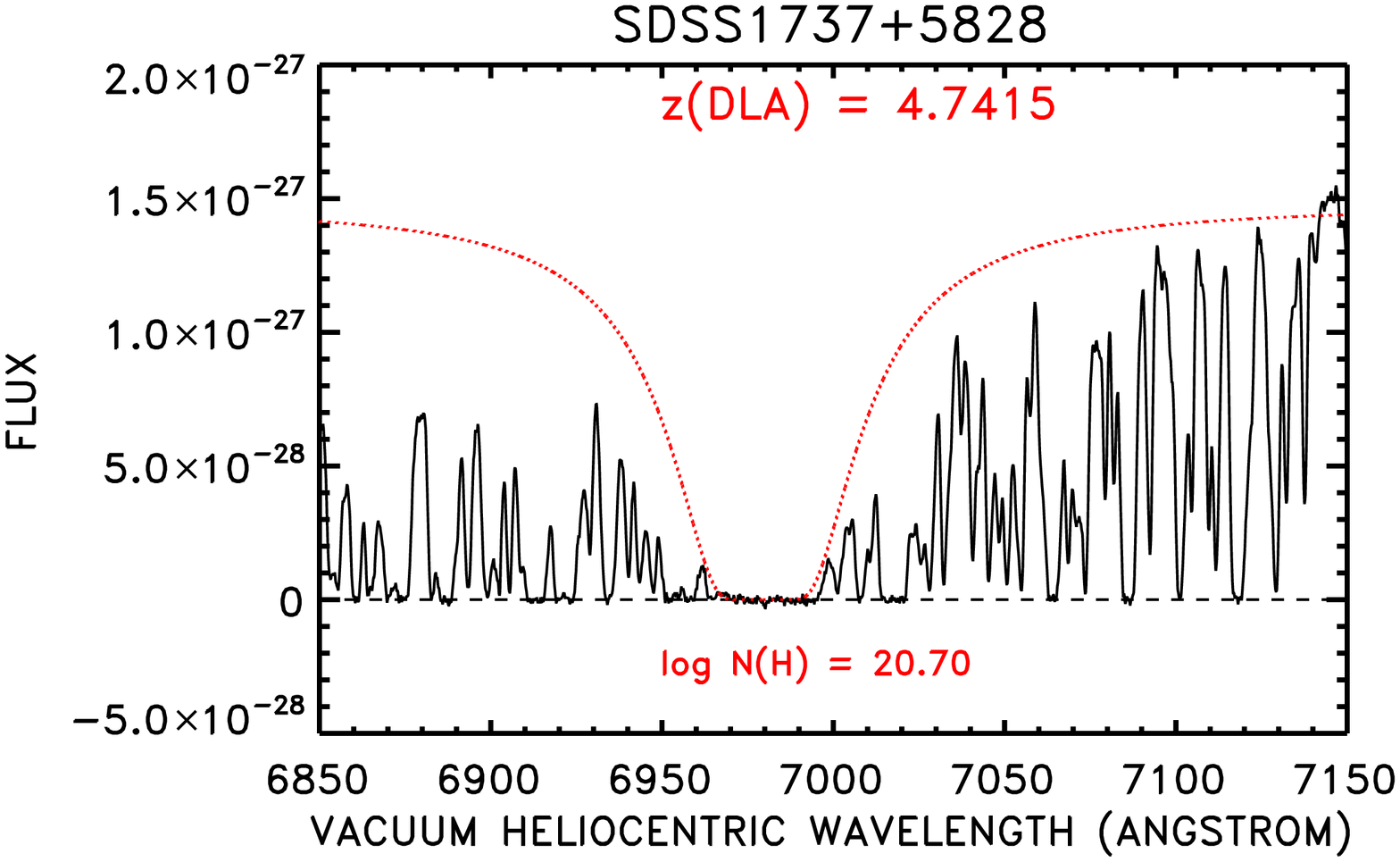}
 \includegraphics[width=8in,angle=90,scale=0.3]{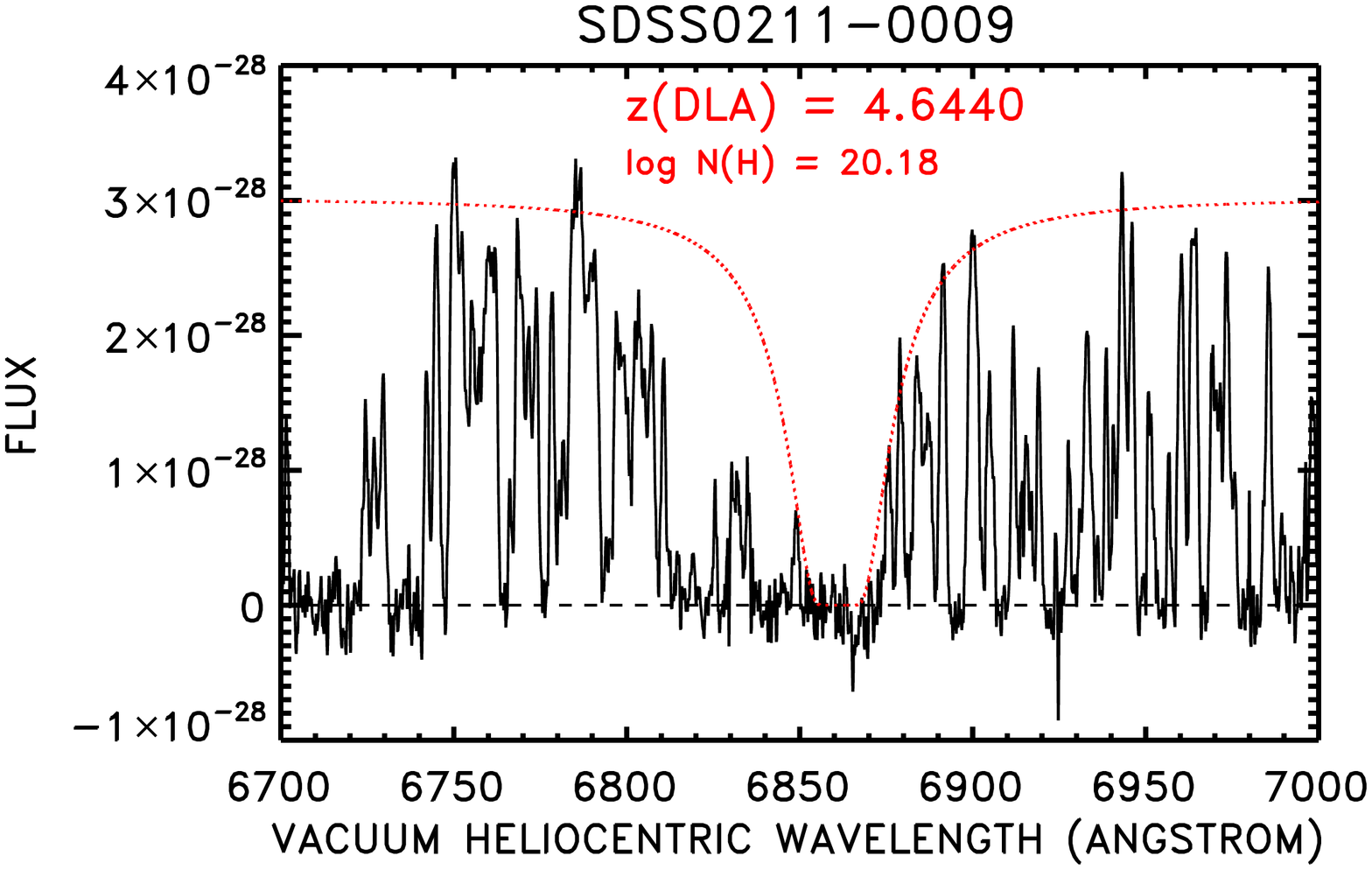}
  \caption{{\it Contd.}
\label{fig:dlas_c}
}
\end{figure*}

\end{document}